\pdfoutput=1
\RequirePackage{ifpdf}
\ifpdf 
\documentclass[pdftex]{sigma}
\else
\documentclass{sigma}
\fi

\numberwithin{equation}{section}

\newtheorem{Theorem}{Theorem}[section]

\newtheorem{Lemma}[Theorem]{Lemma}
\newtheorem{Proposition}[Theorem]{Proposition}
{ \theoremstyle{definition}
\newtheorem{Remark}[Theorem]{Remark} }

\usepackage{eucal}
\usepackage{tikz}
\usetikzlibrary{calc,decorations.pathmorphing,decorations.markings, decorations.pathreplacing,patterns,shapes,arrows}

\DeclareMathAlphabet{\mathpzc}{OT1}{pzc}{m}{it}

\newcommand{\R}{\mathbb{R}}
\newcommand{\C}{\mathbb{C}}
\newcommand{\Z}{\mathbb{Z}}
\newcommand{\N}{\mathbb{N}}
\newcommand{\E}{\mathbb{E}}
\newcommand{\D}{\mathpzc{D}}
\newcommand{\pzcd}{\mathpzc{d}}

\newcommand{\pr}{\mathbb{P}}

\newcommand{\thp}{\theta_p}

\newcommand{\tu}{\tilde{u}}

\newcommand{\gpqq}{\Gamma_{p,q,q}}

\begin{document}

\newcommand{\arXivNumber}{1110.4176}

\renewcommand{\thefootnote}{}

\renewcommand{\PaperNumber}{032}

\FirstPageHeading

\ShortArticleName{Elliptically Distributed Lozenge Tilings of a Hexagon}

\ArticleName{Elliptically Distributed Lozenge Tilings of a Hexagon\footnote{This paper is a~contribution to the Special Issue on Elliptic Hypergeometric Functions and Their Applications. The full collection is available at \href{https://www.emis.de/journals/SIGMA/EHF2017.html}{https://www.emis.de/journals/SIGMA/EHF2017.html}}}

\Author{Dan BETEA}

\AuthorNameForHeading{D.~Betea}

\Address{Paris, France}
\Email{\href{mailto:dan.betea@gmail.com}{dan.betea@gmail.com}}

\ArticleDates{Received October 27, 2017, in final form April 06, 2018; Published online April 12, 2018}

\Abstract{We present a detailed study of a four parameter family of elliptic weights on tilings of a hexagon introduced by Borodin, Gorin and Rains, generalizing some of their results. In the process, we connect the combinatorics of the model with the theory of elliptic special functions. Using canonical coordinates for the hexagon we show how the $n$-point distribution function and transitional probabilities connect to the theory of $BC_n$-symmetric multivariate elliptic special functions and of elliptic difference operators introduced by Rains. In particular, the difference operators intrinsically capture all of the combinatorics. Based on quasi-commutation relations between the elliptic difference operators, we construct certain natural measure-preserving Markov chains on such tilings which we immediately use to obtain an exact sampling algorithm for these elliptic distributions. We present some simulated random samples exhibiting interesting and probably new arctic boundary phenomena. Finally, we show that the particle process associated to such tilings is determinantal with correlation kernel given in terms of the univariate elliptic biorthogonal functions of Spiridonov and Zhedanov.}


\Classification{33E05; 60C05; 05E05}

\renewcommand{\thefootnote}{\arabic{footnote}}
\setcounter{footnote}{0}

\section{Introduction}

\looseness=-1 This paper examines work begun by Borodin, Gorin and Rains in \cite{BGR}. In op.~cit., the authors examined $q$-distributed boxed plane partitions from several perspectives, but the $q$-distributions were obtained as limits of the elliptic distribution briefly appearing in their Appendix. The present paper takes the Appendix of \cite{BGR} and expands upon it, following the steps in \cite{BG,BGR}. However, since we are working at the elliptic (hypergeometric) level (rather than a degeneration as in~\cite{BGR}), new tools are needed to generalize the results of \cite{BGR}. These tools belong to the area of elliptic special functions, an active area of research in algebra and analysis generalizing, among other things, the Askey and $q$-Askey schemes of orthogonal polynomials (as described in~\cite{KS-askey} for example). In some complementary sense, while being a generalization of~\cite{BGR}, the paper is an application of multivariate tools introduced by Rains in \cite{BCn, EHI} (the first is more analytic, the second being more algebraic). They build upon the univariate elliptic biorthogonal functions of Spiridonov and Zhedanov from a few years earlier~\cite{SZ1}. Work in the field of elliptic special functions started with Frenkel and Turaev's discovery of elliptic (theta) hypergeometric series~\cite{FT}~-- the authors of op.\ cit.\ cite Baxter's work (see for example~\cite{baxter}) as the genesis of the theory.

The history of the problem starts with random uniformly distributed boxed plane partitions. Much is known about these: asymptotics and frozen boundary behavior \cite{CKP,CLP,KO_limit}; correlation kernels via orthogonal polynomials (see~\cite{BG,Gor, joh_tilings}); exact sampling algorithms~\cite{BG}. Somewhat central to the subject is the topic of discrete Hahn orthogonal polynomials (which themselves are terminating generalized hypergeometric series). One level up and we arrive at the $q$-distributions on boxed plane partitions in~\cite{BGR} (see also~\cite{KO_limit} for the variational problem used to derive the limit shape for the $q^{\pm\text{Volume}}$ distributions). Central to this subject are certain discrete $q$-orthogonal polynomials ($q$-Racah, $q$-Hahn) from the $q$-Askey scheme, which themselves are terminating $q$-hypergeometric series (see \cite{gasper-rahman} for a full description or \cite{KS-askey} for a distillation of the results).

The present work analyzes the elliptic level. The elliptic distribution was introduced in the Appendix of \cite{BGR}, but also independently from a slightly different perspective in \cite{schlosser}. We look at two aspects: exact sampling algorithms and correlation kernels. The third aspect in \cite{BG,BGR} is obtaining asymptotics of the correlation kernel and through this obtaining the frozen boundary behavior in the large scale limit. While we indeed see a frozen boundary behavior in our case and can characterize it via variational techniques (and we present computer simulations of the results), we cannot yet analyze the asymptotics of elliptic biorthogonal functions. Techniques used in previous works~-- e.g., in~\cite{BGR} -- fail if we replace orthogonal polynomials by elliptic biorthogonal functions. More direct techniques like solving the variational problem described in~\cite{KO_limit} for the $q$-Hahn case and in \cite[Section~2.4]{BGR} seem computationally intractable so far. The reason is the associated complex Burgers equation one has to solve becomes considerably more complicated. Nevertheless, it is a (new) feature of the elliptic model that the apparent frozen boundary can have three nodal points, as seen in the computer simulations.

From a different perspective, we try to create a bridge between elliptic special functions discussed in the references above and combinatorics of tilings of hexagons (equivalently, dimer coverings of the appropriate graph). We give a combinatorial interpretation to several objects appearing in the theory of elliptic special functions: the ($t=q$ case) multivariate elliptic difference operators discovered by Rains~\cite{EHI}, the $\Delta$-symbols of~\cite{BCn} and the (univariate) elliptic biorthogonal functions of Spiridonov and Zhedanov~\cite{SZ1}.

This paper tries to emulate the organization of \cite{BG} and \cite{BGR}, but with notation heavily influenced by~\cite{BCn}. It is organized as follows: in the remainder of the Introduction, we set up most of the important notation and terminology.
We set up the combinatorial and probabilistic aspects in Section~\ref{sec:model}. We study positivity of our a priori complex measure and introduce various coordinate systems used throughout the paper, including the important canonical coordinates which embed our model in a certain square of an elliptic curve.
In Section~\ref{sec:dist} we compute relevant distributions and transition probabilities. Sections~\ref{sec:model} and~\ref{sec:dist} are an in depth expansion of the Appendix in~\cite{BGR}.

Section~\ref{sec:diff_op} recalls some definitions and properties of elliptic tools introduced by Rains \cite{BCn, EHI} (we refer the reader to these works for the proofs we omit) and then connects these with the probability and combinatorics being studied. We show that the constraints of the model are intrinsically captured by the elliptic difference operators under discussion.

Section~\ref{sec:algo} describes a perfect sampling algorithm for such elliptically distributed boxed plane partitions. It is based on the idea of forming a new measure preserving Markov chain out of two old quasi-commuting ones (as in \cite{BF}; see also \cite{DF}). The algorithm starts from a deterministic parallelogram shape and samples relatively easy distributions to successively transform the parallelogram into a hexagon accordingly distributed by increasing one side by one, and decreasing another by one; a parallelogram can be seen as a hexagon with two sides of length zero. We use the quasi-commutation relations for the elliptic difference operators of Section~\ref{sec:diff_op} to construct this algorithm.

Section~\ref{sec:corr} deals with correlations in the model. We start by recalling facts about univariate elliptic biorthogonal functions and show that the time increasing (decreasing) Markov process is determinantal, with correlation kernel given as a determinant of elliptic biorthogonal functions. These replace the orthogonal polynomials discussed above.

We end with two appendices. Appendix~\ref{app:sym_wts} provides a highly symmetric view of the entire picture. In Appendix~\ref{app:sim} we present some computer simulations obtained from the algorithm described in Section~\ref{sec:algo}.

For the remainder of the section, we will set the notation that will appear in the rest of the paper. We define the theta function and elliptic gamma function~\cite{Ruij} as follows
\begin{gather*}
\thp(x) = \prod_{k \geq 0} \big(1-p^k x\big) \big(1-p^{k+1}/x\big), \qquad \Gamma_{p,q}(x) = \prod_{k,l \geq 0} \frac{1-p^{k+1}q^{l+1}/x}{1-p^{k}q^{l} x}.
\end{gather*}

Note the elliptic gamma function is symmetric in $p$ and $q$. The theta-Pochhammer symbol (a generalization of the $q$-Pochhammer symbol) is defined, for $m \ge 0$, as
\begin{gather*}
 \thp(x;q)_m = \prod_{0 \leq i < m} \thp\big(q^i x\big).
\end{gather*}
As is customary in this area, presence of multiple arguments before the semicolon (inside theta or elliptic gamma functions) will mean multiplication. To wit
\begin{gather*}
 \thp\big(u z^{\pm 1};q\big)_m = \thp(uz;q)_m \thp(u/z;q)_m, \qquad \Gamma_{p,q}(a,b) = \Gamma_{p,q}(a) \Gamma_{p,q}(b).
\end{gather*}

We have the following important, if simple, identities (for $n \geq 0$ an integer)
\begin{gather}
 \thp(x) = \thp(p/x), \qquad \thp(px) = \thp(1/x) = -(1/x) \thp(x), \nonumber\\ \Gamma_{p,q}\big(q^n x\big) = \thp(x;q)_n \Gamma_{p,q}(x).\label{thetaid}
\end{gather}

The last identity in \eqref{thetaid} can be extended for $n<0$ or even for non integer $n$ to provide a generalization of the theta-Pochhammer symbol for negative or even non-integer lengths. Theta-Pochhammer symbols satisfy various simple identities, a few of which we list and later use without explicit mention
\begin{gather*}
\thp(a;q)_{n+k} = \thp(a;q)_n \thp\big(a q^n;q\big)_k, \\
\thp(a;q)_n = \thp\big(q^{1-n}/a;q\big)_n (-a)^n q^{\binom{n}{2}}, \\
\thp(a;q)_{n-k} = \frac{\thp(a;q)_n}{\thp\big(q^{1-n}/a;q\big)_k} \left(-\frac{q}{a}\right)^k q^{\binom{k}{2}-n k}, \\
\thp\big(a q^{-n};q\big)_k = \frac{\thp(a;q)_k \thp(q/a;q)_n}{\thp\big(q^{1-k}/a;q\big)_n} q^{-n k}, \\
\thp(a;q)_{-n} = \frac{1} {\thp\big(a q^{-n};q\big)_n} = \frac{1}{\thp(q/a;q)_n} \left(-\frac{q}{a}\right)^n q^{\binom{n}{2}}, \\
\thp\big(a q^n;q\big)_k = \frac{\thp(a;q)_k \thp\big(a q^k;q\big)_n }{\thp(a;q)_n} = \frac{\thp(a;q)_{n+k}}{\thp(a;q)_{n}}.
\end{gather*}

If $f(x_1,\dots,x_n)$ is a function of $n$ variables defined on $(\C^*)^n$, we call it $BC_n$-\textit{symmetric} if it is symmetric (does not change under permutation of the variables) and invariant under $x_k \to 1/x_k$ for all~$k$. It is called a $BC_n$-\textit{symmetric theta function of degree $m$} if in addition, it satisfies the following
\begin{gather*}
 f(p x_1, \dots, x_n) = \left(\frac{1}{p x_1^2}\right)^m f(x_1, \dots, x_n).
\end{gather*}
The prototypical example of a $BC_n$-symmetric theta function of degree one is $\prod\limits_{1 \leq k \leq n} \thp\big(u x_k^{\pm 1}\big)$.

The function $\varphi(z,w) = z^{-1} \thp(zw, z/w)$ plays an important role. It is \textit{$BC_2$-skewsymmetric} (symmetric under reciprocation, skewsymmetric under permutation: $\varphi(z,w) = -\varphi(w,x)$) of degree one. The Weierstrass addition formula for theta functions
\begin{gather*}
 \thp\big(x w^{\pm 1}\big) \thp\big(y z^{\pm 1}\big) - \thp\big(x z^{\pm 1}\big) \thp\big(y w^{\pm 1}\big) = yw^{-1} \thp\big(x y^{\pm 1}\big) \thp\big(w z^{\pm 1}\big)
\end{gather*}
has as consequence that
\begin{gather*}
 \varphi(x,y) = \left ( \frac{\varphi(z,x)}{\varphi(w,x)} - \frac{\varphi(z,y)}{\varphi(w,y)} \right ) \frac{\varphi(w,x) \varphi(w,y)}{\varphi(z,w)}
\end{gather*}
for arbitrary $z$, $w$. We note the expression in parentheses appearing above is a Vandermonde-like factor in transcendental coordinates $X = \frac{\varphi(z,x)}{\varphi(w,x)}$, $Y = \frac{\varphi(z,y)}{\varphi(w,y)}$, so $\varphi(z_k,z_l)$ is an ``elliptic analogue'' of the (Vandermonde) difference $z_k-z_l$. This is indeed the case if one takes the right limit
\begin{gather*}
 \lim_{q \to 1} \frac{\lim\limits_{p \to 0} \varphi\big(i q^{x_k},i q^{x_l}\big)}{i\big(q-q^{-1}\big)} = x_k-x_l.
\end{gather*}

Notationally, for a function $f$ of $n$ variables, we will use the abbreviation $f(\dots x_k \dots)$ to stand for $f(x_1,\dots,x_n)$.

We will make reference to the delta symbols defined in \cite{BCn,EHI} (we are in the case $t=q$ in the notation of both references). We fix $\lambda \in m^n$ a partition (that is, a partition with at most~$n$ parts all bounded by~$m$). Define the partition $2 \lambda^2$ by $\big(2 \lambda^2\big)_i = 2$ $(\lambda_{\lceil i/2 \rceil})$. Then
\begin{gather*}
 \mathcal{C}^0_{\lambda}(x;q) = \prod_{1 \leq i} \thp\big(q^{1-i}x;q\big)_{\lambda_i}, \\
 \mathcal{C}^+_{\lambda}(x;q) = \prod_{1 \leq i \leq j} \frac{\thp\big(q^{2-i-j}x;q\big)_{\lambda_i + \lambda_j}}{\thp\big(q^{2-i-j}x;q\big)_{\lambda_i + \lambda_{j+1}}} = \prod_{i<j} \frac{\thp\big(q^{2-i-j}x\big)}{\thp\big(q^{2-i-j+\lambda_i+\lambda_j}x\big)} \prod_{1 \leq i} \frac{\thp\big(q^{2-2i}x;q\big)_{2 \lambda_i}}{\thp\big(q^{2-i-n}x;q\big)_{\lambda_i}}, \\
 \mathcal{C}^-_{\lambda}(x;q) = \prod_{1 \leq i \leq j} \frac{\thp\big(q^{j-i}x;q\big)_{\lambda_i - \lambda_{j+1}}}{\thp\big(q^{j-i}x;q\big)_{\lambda_i - \lambda_{j}}} = \prod_{i<j} \frac{\thp\big(q^{j-i-1}x\big)}{\thp\big(q^{j-i+\lambda_i-\lambda_j-1}x\big)} \prod_{1 \leq i} \thp\big(q^{n-i}x;q\big)_{\lambda_i}, \\
 \Delta_{\lambda}(a\,|\,\dots b_i\dots ;q)= \frac{\mathcal{C}^{0}(\dots b_i\dots ;q)}{\mathcal{C}^{0}\big(\dots \frac{pqa}{b_i}\dots ;q\big)} \cdot \frac{\mathcal{C}^0_{2 \lambda^2}(pqa;q)}{\mathcal{C}^-_{\lambda}(pq,q;q)\mathcal{C}^+_{\lambda}(pa,a;q)}.
\end{gather*}

Of interest will be the $\Delta$-symbol with six parameters $t_0$, $t_1$, $t_2$, $t_3$, $u_0$, $u_1$ satisfying the balancing condition $q^{2n-2} t_0 t_1 t_2 t_3 u_0 u_1 = q$. Because the usual balancing condition has $pq$ on the right-hand side (the reader should consult the Appendix of~\cite{EHI} for more on why this is necessary), we multiply $u_1$ by $p$ (this choice is arbitrary, so a priori some symmetry is broken, but this will not affect our results). We define the \textit{discrete elliptic Selberg density} as
\begin{gather}
 \Delta_{\lambda}\big(q^{2n-2} t_0^2\,|\,q^n,q^{n-1} t_0 t_1,q^{n-1} t_0 t_2,q^{n-1} t_0 t_3,q^{n-1} t_0 u_0,q^{n-1} t_0 (pu_1);q\big) \nonumber\\
 \qquad{} = {\rm const} \cdot \prod_{i<j} (\varphi(z_i,z_j))^2 \prod_{1 \leq i} q^{l_i(2n-1)} \thp\big(z_i^2\big) \frac{\thp\big(t_0^2,t_0 t_1,t_0 t_2,t_0 t_3,t_0 u_0,t_0 u_1;q\big)_{l_i}}{\thp\big(q,q\frac{t_0}{t_1},q\frac{t_0}{t_2},q\frac{t_0}{t_3},q\frac{t_0}{u_0},q\frac{t_0}{u_1};q\big)_{l_i}} \nonumber\\
 \qquad {}= {\rm const}' \cdot \prod_{i<j}(\varphi(z_i,z_j))^2 \cdot \prod_{i} z_i^{2n-1} \thp\big(z_i^2\big) \frac{\Gamma_{p,q}(t_0 z_i,t_1 z_i,t_2 z_i,t_3 z_i,u_0 z_i,u_1 z_i)}{\Gamma_{p,q}\big(\frac{q}{t_0} z_i,\frac{q}{t_1} z_i,\frac{q}{t_2} z_i,\frac{q}{t_3} z_i,\frac{q}{u_0} z_i,\frac{q}{u_1} z_i\big)}, \label{discrete-selberg}
\end{gather}
where $l_i = n-i+\lambda_i$, $z_i = q^{l_i} t_0$ and the constants are independent of $\lambda$ and present to make the $\Delta$-symbol elliptic in all of its arguments. Their values are explicit~\cite{BCn}. This discrete elliptic Selberg density is the weight function for the discrete elliptic multivariate biorthogonal functions defined in~\cite{BCn}.

We will denote by $\E$ the elliptic curve $\C^* / \langle p \rangle$ for some complex $|p|<1$. An elliptic func\-tion~$f$ (of one variable) will just be a function defined on $\E$ (that is, $f(px) = f(x)$).

Throughout the remainder, constants (by which we mean factors independent of the variables usually denoted by $x_k$, $y_k$, $z_k$) will largely be ignored and we will write ${\rm const}$ wherever this appears; they are there to make measures into probability measures (i.e., normalizing factors) or to make certain functions elliptic (i.e., invariant under $p$-shifts). Their values can often be recovered, and we comment on how to recover them whenever possible.

Finally, throughout this paper we will freely use two different systems of coordinates for our model, related by a simple affine transformation as can be seen in the next section. While this may seem redundant, coordinatizing in two different ways will more aptly reveal different features of the elliptic special functions and difference operators under study.

\section{The model} \label{sec:model}

\subsection{Interpretations} \label{interpretations}

\begin{figure}[t]\centering
 \includegraphics[scale=0.30]{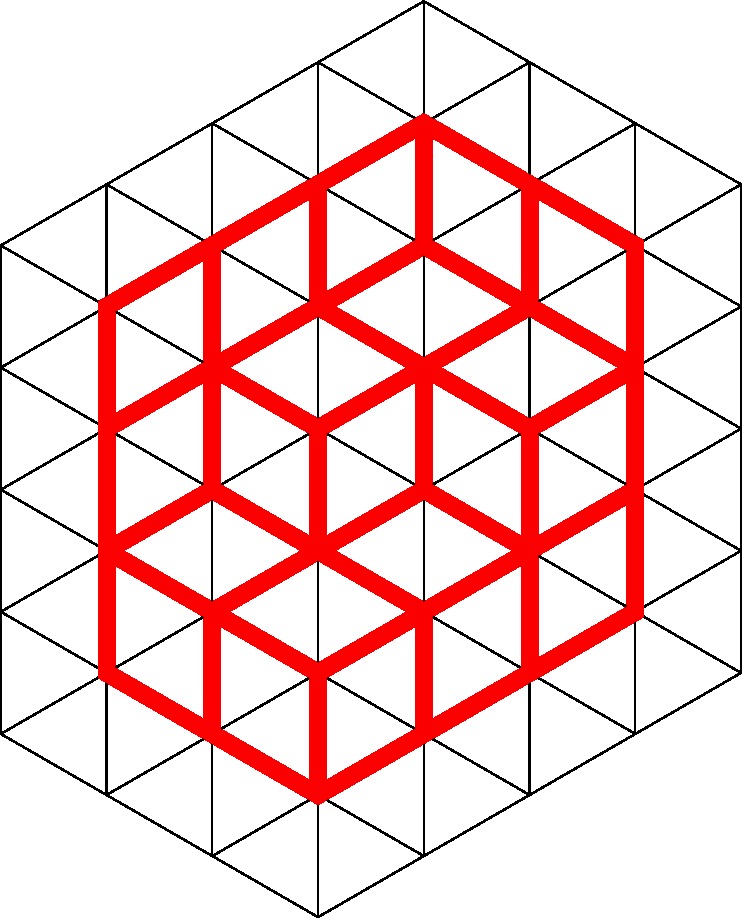}
\qquad
 \includegraphics[scale=0.25]{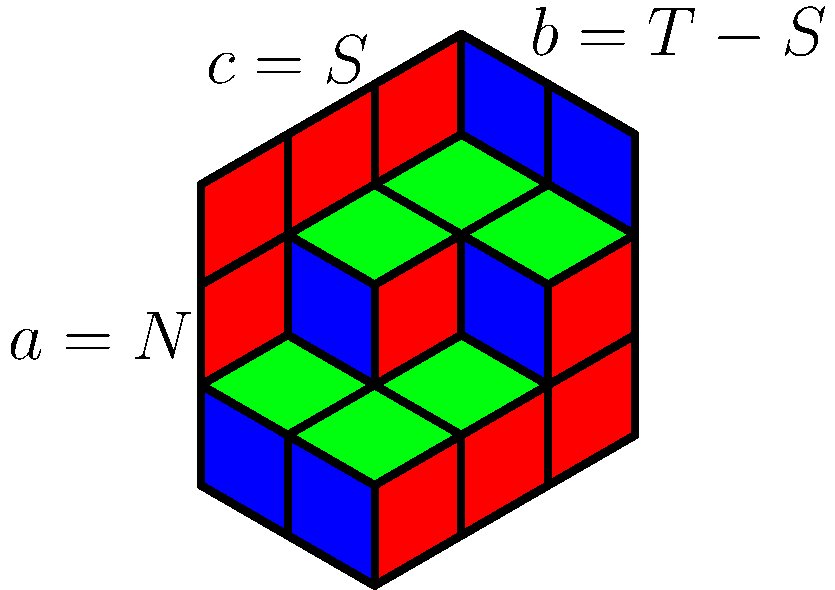}
\caption{A tiling of a $3 \times 2 \times 3$ hexagon and the associated stepped surface.}\label{tiling}
\end{figure}

We consider random tilings of an $a \times b \times c$ regular hexagon embedded in the triangular lattice (with Cartesian coordinates $(i,j)$) by tiles of three types, as can be seen in the Fig.~\ref{tiling}. The probabilistic details are set out in Section~\ref{prob_model}. We will find it more convenient to encode the hexagon via the following three numbers
\begin{gather*}
 N=a, \qquad T=b+c, \qquad S=c.
\end{gather*}

Equivalently, these tilings can be thought of as dimer matchings on the dual honeycomb lattice (every rhombus in a tiling is a line matching two vertices in the dual lattice), stepped surfaces, boxed plane partitions ($b \times c$ rectangles with positive integers $\leq a $ filled in that decrease weakly along rows and columns starting from the top left corner box) or 3D Young diagrams.

\begin{figure}[t]\centering
\includegraphics[scale=0.20]{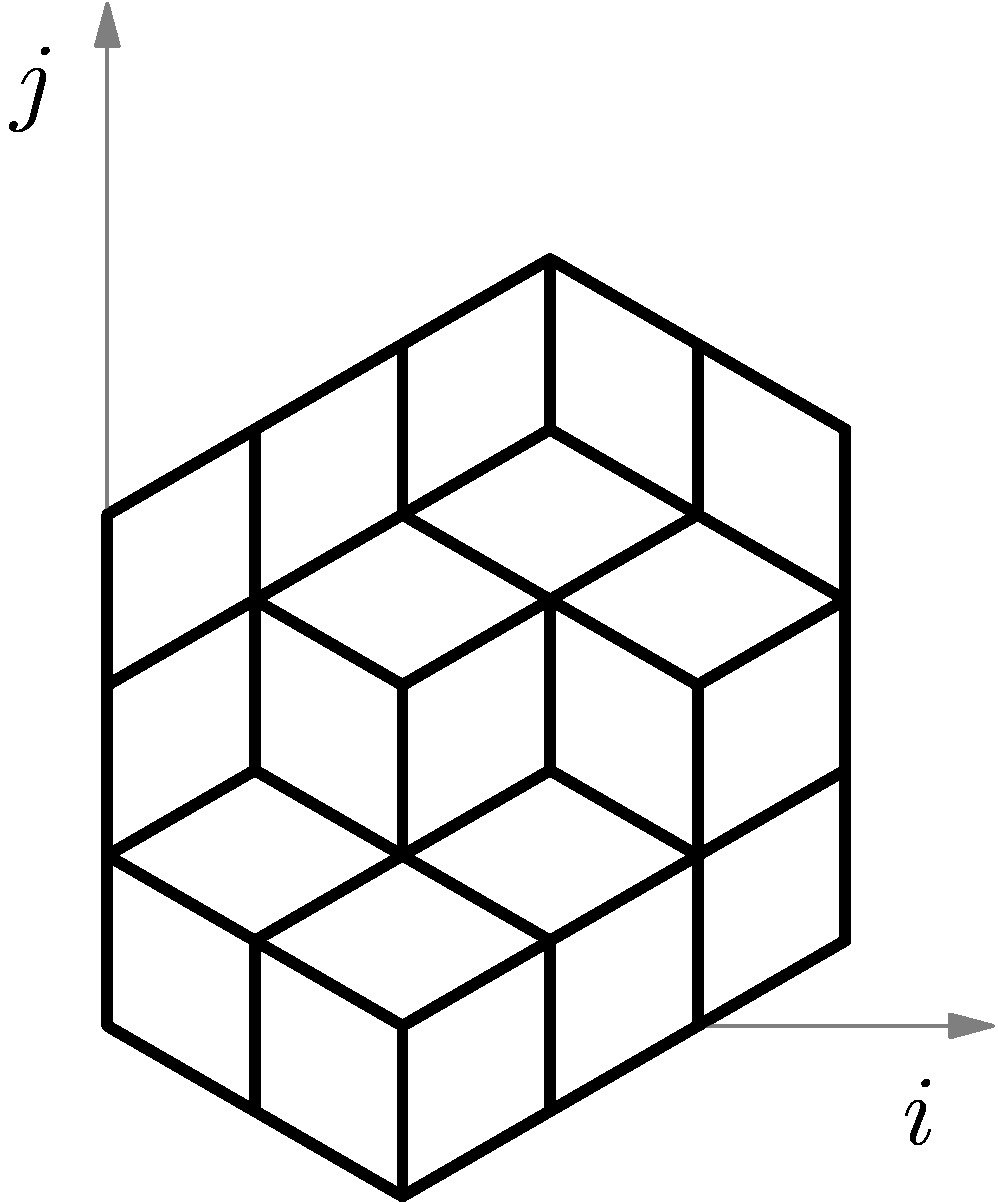}
\includegraphics[scale=0.20]{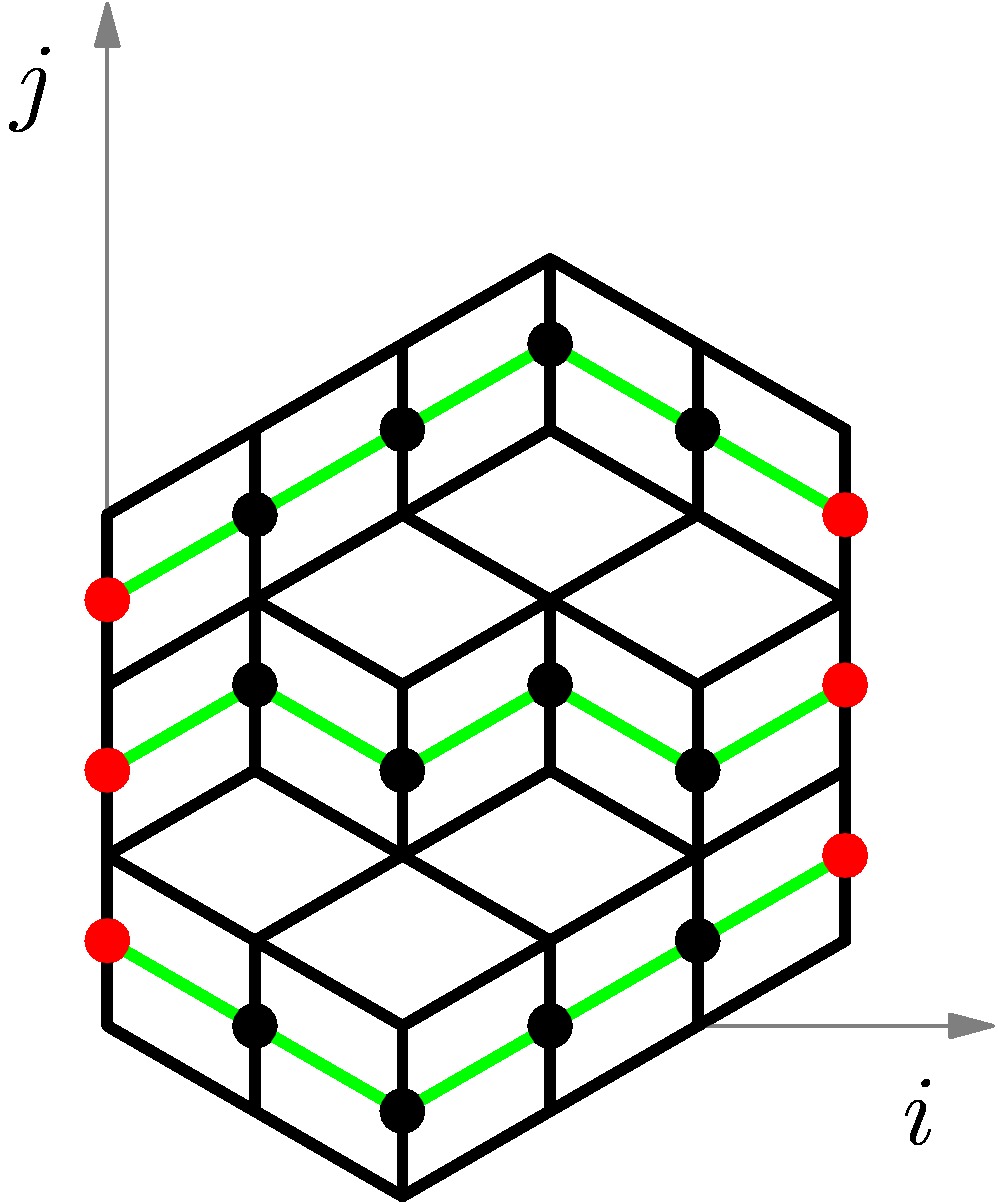}
\includegraphics[scale=0.20]{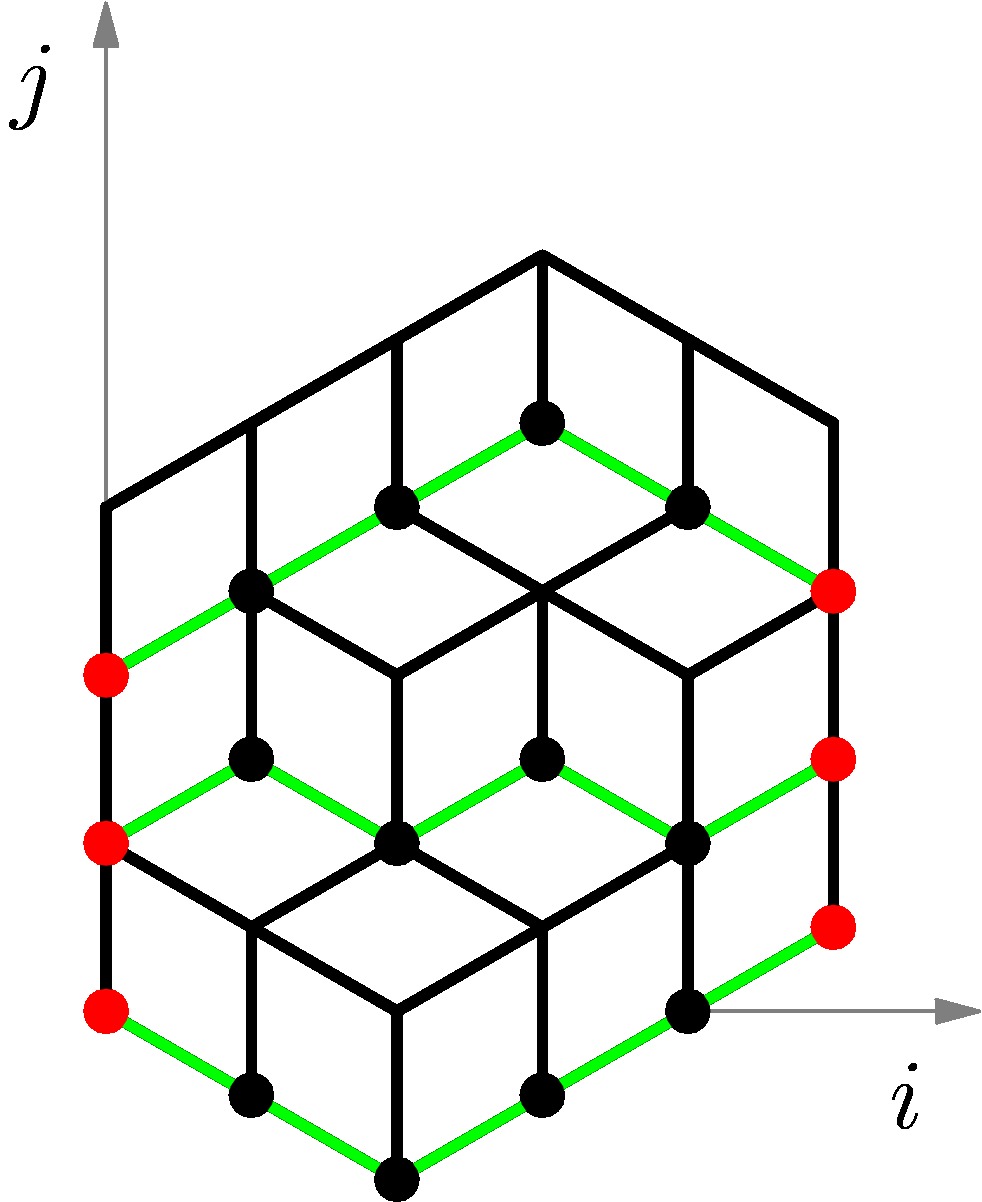}
\includegraphics[scale=0.20]{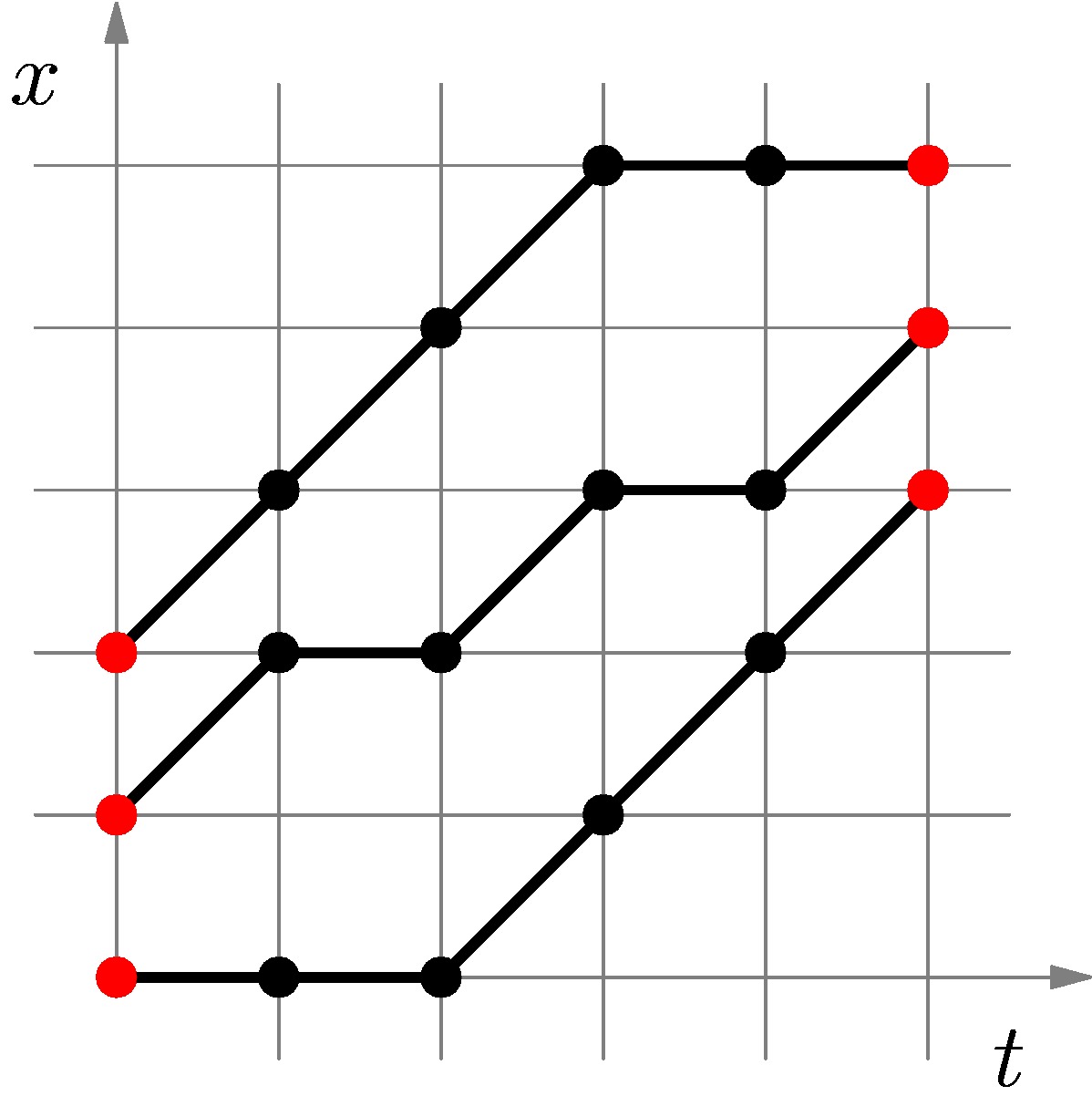}
\caption{Duality between tilings and non-intersecting paths.}\label{coordinates}
\end{figure}

A yet different way of viewing such tilings, important hereinafter, is as collections of non-intersecting paths in the square lattice. The paths start at $N$ consecutive points on the vertical axis (counting from the origin upwards) and end at $N$ consecutive points on the vertical line with coordinate $T$. Each path is composed of horizontal segments or diagonal (Southwest to Northeast, slope one) segments, and the paths are required not to intersect. Fig.~\ref{coordinates} explains this, and also introduces the coordinate frame $(t,x)$ that will be used for computational convenience in various sections to follow
\begin{gather*}
 (i,j) = (t, x-t/2).
\end{gather*}

Following the notation in~\cite{BG}, let $\Omega(N,S,T)$ denote the set of $N$ non-intersecting paths in the lattice~$\N^2$ starting from positions $(0,0),\dots ,(0,N-1)$ and ending at positions $(T,S),\dots ,(T,S+N-1)$. Each path has segments of slope zero or one (paths go either horizontally or diagonally upwards from left to right). Set
\begin{gather*}
\mathfrak{X}^{S,t}_{N,T} = \{ x \in \Z\colon \max(0,t+S-T) \leq x \leq \min(t+N-1,S+N-1) \}, \\
\mathpzc{X}^{S,t}_{N,T} = \big\{ X = (x_1,\dots ,x_N) \in \big(\mathfrak{X}^{S,t}_{N,T}\big)^N \colon x_1<x_2<\dots <x_N \big\}.
\end{gather*}
$\mathfrak{X}^{S,t}_{N,T}$ is the set of all possible particle positions in a vertical section of our hexagon with horizontal coordinate~$t$ (in $(t,x)$ coordinates). $\mathpzc{X}^{S,t}_{N,T}$ is the set of all possible $N$-tuples of particles in the same vertical section.

For $X \in \Omega(N,S,T)$, we have $X = (X(t))_{0 \leq t \leq T}$ and each $X(t) \in \mathpzc{X}^{S,t}_{N,T}$. $X$ is a discrete time Markov chain as it will be shown.

\subsection{Probabilistic model} \label{prob_model}

We will now define the probability measure on $\Omega(N,S,T)$ that will be the object of study. For a tiling $\mathcal{T}$ corresponding to an $X \in \Omega(N,S,T)$ we define its weight to be
\begin{gather*}
w(\mathcal{T}) = \prod_{l \in \{\text{horizontal lozenges}\}} w(l),
\end{gather*}
where by a horizontal lozenge we mean a lozenge whose diagonals are parallel to the $i$ and $j$ axes respectively. The probability of such a tiling is
\begin{gather*}
\pr(\mathcal{T}) = \frac{w(\mathcal{T})}{\sum\limits_{\mathcal{S} \in \Omega(N,S,T)}w(\mathcal{S})}.
\end{gather*}

The weight function $w$ on horizontal lozenges $l$ is defined by
\begin{gather}
w(l) = \frac{(u_1 u_2)^{1/2} q^{j-1/2} \thp\big(q^{2j-1} u_1 u_2\big)}{\thp\big(q^{j-3i/2-1} u_1,q^{j-3i/2} u_1,q^{j+3i/2-1} u_2,q^{j+3i/2} u_2\big)} \nonumber\\
\hphantom{w(l)}{} = \frac{(v_1 v_2)^{1/2} q^{j-S/2-1/2} \thp\big(q^{2j-S-1}v_1 v_2\big)}{\thp\big(q^{j-3i/2-S-1} v_1,q^{j-3i/2-S} v_1, q^{j+3i/2-1} v_2, q^{j+3i/2} v_2\big)},\label{weight}
\end{gather}
where $(i,j)$ is the coordinate of the top vertex of the horizontal lozenge $l$, $u_1$, $u_2$, $q$, $p$ are complex parameters, $|p|<1$ and $u_1 = q^{-S} v_1$, $u_2 = v_2$ -- the reason for this break in symmetry is that it will make other formulas throughout the paper more symmetric.

\begin{Remark} Only considering weights of horizontal lozenges for a tiling of a hexagon is equivalent to considering all types of lozenges but assigning the other two types weight one. This is a~break in symmetry that can easily be fixed~-- see Appendix~\ref{app:sym_wts}. However, for the remainder of the paper we prefer this non-symmetric weight assignment system as it makes computations easier.
\end{Remark}

This weight on lozenge tilings of a hexagon was introduced in \cite{BGR} (see also \cite{schlosser} for an equivalent weight on lattice paths).

The connection with elliptic functions will now be explained. Fix a horizontal coordinate $i$, denote by $w(i,j)$ the weight of the horizontal lozenge with top vertex coordinates $(i,j)$, and observe that for two consecutive vertical positions we have, for $u_1 u_2 u_3 = 1$, the following weight ration
\begin{gather}
r(i,j) = \frac{w(i,j)}{w(i,j-1)} = \frac{q^3 \thp\big(q^{j-3i/2-1} u_1, q^{j+3i/2-1} u_2, q^{-2j-1} u_3\big)}{\thp\big(q^{j-3i/2+1} u_1, q^{j+3i/2+1} u_2, q^{-2j+1} u_3\big)}\nonumber \\
\hphantom{r(i,j)}{} = \frac{q^3 \thp\big(q^{j-3i/2-S-1} v_1, q^{j+3i/2-1} v_2, q^{-2j+S-1}/v_1 v_2\big)}{\thp\big(q^{j-3i/2-S+1} v_1, q^{j+3i/2+1} v_2, q^{-2j+S+1}/v_1 v_2\big)}. \label{wtratio}
\end{gather}

\begin{figure}[t]\centering
 \includegraphics[scale=0.20]{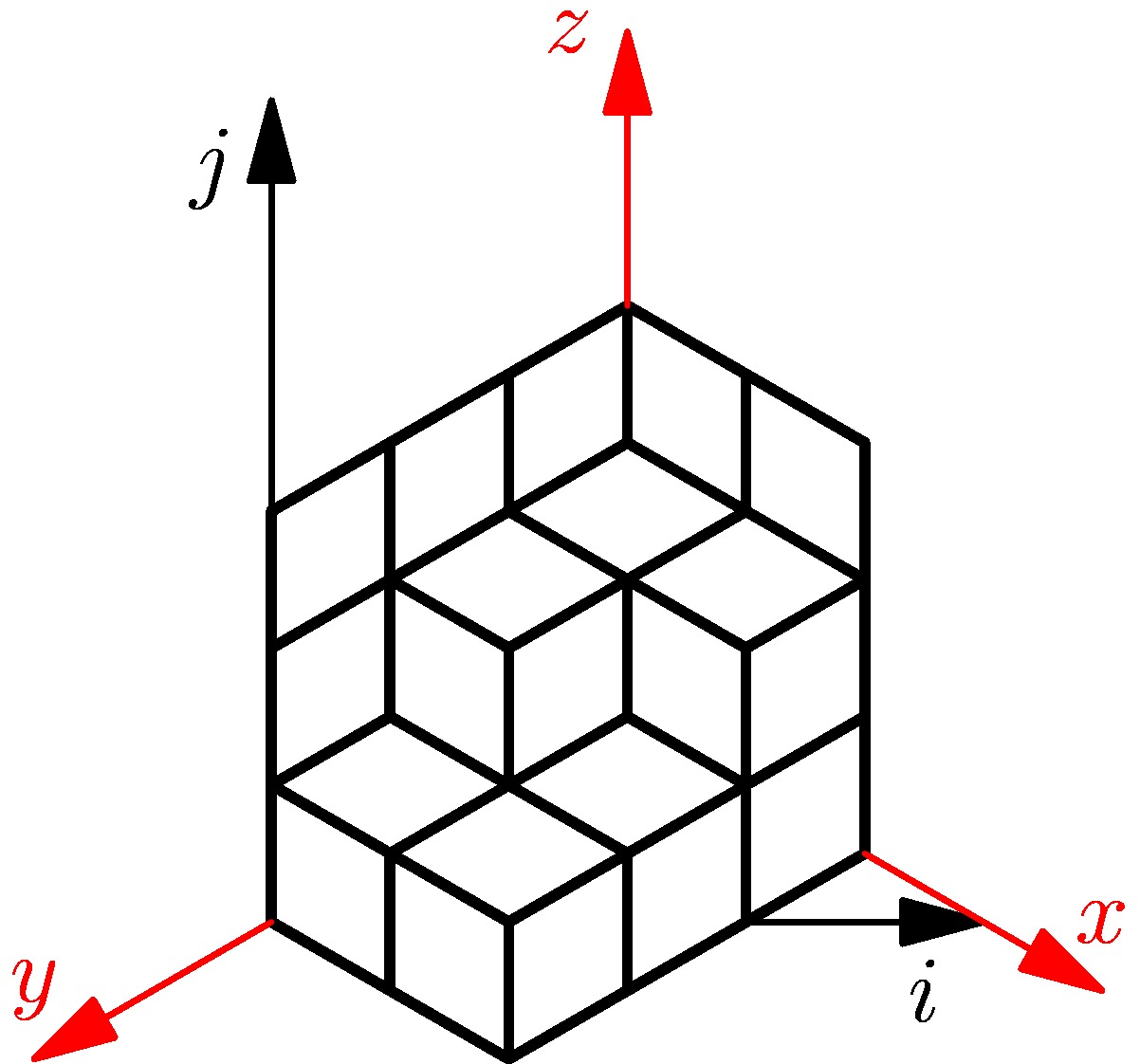}
\caption{From three dimensions to two dimensions.}\label{coord_xyz}
\end{figure}

In three-dimensional coordinates $(x,y,z)$ pictured in Fig.~\ref{coord_xyz} with $i=x-y$, $j=z-(x+y)/2$, the weight ratio looks like
\begin{gather} \label{ellipticratio}
 r(x,y,z) = \frac{w(\text{full box})}{w(\text{empty box})} = \frac{q^3 \thp(\tu_1/q, \tu_2/q, \tu_3 / q)}{\thp(\tu_1 q, \tu_2 q, \tu_3 q)},
\end{gather}
where
\begin{gather*}
 \tu_1 = q^{y+z-2x} u_1, \qquad \tu_2 = q^{x+z-2y} u_2,\qquad \tu_3 = q^{x+y-2z} u_3, \qquad u_1 u_2 u_3 = 1,
\end{gather*}
and $(x,y,z)$ is the three-dimensional centroid of the $1 \times 1 \times 1$ full cube $\includegraphics[scale=0.04]{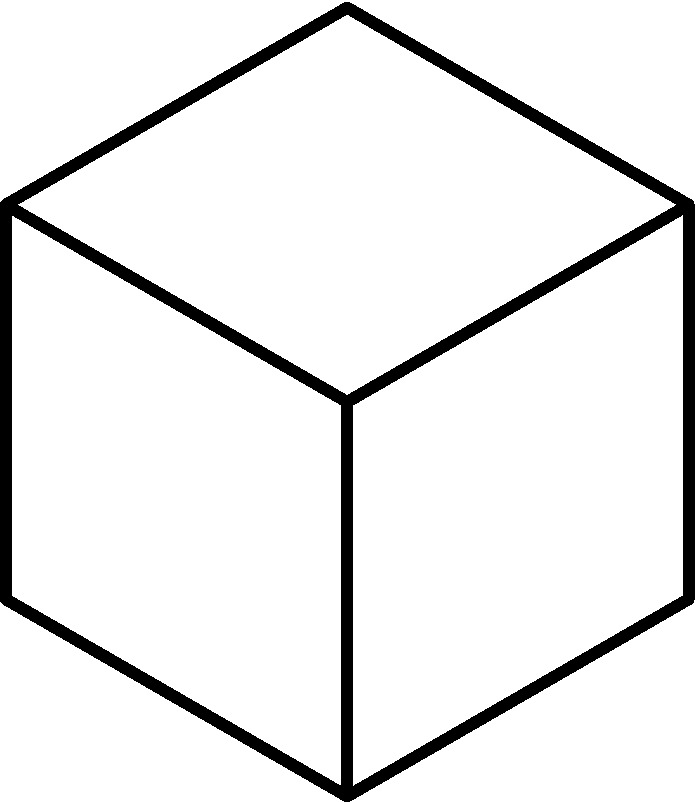}$ with top lid the horizontal lozenge with top vertex coordinate $(i,j)$.

The word \textit{elliptic} now becomes clear as $r$ in \eqref{ellipticratio} is an elliptic function of $q$. Moreover, $r$ is the unique elliptic function of $q$ with zeros at $\tu_1$, $\tu_2$, $\tu_3$ and poles at $1/\tu_1$, $1/\tu_2$, $1/\tu_3$ normalized such that $r(1)=1$. Of interest is also that $r$ is elliptic in $\tu_k$ for $k=1,2,3$ subject to the condition that $\prod\limits_{k=1}^3 \tu_k=1$.

\begin{Remark}$r$ is invariant under the natural action of $S_3$ permuting the $\tu_k$'s (and of course the three axes: $x$, $y$, $z$).
\end{Remark}

We can view our tilings as stepped surfaces composed of $1 \times 1 \times 1$ cubes bounded by the six planes $x=0$, $y=0$, $z=0$, $x=b$, $y=c$, $z=a$. Then the two-dimensional picture in Fig.~\ref{tiling} can be viewed as a projection of the three-dimensional stepped surface onto the plane $x+y+z=0$.

For $\mathcal{T}$ a tiling, we have
\begin{gather*}
 wt(\mathcal{T}) = \prod_{\includegraphics[scale=0.08]{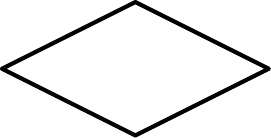} \ \in \ \mathcal{T}} w(i,j),
\end{gather*}
where $(i,j)$ are the coordinates of the top vertex of a horizontal lozenge $\includegraphics[scale=0.08]{hor_lozenge.pdf}$. Grouping all $1 \times 1 \times 1$ cubes into columns in the $z$ direction with fixed $(x,y)$ coordinates (see Fig.~\ref{coord_xyz}), we obtain
\begin{gather*}
 wt(\mathcal{T}) = {\rm const} \cdot \prod_{\includegraphics[scale=0.02]{full_box}}\frac{w(i,j)}{w(i,j-1)},
\end{gather*}
where the product is taken over all cubes (visible and hidden) of the boxed plane partition and $(i,j)$ is the top coordinate of the bounding hexagon of a $1 \times 1 \times 1$ cube. Note to get to this equality we have merely observed that $wt(\text{empty box})$ is a constant independent of $i$ and $j$. We can further refine this as
\begin{gather*}
wt(\mathcal{T}) = {\rm const} \cdot \prod_{v}\left ( \frac{w(i,j)}{w(i,j-1)} \right )^{h(v)} = {\rm const} \cdot \prod_{v} r(i,j)^{h(v)},
\end{gather*}
where $v = (x_0,y_0,z_0)$ ranges over all vertices on the border (but not on the bounding hexagon) of the stepped surface with $x_0$, $y_0$, $z_0$ integers (equivalently, $v$ ranges over all vertices of the triangular lattice inside the hexagon, but we view $v$ in three dimensions). $h(v)$~is the distance from $v$ to the plane $x+y+z=0$ divided by~$\sqrt{3}$.

\subsection{Positivity of the weight} \label{positivity}

The content of the previous subsection shows that in order to make the whole model well defined as a probabilistic model, it suffices to establish positivity of the elliptic weight ratio $r(i,j) = w(i,j)/w(i,j-1)$ defined in \eqref{wtratio}, where $(i,j)$ is the location of a given horizontal tiling and ranges over all possible horizontal tilings inside the hexagon. Recall that
\begin{gather*}
r(i,j) = \frac{q^3 \thp(\tu_1/q,\tu_2/q,\tu_3/q)}{\thp(q \tu_1,q \tu_2,q \tu_3)},
\end{gather*}
where $\tu_1 = q^{j-3i/2} u_1$, $\tu_2 = q^{j+3i/2} u_2$, $\tu_3 = q^{-2j} u_3$ and $u_1 u_2 u_3 = 1$. We recall that $r$ is elliptic in $\tu_k$ for $k=1,2,3$ as well as in~$q$. In order to make $r$ positive, we will first restrict ourselves to the case where $r$ is real valued. This means $r$ is defined over a real elliptic curve, and we have $-1 < p \ne 0 < 1$ (a priori, $p$ is complex of modulus less than~1; $p \in (-1,1)-\{0\}$ is equivalent to $\E$ being defined over~$\R$~-- for more on real elliptic curves, see of \cite[Chapter~5]{silverman}). We can then ensure positivity of $r$ by an explicit computation. We will of course have two cases: $p<0$ and $p>0$. We deal with the case $p>0$ throughout, and make remarks when necessary for $p<0$.

Now that we have restricted ourselves to real elliptic curves $\E$, we first note that $q \in \E$ (i.e., $r$~is elliptic as a function of~$q$). For a chosen $0 < p < 1$ there are two non-isomorphic elliptic curves defined over $\R$ (since $\text{Gal}(\C/\R) = \Z/2\Z$), both homeomorphic to a disjoint union of two circles (every real elliptic curve is topologically homeomorphic to a circle if $p<0$ or with a~disjoint union of two circles if $p>0$ -- one can just see this by plotting the Weierstrass equation in~$\R^2$ and compactifying)
\begin{gather*}
 \E \cong_{\R} \R^* / p^{\Z} \qquad \mathrm{and} \qquad \E \cong_{\R} \big\{ u \in \C^*/ p^{\Z} \colon |u|^2 \in \{1,p\} \big\}.
\end{gather*}

We will call the first case real and the second trigonometric (abusing terminology, since both are real elliptic curves). We will analyze the trigonometric case, but the real case is similar. In the trigonometric case, the curve has two connected components (circles): the identity component (it contains the points $1$ and $-1$) and another component that contains the other 2-torsion points: $\pm \sqrt{p}$. There will be three cases to be analyzed which we list now and motivate after (if $p<0$ there is only one component so the three cases coalesce to only one -- Case 2):
\begin{itemize}\itemsep=0pt
\item \textbf{Case 1:} $q$ lies on the non-identity component: $|q| = \sqrt{p}$;
\item \textbf{Case 2:} $q$ and all the $u_k$'s (and so all the $\tu_k$'s) lie on the identity component ($|q| = |u_1| = |u_2| = |u_3| = 1$);
\item \textbf{Case 3:} $q$ and one of the $u_k$'s lies on the identity component, the other two $u_k$'s lie on the non-identity component.
\end{itemize}

To analyze positivity at a fixed site $(i,j)$ inside the hexagon, we note that $r(q)$ has zeros at~$\tu_k$ and poles at $1 / \tu_k$ ($k=1,2,3$). We note $r=\pm 1$ at $q = \pm 1$ so at least one $u_k$ (along with its reciprocal/complex conjugate $1/u_k$) needs to be on the identity component (so that $r$ can change signs on the identity component). Since $r = -1$ at $q = \pm \sqrt{p}$ and $u_1 u_2 u_3 = 1$, either exactly one or all three of the $u$'s need to be on the identity component. This motivates the three choices above.

\begin{figure}[t]\centering
\includegraphics[scale=0.20]{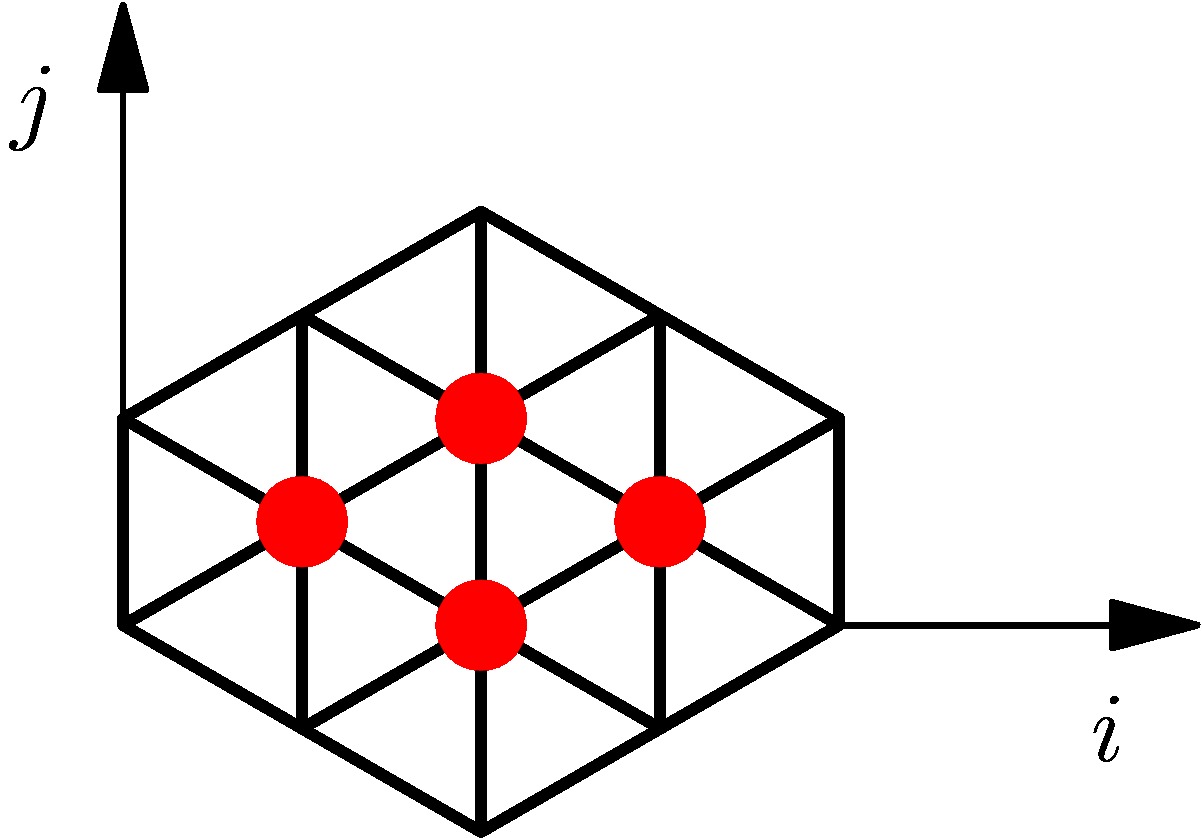}
\caption{The admissible sites $(i,j)$ inside a $1 \times 2 \times 2$ hexagon.}\label{1x2x2_spots}
\end{figure}

Case 1 will never lead to positivity for all four admissible sites $(i,j)$ inside a $1 \times 2 \times 2$ hexagon depicted in Fig.~\ref{1x2x2_spots}). It can thus be eliminated (if a $1 \times 2 \times 2$ hexagon is never positive, much larger ones which are of interest to us will also never be as they contain the $1 \times 2 \times 2$ case). For a proof, we suppose that~$u_1$ is on the identity component, and~$u_2$,~$u_3$ are, along with $q$, on the non-identity component (the case where all three $u$'s are on the identity component is handled similarly). The $\tu$'s differ from the $u$'s by integer powers of~$q$ given in the last three columns of the following table (listed are the four admissible $(i,j)$ pairs in the $1 \times 2 \times 2$ hexagon):
\begin{center}
\begin{tabular}{ccccc}
$j$ & $i$ & $j-\frac{3i}{2}$ & $j+\frac{3i}{2}$ &$-2j$ \\
$1/2$ & $1$ & $-1$ & $2$ & $-1$ \\
$1$ & $2$ & $-2$ & $4$ & $-2$ \\
$0$ & $2$ & $-3$ & $3$ & $0$ \\
$1/2$ & $3$ & $-4$ & $5$ & $-1$
\end{tabular}
\end{center}

\noindent Notice mod 2 (and we only care about mod~2 as $q^2$ is on the identity component), the four vectors (from the last three columns of the table) above are $(1, 0, 1)$, $(0, 0, 0)$, $(1, 1, 0)$, $(0, 1, 1)$. The corresponding $\tu_k$'s we get by multiplying each $u_k$ by $q$ to the power coming from the vector $(0,1,1)$, that is $(\tu_1,\tu_2,\tu_3) = \big(q^{-4} v_1,q^{5} v_2,q^{-1} v_3\big)$, will all be on the identity component, which means the elliptic weight ratio will be negative at the site $(i,j) = (1/2,3)$ as $q$ is on the non-identity component. This is a~contradiction. The other cases are handled similarly, leading to contradictions. This proves $q$ must be on the identity component, so only Cases~2 and~3 above can lead to positive hexagons.

\begin{figure}[t]\centering
\includegraphics[scale=0.20]{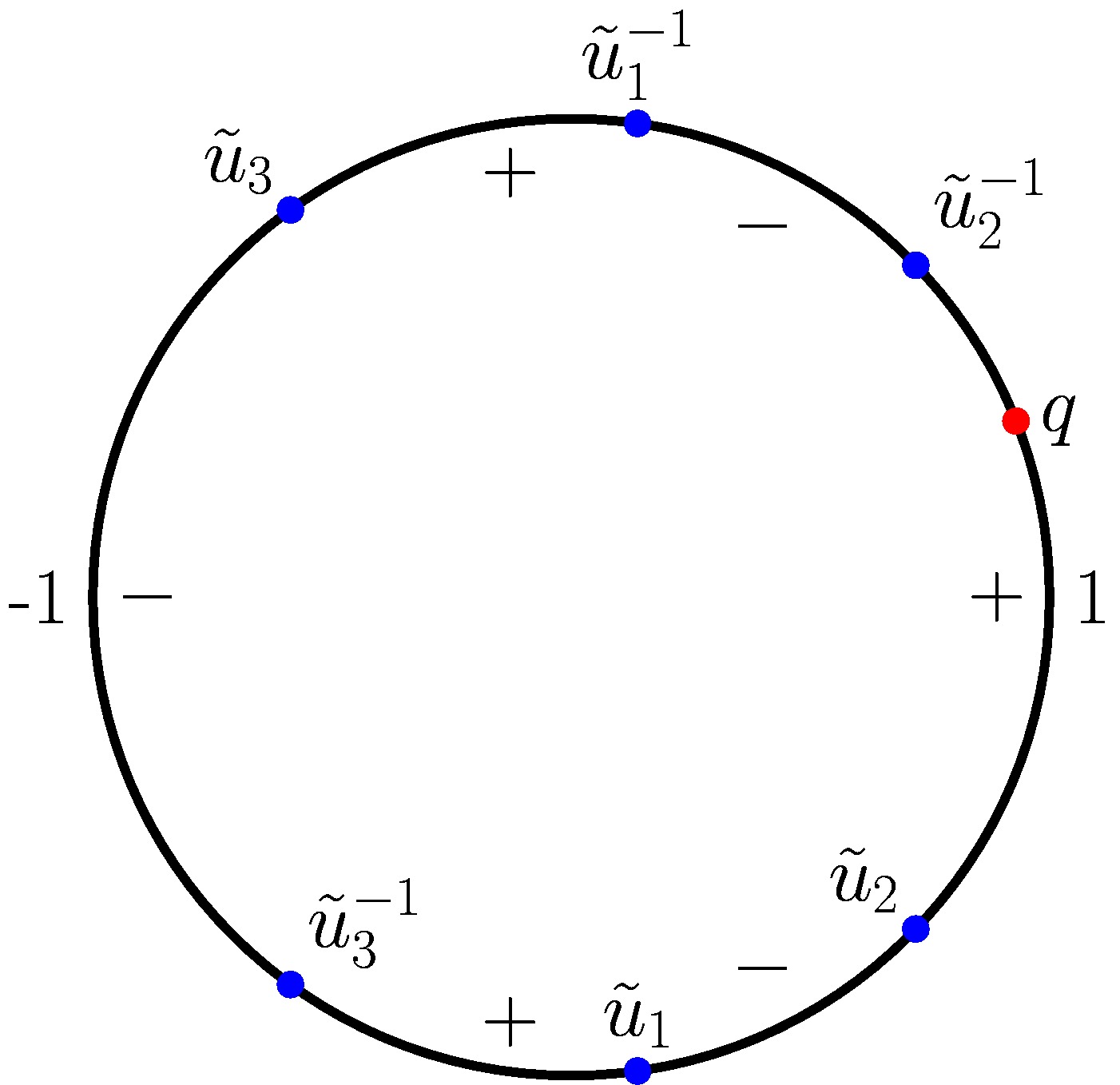}

\caption{The identity component of $\E \cong_{\R} \{ u \in \C^*/ p^{\Z} \colon |u|^2 \in \{1,p \} \}$. For positivity of $r$ throughout the hexagon (i.e., for all admissible $\tu_k$'s), $q$ must always be closer to~1 than any~$\tu_k^{\pm 1}$ as depicted.}\label{unit_circle}
\end{figure}

We will next discuss the case where $q$ and all $u_k$ are on the identity component (Case~2 above; for Case~3 the reasoning is similar). For a fixed site $(i,j)$ inside the hexagon, the three $\tu_k$'s and their reciprocals (complex conjugates) break down the unit circle into six arcs (see Fig.~\ref{unit_circle}) and~$q$ must be on one of the three arcs where $r$ is positive (as depicted in the figure). If we want to ensure positivity of the ratio for all four admissible sites $(i,j)$ within a given $1 \times 2 \times 2$ hexagon (Fig.~\ref{1x2x2_spots}), we first observe that for $|x|=1$ we have
\begin{gather*}
 \thp(x) = (1-x) \prod_{i \geq 1} \big|1-p^i x\big|^2,
\end{gather*}
so we reduce to positivity of the corresponding four functions $\prod \frac{1-\tu_i/q}{1-\tu_i q}$. Through standard trigonometric manipulations we thus want positivity of each of the following functions
\begin{gather*}
 \frac{\sin \pi (\alpha_1-\alpha)}{\sin \pi (\alpha_1+\alpha)} \cdot \frac{\sin \pi (\alpha_2-\alpha)}{\sin \pi (\alpha_2+\alpha)} \cdot \frac{\sin \pi (\alpha_3-\alpha)}{\sin \pi (\alpha_3+\alpha)}, \\
\frac{\sin \pi (\alpha_1)}{\sin \pi (\alpha_1+2 \alpha)} \cdot \frac{\sin \pi (\alpha_2-3 \alpha)}{\sin \pi (\alpha_2- \alpha)} \cdot \frac{\sin \pi (\alpha_3)}{\sin \pi (\alpha_3+2 \alpha)}, \\
\frac{\sin \pi (\alpha_1-3 \alpha)}{\sin \pi (\alpha_1-\alpha)} \cdot \frac{\sin \pi (\alpha_2)}{\sin \pi (\alpha_2+2 \alpha)} \cdot \frac{\sin \pi (\alpha_3+\alpha)}{\sin \pi (\alpha_3+3 \alpha)}, \\
\frac{\sin \pi (\alpha_1+\alpha)}{\sin \pi (\alpha_1+3\alpha)} \cdot \frac{\sin \pi (\alpha_2-2\alpha)}{\sin \pi (\alpha_2)} \cdot \frac{\sin \pi (\alpha_3-2 \alpha)}{\sin \pi (\alpha_3)},
\end{gather*}
where $2 \pi \alpha_i = \arg u_i$, $\alpha_1 + \alpha_2 + \alpha_3 \in \{ 0,1,2 \}$, $2 \pi \alpha = \arg q$ and $(\alpha, \alpha_1, \alpha_2) \in \R^3/\Z^3$. One way to make all of these positive, checked by direct calculation, is depicted in Fig.~\ref{unit_circle}. That is, as $(i,j)$ range over all four sites inside a $1 \times 2 \times 2$ hexagon, there should not be any~$\tu_k$ ($k=1,2,3$) or any~$\tu_k^{-1}$ on the arc subtended by~$1$ and $q$ not containing $-1$. Furthermore, numerical simulations in Mathematica suggest this is the only way.

\begin{Remark} In view of the above, for any reasonably large hexagon (i.e., containing a $1 \times 2 \times 2$ hexagon) and parameters $u_1$, $u_2$, $u_3$ satisfying the balancing condition $\prod u_i = 1$, the set of $q$'s giving rise to nonnegative weights is conjecturally a symmetric closed arc containing~1. This is the only case we shall consider in what follows.
\end{Remark}

\subsection{Degenerations of the weight}

Certain degenerations of the weight have been studied before (among the relevant sources for our purposes are \cite{BG,BGR,Gor,joh_tilings,KO_limit}) from many angles. For example, when $q = 1$ the weight in~\eqref{weight} becomes a constant independent of the position of the horizontal lozenges, and so we are looking at uniformly distributed tilings of the appropriate hexagon. An exact sampling algorithm to sample such a tiling was constructed in~\cite{BG} and the theory behind this is closely connected to the theory of discrete Hahn orthogonal polynomials (see \cite{BG,Gor, joh_tilings}). The frozen boundary phenomenon (the shape of a~``typical boxed plane partition'') was first proven in~\cite{CLP} and then via alternate techniques in~\cite{CKP,KO_limit}.

A more general limit than the above is the following: in \eqref{weight} we let $v_1 = v_2 = \kappa \sqrt{p}$ and then let $p \to 0$. This is the $q$-Racah limit (named so the discrete orthogonal polynomials that appear in the analysis). This limit is the most general limit that can be analyzed by orthogonal polynomials (as $q$-Racah polynomials sit atop the $q$-Askey scheme~-- see~\cite{KS-askey}). Up to gauge equivalence, we obtain the weight of a horizontal lozenge with top corner~$(i,j)$ as
\begin{gather} \label{q-racah}
 w(i,j) = \kappa q^j - \frac{1}{\kappa q^j}.
\end{gather}

This weight was studied in \cite{BGR}. Upon renormalizing, if we take $\kappa$ to 0 or $\infty$, we see the $q$-Racah weight is an interpolation between two types of weights
 \begin{gather*}
 w(i,j) = q^j \qquad \text{and} \qquad w(i,j) = q^{-j}.
 \end{gather*}
A direct alternative limit from the elliptic level is given by $v_1 = v_2 = p^{1/3}$, $p\to 0$ (and then replace $q^2$ by $q$ or $1/q$). These two weights give rise to tilings weighted proportional to~$q^{\text{Volume}}$ or~$q^{\text{-Volume}}$, where Volume = number of $1 \times 1 \times 1$ cubes in the stepped surface representing a tiling. This is the $q$-Hahn weight, as $q$-Hahn orthogonal polynomials appear in its analysis. The frozen boundary phenomenon for this type of weight was first studied in~\cite{KO_limit}, and then via alternative methods in~\cite{BGR}.

Finally, the Racah weight is the limit $q \to 1$ in \eqref{q-racah} (we denote $k = \log_q(\kappa)$ and need $\kappa \to 1$ as $q \to 1$). The weight function becomes
\begin{gather*}
 w(i,j) = k+j.
\end{gather*}

\begin{figure}[t]
 \centering
 \begin{tikzpicture}[scale=0.5]
 \node[draw] at (0, 4) {Elliptic hypergeometric (elliptic weights; elliptic biorthogonal ensembles)};
 \node at (0, 3) {$\downarrow$};
 \node[draw] at (0, 2) {$q$-hypergeometric ($q$-weights; $q$-orthogonal polynomial ensembles)};
 \node at (0, 1) {$\downarrow$};
 \node[draw] at (0, 0) {Hypergeometric (uniform/Racah weight; Hahn/Racah orthogonal polynomial ensembles)};
 \end{tikzpicture}
 \caption{A barebones schematic view of the Askey hierarchy of hypergeometric functions.} \label{fig:hyp}
\end{figure}
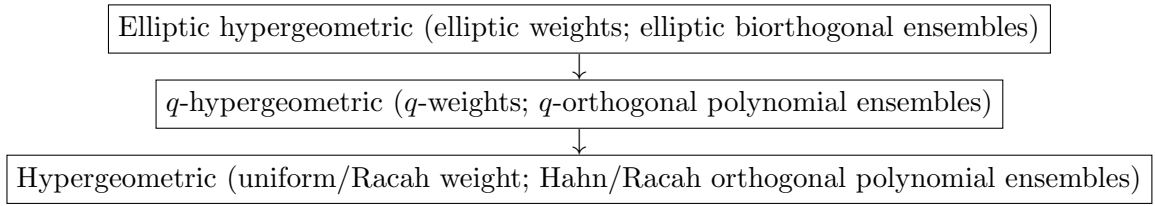

Notice in all these limits the weight of a horizontal lozenge is independent of the horizontal coordinate of its top vertex. They correspond to the hypergeometric hierarchy of special functions involved in the algebra and analysis, depicted in Fig.~\ref{fig:hyp} (down arrows are limits).

As a final sidenote, the most general degeneration of the weight is the top level trigonometric limit $p \to 0$, which gives rise to a three parameter family of weights (the use of the word \textit{trigonometric} here should not be confused with its usage in Section~\ref{positivity}). Being more general (more parameters) than the $q$-Racah limit, its analysis requires~$q$ rational biorthogonal functions rather than orthogonal polynomials. We will not use this limit hereinafter, as we can approximate the trigonometric level by choosing $p$ really small at the elliptic level.

\subsection{Canonical coordinates}

It will be convenient for various computations to express the geometry of an elliptic lozenge tiling in terms of coordinates on a certain product of elliptic curves. First we will introduce six parameters $A$, $B$, $C$, $D$, $E$, $F$ depending on $q$, $t$, $S$, $T$, $N$, $v_1$, $v_2$. Note we have listed, other than $q$, six parameters, of which four are discrete and dictate the geometry: $t$, $S$, $T$, $N$. $t$ here is a discrete time parameter and ranges from $0$ to $T$. It will be explained better in Section~\ref{sec:dist}. It corresponds to the fact that we will be interested in distributions of particles (absence of rhombi) on a certain vertical line: that is, tilings of hexagons that have prescribed positions of particles (or holes) on the vertical line with horizontal coordinate~$t$. The set of parameters is
\begin{alignat}{3}
& A=q^{t/2+S/2-T+1/2} \sqrt{v_1 v_2},\qquad && B=q^{t/2+S/2+T+1/2} \sqrt{\frac{v_2}{v_1}},&\nonumber\\
& C=q^{t/2-S/2-N+1/2} \frac{1}{\sqrt{v_1 v_2}},\qquad && D=q^{-t/2+S/2-N+1/2} \frac{1}{\sqrt{v_1 v_2}},& \nonumber\\
& E=q^{-t/2-S/2+1/2} \sqrt{\frac{v_1}{v_2}},\qquad && F=q^{-t/2-S/2+1/2} \sqrt{v_1 v_2}.& \label{parmatching}
\end{alignat}
Observe that $q^{2N-2}ABCDEF = q$.

Recall that the weight function (to be more precise, the ratio of weights of a full unit box to an empty one in \eqref{ellipticratio}) depends on the geometry of the hexagon via the three parameters $\tu_1$, $\tu_2$, $\tu_3$ ($\prod \tu_k = 1$) which in the $(i,j)$ coordinates are:
\begin{gather*}
\tu_1=q^{j-3i/2-S}v_1, \qquad \tu_2=q^{j+3i/2} v_2, \qquad \tu_3=q^{-2j+S} /v_1 v_2.
\end{gather*}

We want to change coordinates from $(i,j)$ (two-dimensional) or $(x,y,z)$ (three-dimensional) to $(\tu_1,\tu_2,\tu_3)$ via the above formulas. We call these new coordinates \textit{canonical}. In practice each line of interest in the geometry has an equation in the $(i,j)$ plane which can then be translated in terms of the $\tu_k$'s by solving in \eqref{parmatching} for $t$, $S$, $T$, $N$, $v_1$, $v_2$ in terms of $A$, $B$, $C$, $D$, $E$, $F$. We thus find the following equations for the relevant edges of our hexagon:
\begin{gather}
\text{left vertical edge (corresp.\ eq.: $i=0$)}\colon \ \frac{\tu_1}{\tu_2} = q^{-S} v_1/v_2 = \left(\frac{ABC}{DEF}\right )^{1/2} E^3 q^{-3/2}, \nonumber\\
\text{right vertical edge (corresp.\ eq.: $i=T$)}\colon \ \frac{\tu_1}{\tu_2} = q^{-3T-S} v_1/v_2 = \left(\frac{ABC}{DEF}\right )^{1/2} B^{-3} q^{3/2}, \nonumber\\
\text{NW edge (corresp.\ eq.: $j=i/2+N$)}\colon \ \frac{\tu_3}{\tu_1} = q^{2S-3N} 1/v_1^2 v_2 = \left(\frac{ABC}{DEF}\right )^{1/2} D^3 q^{-3/2}, \nonumber\\
\text{SE \ edge \ (corresp.\ eq.: $j=i/2-(T-S)$)} \colon \ \frac{\tu_3}{\tu_1} = q^{3T-S} 1/v_1^2 v_2 = \left(\frac{ABC}{DEF}\right )^{1/2} A^{-3} q^{3/2}, \nonumber\\
\text{NE edge (corresp.\ eq.: $j=-i/2+S+N$)}\colon \ \frac{\tu_2}{\tu_3} = q^{2S+3N} v_1 v_2^2 = \left(\frac{ABC}{DEF}\right )^{1/2} C^{-3} q^{3/2}, \nonumber\\
\text{SW edge (corresp.\ eq.: $j=-i/2$)}\colon \ \frac{\tu_2}{\tu_3} = q^{-S} v_1 v_2^2 = \left(\frac{ABC}{DEF}\right )^{1/2} F^3 q^{-3/2}, \nonumber\\
\text{vertical particle line (corresp.\ eq.: $i=t$)}\colon \ \frac{\tu_1}{\tu_2} = q^{-3t-S} v_1/v_2 = \frac{D E F}{ABC}.\label{canonical-edges}
\end{gather}

\begin{figure}[t]\centering
\includegraphics[scale=0.12]{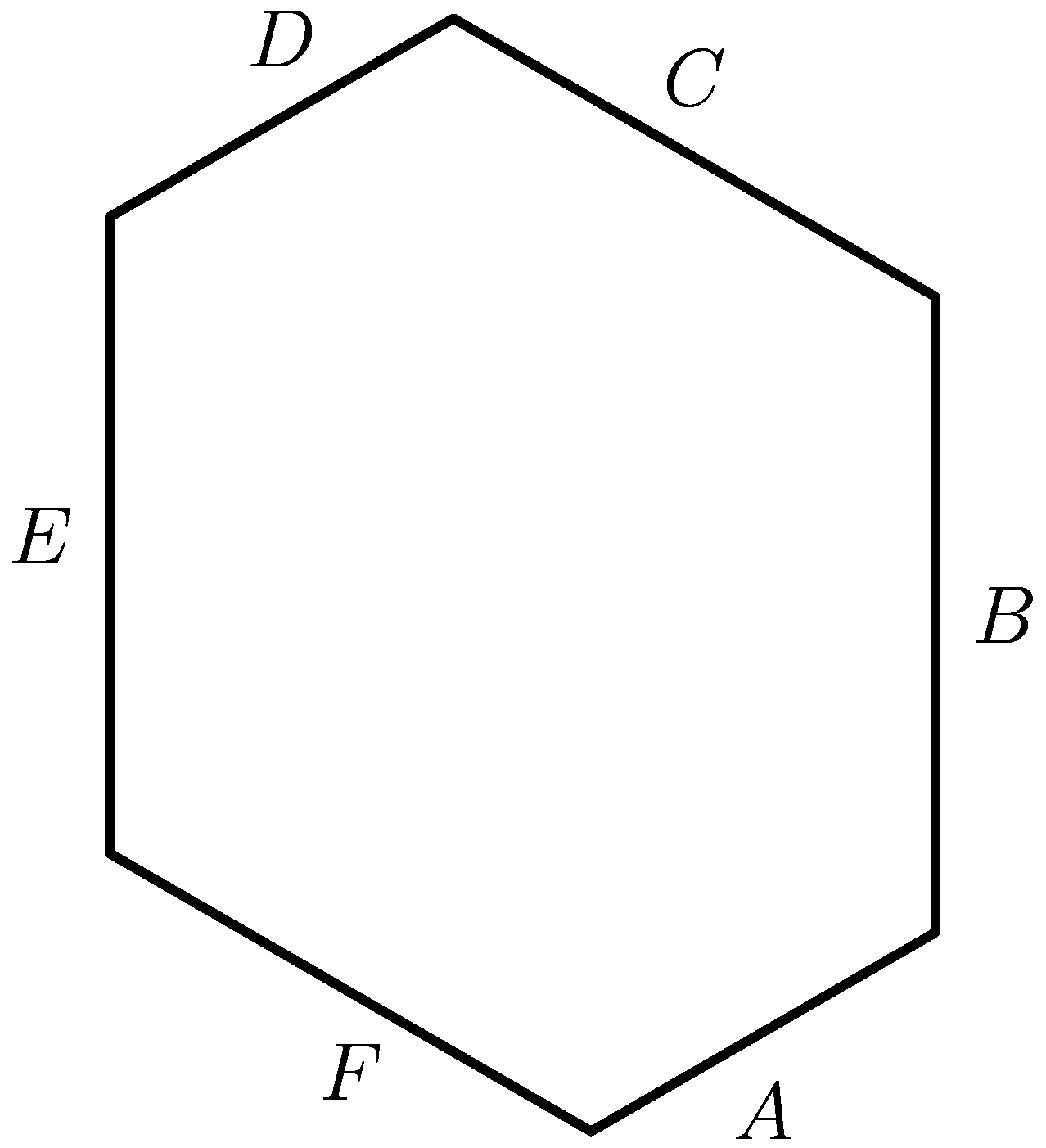}

\caption{Correspondence between edges and the six parameters.}\label{hex_par}
\end{figure}

\begin{Remark} We can see from above that there exists a bijection, depicted in Fig.~\ref{hex_par}, between the six bounding edges of our hexagon and the six parameters $A$, $B$, $C$, $D$, $E$, $F$: to an edge we assign the parameter that appears to the power $\pm 3$ above. The six parameters are not independent: they satisfy one balancing condition $ABCDEF = q^{3-2N}$, but then neither are the six edges: they must satisfy the condition that the hexagon they form is tilable by the three types of rhombi.
\end{Remark}

With \eqref{canonical-edges} in mind we have a (local) map $\R^2 \to \E^2$, where $\E^2$ is isomorphic to the subvariety of $\E^3$ with coordinates $(\tu_1,\tu_2,\tu_3)$ and relation $\prod \tu_i = 1$, which embeds our hexagon in $\E^2$
\begin{gather*}
 (i,j) \mapsto (\tu_1,\tu_2,\tu_3).
\end{gather*}

Note that $\E^2$ is the square of a real elliptic curve if parameters are chosen so that the weight ratio (of full to empty unit boxes) is real positive. Hence as $\E$ is homeomorphic to a circle or a~disjoint union of two circles, the above embeds our hexagon in a two-dimensional real torus.

\section{Distributions and transition probabilities} \label{sec:dist}

In this section we compute the $N$-point correlation function and transitional probabilities for the model under study. We refer the reader to the Appendix of~\cite{BGR} for the relevant application of Kasteleyn's theorem which makes these computations easy and to Kasteleyn's original paper for the theory itself~\cite{kasteleyn}.

\begin{figure}[t]\centering
\includegraphics[scale=0.40]{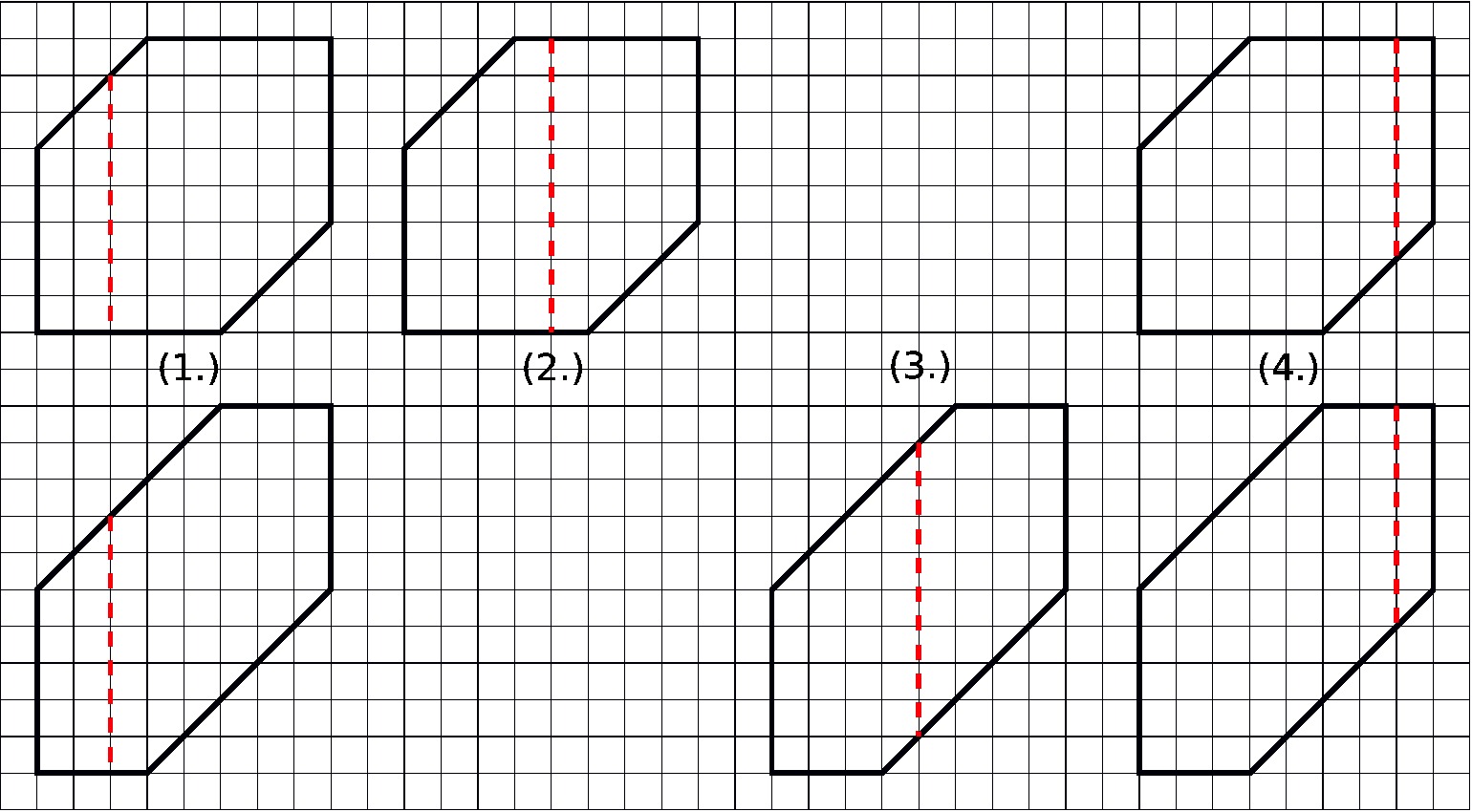}
\caption{The four ways of choosing a vertical particle line (dashed) inside a hexagon. In all cases $N=5$ particles, $T = 8$, $S \in \{3,5\}$.}
\label{4cases_h}
\end{figure}

Take a collection of $N$ non-intersecting lattice paths in $\Omega(N,S,T)$. Fix a vertical line inside the hexagon with integer abscissa~$t$ ($ 0 \leq t \leq T$). This vertical line will contain $N$ particles $X = (x_1<\dots <x_N) \in \mathpzc{X}^{S,t}_{N,T}$. Depending on the geometry of our hexagon, there are four ways in which we can fix this vertical line. They are described below and in Fig.~\ref{4cases_h}:
\begin{gather}
\text{\textbf{Case~1:}} \ t<S, \ t<T-S, \ 0 \leq x_k \leq t+N-1;\nonumber \\
\text{\textbf{Case~2:}} \ S\leq t \leq T-S, \ 0 \leq x_k \leq S+N-1; \nonumber\\
\text{\textbf{Case~3:}} \ T-S \leq t < S, \ t+S-T \leq x_k \leq t+N-1; \nonumber\\
\text{\textbf{Case~4:}} \ t\geq T-S, \ t\geq S, \ t+S-T \leq x_k \leq S+N-1.\label{4cases}
\end{gather}

We make use of the following notations:
\begin{description}\itemsep=0pt
\item $L_t(X) = $ sum of products of weights corresponding to holes (horizontal lozenges) to the left of the vertical line with coordinate $t$. The sum is taken over all possible ways of tiling the region to the left of this line. Equivalently, it is taken over all families of paths starting at $((0,0),\dots ,(0,N-1))$ and ending at $((t,x_1),\dots ,(t,x_N))$.
\item $R_t(X) = $ sum of products of weights corresponding to holes to the right of the vertical line with coordinate $t$. The sum is taken over all possible ways of tiling the region to the right of this line. Equivalently, it is taken over all families of paths starting at $((t,x_1),\dots ,(t,x_N))$ and ending at $((T,S),\dots ,(T,S+N-1))$.
\item $C_t(X) = $ product of weights corresponding to the holes on this vertical line.
\end{description}

Furthermore let
\begin{gather} \label{ell-vand}
 \varphi_{t,S}(x_k,x_l) = q^{-x_k} \thp\big(q^{x_k-x_l},q^{x_k+x_l+1-t-S} v_1 v_2\big).
\end{gather}

\begin{Remark}
 The product $\prod\limits_{k < l} \varphi_{t,S}(x_k,x_l)$ is an elliptic analogue of the Vandermonde product $\prod\limits_{k < l} (x_k-x_l)$.
\end{Remark}

\begin{Proposition} \label{L_t}
We have
\begin{gather*}
L_t(X = (x_1,\dots ,x_N)) = {\rm const} \cdot \prod_{k < l} \varphi_{t,S}(x_k,x_l) \\
\qquad {} \times \prod_{1 \leq k \leq N} q^{N x_l} \thp\big( q^{2 x_l +1-t-S} v_1 v_2\big) \frac { \thp\big(q^{1-N-t},q^{1-t-S} v_1,q^{t} v_2,q^{1-t-S} v_1 v_2;q\big)_{x_l} } { \thp\big(q,q^{2-2t-S} v_1,q v_2,q^{1+N-S} v_1 v_2;q\big)_{x_l} }.
\end{gather*}
\end{Proposition}

\begin{proof}This follows from an elaborate calculation and Lemma 10.2 in Appendix A of~\cite{BGR} which itself follows from Kasteleyn's theorem.

First, we restrict ourselves to the case $S<t<T-S$ (Case 2 in~\eqref{4cases}; computations are similar for the other three cases). Note in such a case we have $N$ particles and $S$ holes on the line with abscissa $t$. We then need to apply a particle--hole involution, as the weight in Lemma~10.2 in Appendix~A of~\cite{BGR} is given in terms of the positions of the holes (horizontal lozenges on the $t$-line). There are two types of products appearing in the total weight in question: a univariate one over the holes and a bivariate Vandermonde-like (again over the holes). For the first product, we just reciprocate to turn it into a product over particles (as the total product over holes and particles of the functions involved is a constant dependent only on $t$, $S$, $T$, $N$, $q$, $p$, $v_1$, $v_2$). For the Vandermonde-like product, we note for a function $f$ satisfying $f(y_i,y_j) = -f(y_j, y_i)$ we have
\begin{gather*}
 \prod_{1 \leq i < j \leq S} f(y_i,y_j) = \prod_{1 \leq i < j \leq N} f(x_i,x_j) \prod_{0 \leq u < v \leq S+N-1} f(u,v) \\
\hphantom{\prod_{1 \leq i < j \leq S} f(y_i,y_j) =}{} \times \prod_{1 \leq i \leq N} \frac{1}{\prod\limits_{0 \leq u < x_i} f(x_i,u) \prod\limits_{x_i < u \leq S+N-1} f(u,x_i)},
\end{gather*}
where $y$'s represent locations of holes (top vertices of horizontal lozenges) and $x$'s locations of particles. We take $f = \varphi_{t,S}$ as defined in \eqref{ell-vand}. Finally, in Appendix~A of~\cite{BGR}, the convention is that particles and holes are counted from the top going down. This is opposite to the convention in this paper, so we substitute~$x_l \mapsto S+N-1-x_l$. After standard manipulations with theta-Pochhammer symbols we arrive at the desired result.
\end{proof}

\begin{Proposition} \label{R_t}
We have
\begin{gather*}
 R_t(X = (x_1,\dots ,x_N)) = {\rm const} \cdot \prod_{k < l} \varphi_{t,S}(x_k,x_l) \\
\qquad ХЇ \times \prod_{1 \leq k \leq N} q^{N x_l} \thp\big( q^{2 x_l +1-t-S} v_1 v_2\big) \frac {\thp\big(q^{1-N-S},q^{-2 t-S} v_1,q^{1+T} v_2,q^{1-T} v_1 v_2;q\big)_{x_l} } { \thp\big(q^{1-S-t+T},q^{1-t-S-T} v_1,q^{2+t} v_2,q^{1+N-t} v_1 v_2;q\big)_{x_l} }.
\end{gather*}
\end{Proposition}

\begin{proof}Similar to the previous proof except we use Lemma~10.3 in Appendix~A of~\cite{BGR}.
\end{proof}

\begin{Proposition} \label{C_t}
We have
\begin{gather*}
C_t(X = (x_1,\dots ,x_N)) = {\rm const} \cdot \prod_{1 \leq k \leq N} \frac{\thp\big(q^{x_l-2t-S} v_1, q^{x_l-2t-S+1} v_1, q^{x_l+t} v_2, q^{x_l+t+1} v_2\big)} { q^{x_l} \thp\big(q^{2 x_l + 1 - t-S} v_1 v_2\big)}.
\end{gather*}
\end{Proposition}

\begin{proof}This weight is (up to a constant not depending on holes or particles) the reciprocal of the total weight of the $S$ holes (horizontal lozenges) on the $t$-line and the latter is readily computed from the definition \eqref{weight}.
\end{proof}

\begin{Theorem} \label{thmtline}
 We have
\begin{gather*} 
\pr(X(t) = (x_1,\dots ,x_N)) = {\rm const} \cdot \prod_{k < l} (\varphi_{t,S}(x_k,x_l))^2 \prod_{1 \leq k \leq N} q^{(2N-1) x_k} \thp\big(q^{2x_k+1-t-S} v_1 v_2\big) \\
 \quad {}\times \prod_{1 \leq k \leq N} \frac{\thp\big(q^{1-N-t},q^{1-N-S},q^{1-t-S} v_1,q^{1+T} v_2,q^{1-T} v_1 v_2,q^{1-t-S} v_1 v_2;q\big)_{x_k}}{\thp\big(q,q^{1-S-t+T},q^{1-t-T-S} v_1,q v_2,q^{1+N-S} v_1 v_2,q^{1+N-t} v_1 v_2;q\big)_{x_k}} \\
= {\rm const} \cdot \prod_{k < l} (\varphi_{t,S}(x_k,x_l))^2 \prod_{1 \leq k \leq N} q^{(2N-1) x_k} \thp\big(q^{2x_k} F^2\big) \frac{\thp\big(AF,BF,CF,DF,EF,F^2;q\big)_{x_k}}{\thp\big(q,q \frac{A}{F},q \frac{B}{F},q \frac{C}{F},q \frac{D}{F},q \frac{E}{F};q\big)_{x_k}}.
\end{gather*}
\end{Theorem}

\begin{proof} It follows from $\pr(X(t) = (x_1,\dots ,x_N)) \propto L_t(X) C_t(X) R_t(X).$
\end{proof}

\begin{Remark} The above distribution is what was called in the Introduction the discrete elliptic Selberg density. That is to say,
 \begin{gather*}
 \pr(X(t) = (x_1,\dots ,x_N)) \\
 \qquad {} = \Delta_{\lambda}\big(q^{2N-2} F^2\,|\,q^N,q^{N-1}AF,q^{N-1}(pB)F,q^{N-1}CF,q^{N-1}DF,q^{N-1}EF\big),
 \end{gather*}
where $\lambda \in m^n$ ($m=S+N-1,n=N$) and $\lambda_i+N-i = x_{N+1-i}$ (to account for the fact that $x_1<x_2<\dots <x_N$ whereas parts of partitions are always non-increasing order). The particle--hole involution invoked in Proposition~\ref{L_t} then takes the following form: if $\lambda_p$ is the partition associated to the particle positions (at time $t$) via the above equation and $\lambda_h$ is the partition associated to the whole positions at the same time (in the case above, there are~$S$ holes), then
\begin{gather*}
 \lambda_h = \big(\lambda_p^c\big)',
\end{gather*}
where $\lambda^c$ denotes the complemented partition corresponding to $\lambda \in m^n$ ($\lambda^c_i = m-\lambda_{n+1-i}$) and~$\lambda'$ denotes the dual (transposed) partition ($\lambda_i' =$ number of parts of $\lambda$ that are $\geq i$). The fact that both probabilities (in terms of holes and in terms of particles) are $\Delta$-symbols can be observed directly as shown in Proposition~\ref{L_t} or using the following relations appearing in~\cite{BCn}
\begin{gather*}
\Delta_{\lambda'}(a\,|\,\dots b_i\dots ;1/q)= \Delta_{\lambda}\big(a/q^2\,|\,\dots b_i\dots ;q\big), \\
\frac{\Delta_{\lambda^c}(a\,|\,\dots b_i\dots ;q)}{\Delta_{m^n}(a\,|\,\dots b_i\dots ;q)} = \Delta_{\lambda}\left(\frac{q^{2m-2}}{q^{2n} a}\,\Big|\,\dots \frac{q^{n-1}b_i}{q^m a}\dots ,q^n,pq^n,q^{-m},pq^{-m};q\right).
\end{gather*}
\end{Remark}

We will for brevity denote the measure described in Theorem~\eqref{thmtline} by $\rho_{S,t}$ (note it also depends on $N$, $T$, $v_1$, $v_2$, $p$, $q$, but it is the dependence on~$S$ and~$t$ that will be of most interest to us). Observe we can transform the factor
\begin{gather*}
 q^x q^{(2N-2) x} \frac{\thp\big(q^{1-t-S} v_1, q^{1+T} v_2\big)}{\thp\big(q^{1-t-S-T} v_1, q v_2\big)}
\end{gather*}
 appearing in the univariate product of the above probability into something proportional to
\begin{gather*}
 q^x \frac{\thp\big(q^{N-t-S} v_1, q^{N+T} v_2\big)}{\thp\big(q^{2-N-t-S-T} v_1, q^{2-N} v_2\big)} \frac{1}{\thp\big(q^{x+1-t-S} v_1, q^{-x+t+S+T} /v_1, q^{x+1+T} v_2, q^{-x} /v_2\big)_{N-1}},
\end{gather*}
by using
\begin{gather*}
 \thp\big(A q^{N-1};q\big)_x = \frac{\thp(A;q)_x \thp\big(Aq^x;q\big)_{N-1}}{\thp(A;q)_{N-1}}, \\
 \thp\big(A q^{1-N};q\big)_x = \frac{q^{(1-N)x}\thp(A;q)_x \thp(q/A;q)_{N-1}}{\thp\big(q^{1-x}/A;q\big)_{N-1}}
\end{gather*}
and absorbing into the initial constant anything independent of $x$ (of the particle positions $x_k$). After using~\eqref{parmatching} our probability distribution becomes
\begin{gather}
 \pr(X(t) = (x_1,\dots ,x_N))= {\rm const} \cdot \prod_{k < l} (\varphi_{t,S}(x_k,x_l))^2 \nonumber \\
\qquad{}\times \prod_{1 \leq k \leq N} \frac{1}{\thp\big(B (F q^{x_k})^{\pm 1}, E (F q^{x_k})^{\pm 1};q\big)_{N-1}} \prod_{1 \leq k \leq N} w(x_k),\label{tline2}
\end{gather}
where
\begin{gather*}
 w(x) = \! \frac{q^{x} \thp\big(q^{2x{+}1{-}t{-}S}\! v_1 v_2\big) \thp\big(q^{1{-}N{-}t}\!,q^{1{-}N{-}S}\!,q^{N{-}t{-}S}\! v_1,q^{N{+}T}\! v_2,q^{1{-}T}\! v_1 v_2,q^{1{-}t{-}S}\! v_1 v_2;q\big)_{x}}{\thp\big(q^{1{-}t{-}S}\! v_1 v_2\big) \thp\big(q,q^{1{-}S{-}t{+}T}\!,q^{2{-}N{-}t{-}T{-}S}\! v_1,q^{2{-}N}\! v_2,q^{1{+}N{-}S}\! v_1 v_2,q^{1{+}N{-}t}\! v_1 v_2;q\big)_{x}}\!\! \\
 \hphantom{w(x)}{} =\! \frac{q^x \thp\big(F^2 q^{2x}\big) \thp\big(AF,BF \big( \frac{q}{ABCDEF} \big )^{\frac{1}{2}}, CF,DF, EF \big( \frac{q}{ABCDEF} \big)^{\frac{1}{2}},F^2;q\big)_{x}}{\thp\big(F^2\big) \thp\big(\frac{F}{A} q,\frac{F}{B}q \big( \frac{ABCDEF}{q} \big)^{\frac{1}{2}},\frac{F}{C} q,\frac{F}{D} q,\frac{F}{E}q \big( \frac{ABCDEF}{q} \big)^{\frac{1}{2}},q;q\big)_{x}}.
\end{gather*}

Here we note $w$ is the weight function for the discrete elliptic univariate biorthogonal functions of Spiridonov and Zhedanov, see \cite{SZ1,SZ2}. It is of course also the discrete elliptic Selberg density for $N=1$, hence a $\Delta$-symbol in one variable as seen in~\eqref{discrete-selberg}. Notice in~\eqref{tline2} above~$B$ and~$E$ play a special role, as does $F$. This will become more transparent in Section~\ref{sec:corr}. The weight $w$ is elliptic in $q$, $v_1$, $v_2$ and $q^{\{t,S,T,N\}}$, or, analogously, in $A$, $B$, $C$, $D$, $E$, $F$, $q$.

\begin{Remark}
Note that in the definition of $w$ above, the first line is given in terms of the geometry of the hexagon and the choice of the particular particle line (Case 2 in \eqref{4cases} as previously discussed), while the second line is intrinsic and the geometry of the hexagon only comes in after using \eqref{parmatching}. We can also define the equivalent of \eqref{parmatching} in the other three cases described in~\eqref{4cases} (and the three other choices of six parameters differ from \eqref{parmatching} by~(a): interchanging~$S$ ant~$t$, (b):~shifting the six parameters in~\eqref{parmatching} by $q^{\pm(t+S-T)})$, or (c): a combination of both~(a) and~(b)). We will not use this any further, as all calculations will be done in Case 2 from \eqref{4cases}.
\end{Remark}

\begin{Remark}The limit $v_1 = v_2 = \kappa \sqrt{p}$, $p \to 0$ gives the distributions present in~\cite{BGR} at the $q$-Racah level. Such probabilities are also structurally a product of a Vandermonde-like determinant squared (the first two products in \eqref{tline2}) and a product over the particles of univariate weights. Indeed, under the appropriate limits, one can arrive from~\eqref{tline2} to a much simpler, prototypical such $N$-point function: the joint density of the~$N$ eigenvalues of a GUE $N \times N$ random matrix.
\end{Remark}

The transition and co-transition probabilities for the Markov chain $X(t)$ are given by the next two statements.
\begin{Theorem} \label{thmt+}
If $Y=(y_1,\dots ,y_N)$ and $X=(x_1,\dots ,x_N)$ such that $y_k - x_k \in \{0,1\}$ $\forall\, k$, then
\begin{gather*}
 \pr(X(t+1) = Y\,|\,X(t) = X)={\rm const} \cdot \prod_{k < l} \frac{\varphi_{t+1,S}(y_k,y_l)}{\varphi_{t,S}(x_k,x_l)} \prod_{k\colon y_k=x_k+1} w_1(x_k) \prod_{k\colon y_k=x_k} w_0(x_k),
\end{gather*}
where
\begin{gather*}
 w_0(x) = \frac{q^{-x-N+1} \thp\big(q^{x+T-t-S}, q^{x-T-t-S} v_1, q^{x+t+1} v_2, q^{x+N-t} v_1 v_2\big)} { \thp\big(q^{2x+1-t-S}v_1 v_2\big)}, \\
w_1(x) = -\frac{q^{-x} \thp\big(q^{x+1-N-S}, q^{x-2t-S} v_1, q^{x+T+1} v_2, q^{x-T+1} v_1 v_2\big)} { \thp\big(q^{2x+1-t-S}v_1 v_2\big)}.
\end{gather*}
\end{Theorem}

\begin{proof} The formula
 \begin{gather*}
 \pr(X(t+1) = Y\,|\,X(t) = X) = \frac{L_t(X) C_t(X) C_{t+1}(Y) R_{t+1}(Y)}{L_t(X) C_t(X) R_t(X)} = \frac{C_{t+1}(Y) R_{t+1}(Y)}{R_t(X)}
 \end{gather*}
 along with the formulas for $L$, $R$ and $C$ yield the result.
\end{proof}

\begin{Theorem} \label{thmt-}
If $Y=(y_1,\dots ,y_N)$ and $X=(x_1,\dots ,x_N)$ such that $y_k - x_k \in \{0,-1\}$ $\forall\, k$, then
\begin{gather*}
 \pr(X(t-1) = Y\,|\,X(t) = X)={\rm const} \cdot \prod_{k < l} \frac{\varphi_{t-1,S}(y_k,y_l)}{\varphi_{t,S}(x_k,x_l)} \prod_{k\colon y_k=x_k-1} w'_1(x_k) \prod_{k\colon y_k=x_k} w'_0(x_k),
\end{gather*}
where
\begin{gather*}
w'_0(x) = -\frac{q^{-x} \thp\big(q^{x-N-t+1}, q^{x-t-S+1} v_1, q^{x+t} v_2, q^{x-t-S+1} v_1 v_2\big)} { \thp\big(q^{2x+1-t-S}v_1 v_2\big)}, \\
w'_1(x) = \frac{q^{-x-N+1} \thp\big(q^{x}, q^{x-2t-S+1} v_1, q^{x} v_2, q^{x+N-S} v_1 v_2\big)} { \thp\big(q^{2x+1-t-S}v_1 v_2\big)}.
\end{gather*}
\end{Theorem}

\begin{proof} As before
 \begin{gather*}
 \pr(X(t-1) = Y\,|\,X(t) = X) = \frac{L_{t-1} (X) C_{t-1}(X) C_{t}(Y) R_{t}(Y)}{L_t(X) C_t(X) R_t(X)} = \frac{L_{t-1}(Y) C_{t-1}(Y)}{L_t(X)}. \tag*{\qed}
\end{gather*}\renewcommand{\qed}{}
\end{proof}

We are now in a position to define six stochastic matrices (Markov chains) needed in what will follow. Their stochasticity along with other properties will be proven in Section~\ref{sec:diff_op}, although we know the first two are stochastic as they represent the transition probabilities obtained in this section. To condense notation, we denote $z_k = Fq^{x_k}$. Let
\begin{alignat*}{5}
& P^{S,t}_{t+}\colon \ && \mathpzc{X}^{S,t} \times \mathpzc{X}^{S,t+1} \to [0,1], \qquad && P^{S,t}_{t-} \colon \ && \mathpzc{X}^{S,t} \times \mathpzc{X}^{S,t-1} \to [0,1], &\\
& _{t+}P^{S,t}_{S+} \colon \ && \mathpzc{X}^{S,t} \times \mathpzc{X}^{S+1,t} \to [0,1], \qquad && _{t+}P^{S,t}_{S-} \colon \ && \mathpzc{X}^{S,t} \times \mathpzc{X}^{S-1,t} \to [0,1], &\\
& _{t-}P^{S,t}_{S+} \colon \ && \mathpzc{X}^{S,t} \times \mathpzc{X}^{S+1,t} \to [0,1], \qquad && _{t-}P^{S,t}_{S-} \colon \ && \mathpzc{X}^{S,t} \times \mathpzc{X}^{S-1,t} \to [0,1]&
 \end{alignat*}
be defined by
\begin{gather} \allowdisplaybreaks \label{t+}
 P^{S,t}_{t+}(X,Y) = \begin{cases}
\displaystyle{\rm const} \cdot \prod_{k < l} \frac{\varphi_{t+1,S}(y_k,y_l)}{\varphi_{t,S}(x_k,x_l)} \\
\displaystyle \quad{}\times \prod_{k\colon y_k=x_k+1} - \frac{q^{-x_k} \thp\big(A z_k, B z_k, C z_k, q^{1-N} z_k/ABC\big)} { \thp\big(z_k^2\big)} \\
\displaystyle\quad{} \times \prod_{k\colon y_k=x_k} \frac{q^{-x_k-N+1} \thp\big(z_k/A, z_k/B, z_k/C, q^{N-1} z_k ABC\big)} { \thp\big(z_k^2\big)}, \\
\qquad \text{if\ } y_k - x_k \in \{0,1\} \ \forall\, k, \\
0, \quad \text{otherwise};
 \end{cases}
\\ \label{t-}
 P^{S,t}_{t-}(X,Y) = \begin{cases}
\displaystyle{\rm const} \cdot \prod_{k < l} \frac{\varphi_{t-1,S}(y_k,y_l)}{\varphi_{t,S}(x_k,x_l)} \\
\displaystyle\quad{} \times \prod_{k\colon y_k=x_k-1} \frac{q^{-x_k-N+1} \thp\big(z_k/D, z_k/E, z_k/F, q^{N-1} z_k DEF\big)} { \thp\big(z_k^2\big)} \\
\displaystyle\quad{} \times \prod_{k\colon y_k=x_k} - \frac{q^{-x_k} \thp\big(D z_k,E z_k,F z_k, q^{1-N} z_k/DEF\big)} { \thp\big(z_k^2\big)},\\
\qquad \text{if\ } y_k - x_k \in \{0,-1\} \ \forall\, k, \\
0, \quad \text{otherwise};
 \end{cases}
\\\label{t+S+}
 _{t+}P^{S,t}_{S+}(X,Y) = \begin{cases}
\displaystyle{\rm const} \cdot \prod_{k < l} \frac{\varphi_{t,S+1}(y_k,y_l)}{\varphi_{t,S}(x_k,x_l)} \\
\displaystyle\quad{} \times \prod_{k\colon y_k=x_k+1} - \frac{q^{-x_k} \thp\big(A z_k, B z_k, D z_k, q^{1-N} z_k/ABD\big)} { \thp\big(z_k^2\big)} \\
\displaystyle\quad{} \times \prod_{k\colon y_k=x_k} \frac{q^{-x_k-N+1} \thp\big(z_k/A, z_k/B, z_k/D, q^{N-1} z_k ABD\big)} { \thp\big(z_k^2\big)},\\
\qquad \text{if\ } y_k - x_k \in \{0,1\} \ \forall\, k, \\
0, \quad \text{otherwise};
 \end{cases}
\\ \label{t+S-}
 _{t+}P^{S,t}_{S-}(X,Y) = \begin{cases}
\displaystyle{\rm const} \cdot \prod_{k < l} \frac{\varphi_{t,S-1}(y_k,y_l)}{\varphi_{t,S}(x_k,x_l)} \\
\displaystyle\quad{} \times\prod_{k\colon y_k=x_k+1} - \frac{q^{-x_k} \thp\big(B z_k, C z_k, F z_k, q^{1-N} z_k/BCF\big)} { \thp\big(z_k^2\big)} \\
\displaystyle\times \prod_{k\colon y_k=x_k} \frac{q^{-x_k-N+1} \thp\big(z_k/B, z_k/C, z_k/F, q^{N-1} z_k BCF\big)} { \thp\big(z_k^2\big)},\\
\qquad \text{if\ } y_k - x_k \in \{0,-1\} \ \forall\, k, \\
0, \quad \text{otherwise};
 \end{cases}
\\ \label{t-S+}
 _{t-}P^{S,t}_{S+}(X,Y) = \begin{cases}
\displaystyle {\rm const} \cdot \prod_{k < l} \frac{\varphi_{t,S+1}(y_k,y_l)}{\varphi_{t,S}(x_k,x_l)} \\
 \displaystyle \quad{}\times \prod_{k\colon y_k=x_k-1} \frac{q^{-x_k-N+1} \thp\big(z_k/D, z_k/E, z_k/A, q^{N-1} z_k DEA\big)} { \thp\big(z_k^2\big)} \\
\displaystyle \quad \times \prod_{k\colon y_k=x_k} - \frac{q^{-x_k} \thp\big(D z_k,E z_k,A z_k, q^{1-N} z_k/DEA\big)} { \thp\big(z_k^2\big)},\\
\qquad \text{if\ } y_k - x_k \in \{0,1\} \ \forall\, k, \\
 0, \quad \text{otherwise};
 \end{cases}
\\ \label{t-S-}
 _{t-}P^{S,t}_{S-}(X,Y) = \begin{cases}
\displaystyle {\rm const} \cdot \prod\limits_{k < l} \frac{\varphi_{t,S-1}(y_k,y_l)}{\varphi_{t,S}(x_k,x_l)} \\
\displaystyle\quad {} \times \prod_{k\colon y_k=x_k-1} \frac{q^{-x_k-N+1} \thp\big(z_k/E, z_k/F, z_k/C, q^{N-1} z_k EFC\big)} { \thp\big(z_k^2\big)} \\
\displaystyle \quad {} \times \prod\limits_{k\colon y_k=x_k} - \frac{q^{-x_k} \thp\big(E z_k,F z_k,C z_k, q^{1-N} z_k/EFC\big)} { \thp\big(z_k^2\big)}, \\
\qquad \text{if\ } y_k - x_k \in \{0,-1\} \ \forall\, k, \\
0, \quad \text{otherwise}.
 \end{cases}
\end{gather}
The normalizing constants are independent of the $x_k$'s and the $y_k$'s. They will become explicit in Section~\ref{sec:diff_op}. Note that $_{t-}P^{S,t}_{S-}$, under interchanging $t$ and $S$, becomes $P^{S,t}_{t-}$. Under the same procedure $_{t+}P^{S,t}_{S+}$ becomes $P^{S,t}_{t+}$. We can think of $P^{S,t}_{t+}$ ($P^{S,t}_{t-}$) as a Markov chain that increases (decreases) $t$, while $_{t \pm}P^{S,t}_{S+}$ ($_{t \pm}P^{S,t}_{S-}$) increases (decreases) $S$.
\begin{Remark}
In the $q$-Racah limit $v_1 = v_2 = \kappa \sqrt{p}, \ p \to 0$, the chains ${}_{t \pm}P^{S,t}_{S+}$ coalesce into one ($P^{S,t}_{S+}$ in \cite{BGR}). Likewise for ${}_{t \pm}P^{S,t}_{S-}$.
\end{Remark}

\section{Elliptic difference operators} \label{sec:diff_op}

In this section we explain how recent constructions in the field of elliptic special functions intrinsically capture the model we described thus far. The main two references are \cite{BCn,EHI} and we will state results from these without proofs (with a few exceptions where the proofs are short). The focus will be on certain elliptic difference operators satisfying normalization, quasi-commutation and quasi-adjointness relations. We define them abstractly in the first subsection. We then turn to motivating the definitions and interpreting the operators probabilistically.

\subsection{Definitions and some properties}

In \cite{BCn,EHI} Rains has introduced a family of difference operators acting on various classes of $BC_n$-symmetric functions. To define them, we let $r_0, r_1, r_2, r_3 \in \C^*$ satisfy $r_0 r_1 r_2 r_3 = pq^{1-n}$. Then define $\D(r_0, r_1, r_2, r_3)$ (also depending on $q$, $p$, $n$) by
\begin{gather}
 (\D(r_0, r_1, r_2, r_3) f) (\dots z_k\dots ) \nonumber\\
 \qquad {} = \sum_{\sigma \in \{ \pm 1 \}^n} \prod_{1\leq k \leq n} \frac{\prod\limits_{0 \leq s \leq 3}\thp\big(r_s z_k^{\sigma_k}\big)}{\thp\big(z_k^{2 \sigma_k}\big)} \prod_{1 \leq k < l \leq n} \frac{\thp\big(q z_k^{\sigma_k} z_l^{\sigma_l}\big)}{\thp\big(z_k^{\sigma_k} z_l^{\sigma_l}\big)} f\big(\dots q^{\sigma_k/2} z_k\dots \big). \label{diffop}
\end{gather}

\begin{Remark}The difference operator above described is the special case $t=q$ of the more general elliptic $(q,t)$ difference operator mentioned in the references -- note in \textit{this remark alone}~$t$ has the meaning of Macdonald's~\cite[Chapter VI]{mac} parameter $t$ and not of time.
\end{Remark}

In view of $r_0 r_1 r_2 r_3 = pq^{1-n}$ we will break symmetry and denote the difference operator by $\D(r_0,r_1,r_2)$, the fourth parameter being implied by the balancing condition.

\begin{Remark}$\D$ takes $BC_n$-symmetric functions to $BC_n$-symmetric functions.
\end{Remark}

By letting $\D$ act on the function $f \equiv 1$, we obtain the following important lemma, whose proof we sketch following~\cite{EHI}.
\begin{Lemma} \label{sum}
For $r_0 r_1 r_2 r_3 = pq^{1-n}$ we have
\begin{gather*}
 \sum_{\sigma \in \{ \pm 1 \}^n} \prod_{1\leq k \leq n} \frac{\prod\limits_{0 \leq s \leq 3}\thp\big(r_s z_k^{\sigma_k}\big)}{\thp\big(z_k^{2 \sigma_k}\big)} \prod_{1 \leq k < l \leq n} \frac{\thp\big(q z_k^{\sigma_k} z_l^{\sigma_l}\big)}{\thp\big(z_k^{\sigma_k} z_l^{\sigma_l}\big)} = \prod_{0 \leq k < n} \thp\big(q^k r_0 r_1, q^k r_0 r_2, q^k r_1 r_2\big).
\end{gather*}
\end{Lemma}

\begin{proof}By direct computation the left-hand side above is invariant under $z_k \to pz_k$ for all $k$ (this is insured by the fact $r_0 r_1 r_2 r_3 = pq^{1-n}$). It is also $BC_n$-symmetric. Finally, by multiplying the left-hand side by $R = \prod_{k} z_k^{-1} \thp\big(z_k^2\big) \prod\limits_{k<l} \varphi(z_k,z_l)$ we will have cleared its potential poles. Because~$R$ is $BC_n$-skewsymmetric the result will end up being a multiple of~$R$: $R \cdot \mathrm{LHS} = {\rm const} \cdot R$ showing the left-hand side has no singularities in the variables and is thus independent of the~$z_i$'s. Evaluating at $z_i = r_0 q^{n-i}$ yields the result.
\end{proof}

Hereinafter we will use $\D$ for the normalized difference operator, so that $\D(r_0,r_1,r_2) 1 = 1$, following Lemma~\ref{sum}.

The difference operators described above satisfy a number of identities, including a series of quasi-commutation relations. For an elegant proof which relies on the action of these operators on a suitably large space of functions see~\cite{EHI} or~\cite{BCn}.

\begin{Lemma} \label{commrel}
If $U$, $V$, $W$, $Z$ are four parameters, then
\begin{gather*}
 \D(U,V,W) \D\big(q^{1/2} U, q^{1/2} V, q^{-1/2} Z\big) = \D(U,V,Z) \D\big(q^{1/2} U, q^{1/2} V, q^{-1/2} W\big).
\end{gather*}
\end{Lemma}

Next we look at the action of the difference operators on special classes of functions. For $\lambda \in m^n$ a partition, let
\begin{gather*}
 \pzcd_{\lambda}(\dots x_k\dots ) = \prod_{1 \leq k \leq n} \frac{\prod\limits_{1 \leq l \leq m+n} \thp\big(u q^{l-1} x_k^{\pm 1}\big)}{\prod\limits_{1 \leq l \leq n} \thp\big(u q^{\lambda_l+n-l} x_k^{\pm 1}\big)}.
\end{gather*}
By direct computation, we see that $ \pzcd_{\lambda}\big(\dots u q^{\mu_k+n-k}\dots \big) = \delta_{\lambda, \mu} c_{\lambda}$.
\begin{Remark}
 $\pzcd_{\lambda}$ is a special version of the interpolation theta functions
 \begin{gather*}P_{\lambda}^{*(m,n)}(\dots x_k\dots ;a,b;q;p)
 \end{gather*}
 defined in \cite{BCn} (matching the notation in the reference with ours, $a=u$, $b = q^{-m-n+1}/a)$). They are defined, up to normalization, by two properties: being $BC_n$-symmetric of degree $m$ and vanishing at $\mu \ne \lambda$.
\end{Remark}

If we now define $\mathfrak{d}_{\lambda} = \frac{\pzcd_{\lambda}} {c_{\lambda}}$, we see that
\begin{gather*} 
 \mathfrak{d}_{\lambda} \big(\dots u q^{\mu_k+n-k}\dots \big) = \delta_{\lambda, \mu},
\end{gather*}
so that in a precise way, $\mathfrak{d}_{\lambda}$ is an interpolation Kronecker delta theta function. We then immediately have the following proposition.
\begin{Proposition} \label{deltafunction}
Fix $\tau \in \{ \pm 1 \}^n$. Let $z_k = u q^{\lambda_k+n-k}$. Then
\begin{gather*}
 (\D(r_0,r_1,r_2) \mathfrak{d}_{\lambda})\big(\dots q^{-\tau_k/2} z_k\dots \big) \\
 \qquad {}= \prod_{k} \frac{\thp\big(r_0 z_k^{\tau_k},r_1 z_k^{\tau_k}, r_2 z_k^{\tau_k}, \big(pq^{1-n}/r_0r_1r_2\big) z_k^{\tau_k}\big)} {\thp\big(z_k^{2 \tau_k}\big)} \prod_{k<l} \frac{\thp\big(q z_k^{\tau_k} z_l^{\tau_l}\big)}{\thp\big(z_k^{\tau_k} z_l^{\tau_l}\big)}.
\end{gather*}
\end{Proposition}

\begin{proof}Immediate by substituting into the definition of the difference operator \eqref{diffop}. For any $\sigma \ne \tau$, $q^{\sigma_k/2 - \tau_k/2} z_k$ will be of the form $u q^{\mu_k+n-k}$ with $\mu \ne \lambda$ and the corresponding summand will be zero.
\end{proof}

A useful final property of the difference operators is their quasi-adjointness. It was shown in~\cite{EHI} that the $\D$'s satisfy a certain adjointness relation that we will need in the next section. We start with six parameters $t_0$, $t_1$, $t_2$, $t_3$, $u_0$, $u_1$ satisfying the balancing condition
\begin{gather*}
 q^{2n-2} t_0 t_1 t_2 t_3 u_0 u_1 = pq.
\end{gather*}

We fix the number of variables at $n$ and $\lambda$ will be a partition in $m^n$. As in the introduction, we denote $l_i = \lambda_i+n-i$. We define the discrete Selberg inner product $\langle \cdot, \cdot \rangle$ (depending on $p$,~$q$ and the six parameters) by
\begin{gather}
\langle f,g \rangle = \frac{1}{Z}\sum_{\lambda \subseteq m^n} f\big(\dots t_0 q^{l_i}\dots \big) g\big(\dots t_0 q^{l_i}\dots \big) \nonumber\\
\hphantom{\langle f,g \rangle =}{} \times \Delta_{\lambda}\big(q^{2n-2} t_0^2\,|\,q^n, q^{n-1} t_0 t_1,q^{n-1} t_0 t_2,q^{n-1} t_0 t_3,q^{n-1} t_0 u_0,q^{n-1} t_0 u_1;q\big),\label{inner-product}
\end{gather}
where $f$, $g$ belong to some sufficiently nice set of functions (we will assume they are $BC_n$-symmetric) and $Z$ is an explicit constant that makes $\langle 1,1 \rangle = 1$. This is a discrete analogue of the continuous inner product introduced in~\cite{EHI} and can be obtained from that by residue calculus.

If the above conditions are satisfied, then \cite{EHI}
\begin{gather*} 
\langle \D(u_0,t_0,t_1)f,g \rangle = \big\langle f,\D(u_1',t_2',t_3') g \big\rangle',
\end{gather*}
where
\begin{gather*}
 (t_0',t_1',t_2',t_3',u_0',u_1') = \big(q^{1/2} t_0,q^{1/2} t_1,q^{-1/2} t_2,q^{-1/2} t_3,q^{1/2} u_0,q^{-1/2} u_1\big)
\end{gather*}
and $\langle \,,\, \rangle'$ is the inner product defined in \eqref{inner-product} with primed parameters inserted throughout.

\subsection{Interpretation of difference operators and their properties}

We now show how the difference operators and their properties discussed in the previous section can be given probabilistic interpretations. First, observe from~\eqref{parmatching} that $q^{2n-3} ABCDEF = 1$.

In what follows $h_k$ ($h'_k$) is the location of the $k$-th particle on the vertical line $i=t$ ($i=t+1$) in the $(i,j)$ frame (note according to the $t \to t+1$ dynamics the particles move either up or down by $1/2$). The following proposition links difference operators with combinatorics.

\begin{Proposition} \label{parameterprop}
For $A$, $B$, $C$, $D$, $E$, $F$ and $z_k = F q^{h_k}$ given by \eqref{parmatching}, the summands in
\begin{gather*}
 (\D(A,B,C) 1) (\dots z_k\dots )
\end{gather*}
$($see equation \eqref{diffop}$)$, appropriately normalized using \eqref{sum}, are equal to the transition probabilities $($entries in the stochastic matrix$)$ $P^{S,t}_{t+} (H,H')$ defined in~\eqref{t+}, after switching coordinates from $(x,y)$ back to $(i,j)$. This statement also holds for
\begin{gather*}
\D(D,E,F) \quad \text{and} \quad P^{S,t}_{t-}, \\
\D(A,B,D) \quad \text{and} \quad _{t+}P^{S,t}_{S+}, \\
\D(B,C,F) \quad \text{and} \quad _{t+}P^{S,t}_{S-}, \\
\D(D,E,A) \quad \text{and} \quad _{t-}P^{S,t}_{S+}, \\
\D(E,F,C) \quad \text{and} \quad _{t-}P^{S,t}_{S-}.
\end{gather*}
\end{Proposition}

\begin{proof}
We will only prove the statement for $\D(A,B,C)$ and $t+$ (the equivalent statement for $\D(D,E,F)$ and $t-$ is proved much the same way). The proof is immediate in view of~\eqref{parmatching}, the change of variables $(X,Y) \mapsto (H,H')$ in \eqref{t+} (to the $(i,j)$ coordinates) and the following observations.

First, a choice of $\sigma_k \in \{ \pm 1\}$ for all $k$ in the definition of $\D(A,B,C)$ is equivalent to a~choice of which particles move up/down from the position vector $H$ (at vertical line $t$) to the position vector $H'$ (at vertical line~$t+1$). If $\sigma_k = 1$, the corresponding $k$-th particle at vertical position $h_k$ moves up to $h'_k = h_k+1/2$ (and if $\sigma_k=-1$, the $k$-th particle moves down). Next observe that in the univariate product appearing in any term of $(\D(A,B,C) 1) (\dots z_k\dots )$, we can change $\thp\big(u z_i^{-b}\big)$ ($b=1,2$) to $\thp\big(z_i^{b}/u\big)$ by the reflection formula for theta functions and it will now match with the univariate product appearing in~$P_{t+}^{S,t}$. The product $\prod\limits_{k\colon y_k=x_k+1} (\dots ) \prod\limits_{y_k=x_k} (\dots )$ now indeed is identical (modulo constants independent of the particle positions) to $\prod\limits_{k\colon h'_k=h_k+1/2} (\dots ) \prod\limits_{k\colon h'_k=h_k-1/2} (\dots )$ which is nothing more than \linebreak $\prod\limits_{k\colon \sigma_k=1} (\dots ) \prod\limits_{k\colon \sigma_k=-1} (\dots )$ in \eqref{diffop}.

The elliptic Vandermonde product $\prod\limits_{k<l}$ appearing in \eqref{t+} is the same product (modulo constants) as the Vandermonde-like product in any term of $(\D(A,B,C) 1) (\dots z_k\dots )$ once we have transformed (in the latter product) $\thp(z_l/z_k)$ into $\thp(z_k/z_l)$ and $\thp(1/z_k z_l)$ into $\thp(z_k z_l)$, picking up appropriate multipliers in front that will be powers of $q$ appearing the Vandermonde-like product in~\eqref{t+}. The extra powers of $q$ appearing in \eqref{t+} will also surface in the difference operator once we have performed the aforementioned transformations. Finally observe that the ratio $\frac{\varphi_{t+1,S}(h_k',h_l')}{\varphi_{t,S}(h_k,h_l)}$ reduces (modulo the power of $q$ up front already accounted for) to a ratio of only two theta functions (of the four initially present) because either $h_k'-h_l' = h_k-h_l$ or $h_k'+h_l' = h_k+h_l$ (depending whether particles $k$ and $l$ moved both in the same or in different directions).
\end{proof}

\begin{Remark}We describe how the difference operators capture the particle interpretation of the model intrinsically. In their definition specialized appropriately as in the statement of the above proposition, if two consecutive particles $k$, $k+1$ are one unit apart ($h_{k+1}-h_k=1$), the bottom one cannot move up and the top one down to collide because the summand in the difference operator is zero (indeed $\thp\big(q z_k z_{k+1}^{-1}\big) = \thp(1) = 0$ in the cross terms). Thus, the non-intersecting condition on the paths is intrinsically built into the difference operator. A~similar reasoning shows that top-most and bottom-most particles are not allowed to leave the bounding hexagon either. To exemplify, for the difference operator $\D(A,B,C)$ corresponding to the $t \to t+1$ transition (particles moving from left most vertical line to the right), we observe that the restriction on top (bottom) particle is not to cross the NE (SE) edge labeled~$C$~($A$) in Fig.~\ref{hex_par} (or indeed not to ``walk too far'' to the right by crossing the $B$ edge). However $A$ and $C$ are two of the parameters of the difference operator, and the corresponding terms in the univariate product in the appropriate summand in~\eqref{diffop} become zero once the top (bottom) particle tries to leave the hexagon. Same reasoning applies to the particles not being able to ``walk too far right''. Hence the difference operators intrinsically capture the boundary constraints of our model.
\end{Remark}

\begin{Remark} Proposition~\ref{parameterprop} is even more general, as we obtain $\binom{6}{3} = 20$ different stochastic matrices (Markov chains) from the twenty different difference operators, six of which we have already described.
\end{Remark}

We are now in a position to prove that the six matrices defined in Section~\ref{sec:dist} are indeed stochastic and measure preserving.

\begin{Theorem} \label{stochasticity}We have
\begin{gather*}
\sum_{Y} P_{t \pm}^{S,t}(X,Y) = 1, \\
\sum_{Y} {}_{t \pm}P_{S \pm}^{S,t} (X,Y) = 1, \\
\rho_{S,t \pm 1}(Y) = \sum_{X} P_{t \pm}^{S,t}(X,Y) \cdot \rho_{S,t}(X), \\
\rho_{S \pm 1,t}(Y) = \sum_{X} {}_{t \pm}P_{S \pm}^{S,t}(X,Y) \cdot \rho_{S,t}(X).
\end{gather*}
\end{Theorem}

\begin{proof} There is one way to prove these statements which works for four of the six matrices. Observe that the results for $t \pm$ follow from Theorems~\ref{thmt+} and~\ref{thmt-}, and then to observe that under $t \leftrightarrow S$, we have
\begin{gather*}
 \mathpzc{X}^{S,t} = \mathpzc{X}^{t,S} \qquad \text{and} \qquad \rho_{S,t} = \rho_{t,S},
\end{gather*}
 and then under interchanging $S$ and $t$, $P^{S,t}_{t+}$ becomes ${}_{t+}P^{S,t}_{S +}$ (and $P^{S,t}_{t-}$ becomes ${}_{t-}P^{S,t}_{S-}$, respectively). This idea also worked in the $q$-Racah and Hahn limits (see \cite{BG,BGR}).

Alternatively we can observe that the first two equalities are, by using \eqref{parmatching} and Proposition~\ref{parameterprop}, restatements of Lemma~\ref{sum} for difference operators corresponding to parameters $(A,B,C)$ (for $P^{S,t}_{t+}$), $(D,E,F)$ (for $P^{S,t}_{t-}$), $(A,B,D)$ (for ${}_{t+}P^{S,t}_{S+}$), $(B,C,F)$ (for ${}_{t+}P^{S,t}_{S-}$), $(D,E,A)$ (for ${}_{t-}P^{S,t}_{S+}$), $(E,F,C)$ (for ${}_{t-}P^{S,t}_{S-}$). Moreover, the normalizing constants that we omitted in defining the transition matrices can be recovered easily from Proposition~\ref{parameterprop}.

The last two statements are special cases of the adjointness relation. We will prove the third statement for the $t+$ operator. Similar results exist for the other five operators. We recall that $\rho_{S,t}(X)$ is nothing more than the discrete elliptic Selberg density
\begin{gather*}
 \Delta_{\lambda_X}\big(q^{2N-2}F^2\,|\,q^{N},q^{N-1} AF,q^{N-1} (pB)F,q^{N-1} CF,q^{N-1} DF,q^{N-1} EF\big)
\end{gather*}
defined in the introduction, with $\lambda_{X,k}+n-k = x_{n+1-k}$. We also define the partition $\lambda_Y$ to be the one corresponding to vertical line with abscissa $t+1$ and particle positions given by $Y$: $\lambda_{Y,k}+n-k = y_{n+1-k}$. Then one sees $\rho_{S,t + 1}(Y) = \sum_{X} P_{t+}^{S,t}(X,Y) \cdot \rho_{S,t}(X)$ is equivalent to
\begin{gather} \label{adjoint_app}
\big\langle \D(A,B,C) \mathfrak{d}_{\lambda_Y},1 \big\rangle = \big\langle \mathfrak{d}_{\lambda_Y}, \D(D',E',F') 1 \big\rangle',
\end{gather}
where the prime parameters and $\langle \cdot, \cdot \rangle '$ are defined in the previous section.

The right-hand side in \eqref{adjoint_app} equals $\sum_{\mu} \mathfrak{d}_{\lambda_Y}(\dots Fq^{\mu_k+n-k}\dots ) \Delta_{\mu}' = \Delta_{\lambda_Y}' = \rho_{S,t+1}(Y)$ (observe $\Delta'$ = $\Delta$ with prime parameters corresponds to the distribution of particles at the line $t+1$) while the left-hand side equals $\sum_{\lambda_X} \pr(\lambda_Y\,|\,\lambda_X) \cdot \Delta_{\lambda_X} = \sum_{X}P_{t+}^{S,t}(Y\,|\,X) \cdot \rho_{S,t}(X)$. The result follows.
\end{proof}

We now give a graphical description of the six Markov processes described thus far. The key is to look at the domain and codomain of the difference operators in canonical coordinates. We will exemplify with the difference operator $\D(A,B,D)$, corresponding to Markov chain $_{t+}P^{S,t}_{S+}$. Recall this Markov chain quasi-commutes with the $t \to t+1$ chain. The key is the following relation, a restatement of Theorem~\ref{stochasticity}:
\begin{gather*}
 \sum_{X} \pr(Y\,|\,X; A,B,D) \pr(X; A,B,C,D,E,F) = \pr(Y;A',B',C',D',E',F'),
 \end{gather*}
 where
\begin{gather*}
 (A',B',C',D',E',F') = \big(q^{\frac{1}{2}} A, q^{\frac{1}{2}} B, q^{-\frac{1}{2}} C, q^{\frac{1}{2}} D, q^{-\frac{1}{2}} E, q^{-\frac{1}{2}} F\big).
\end{gather*}

\begin{figure}[t]\centering
 \includegraphics[scale=0.30]{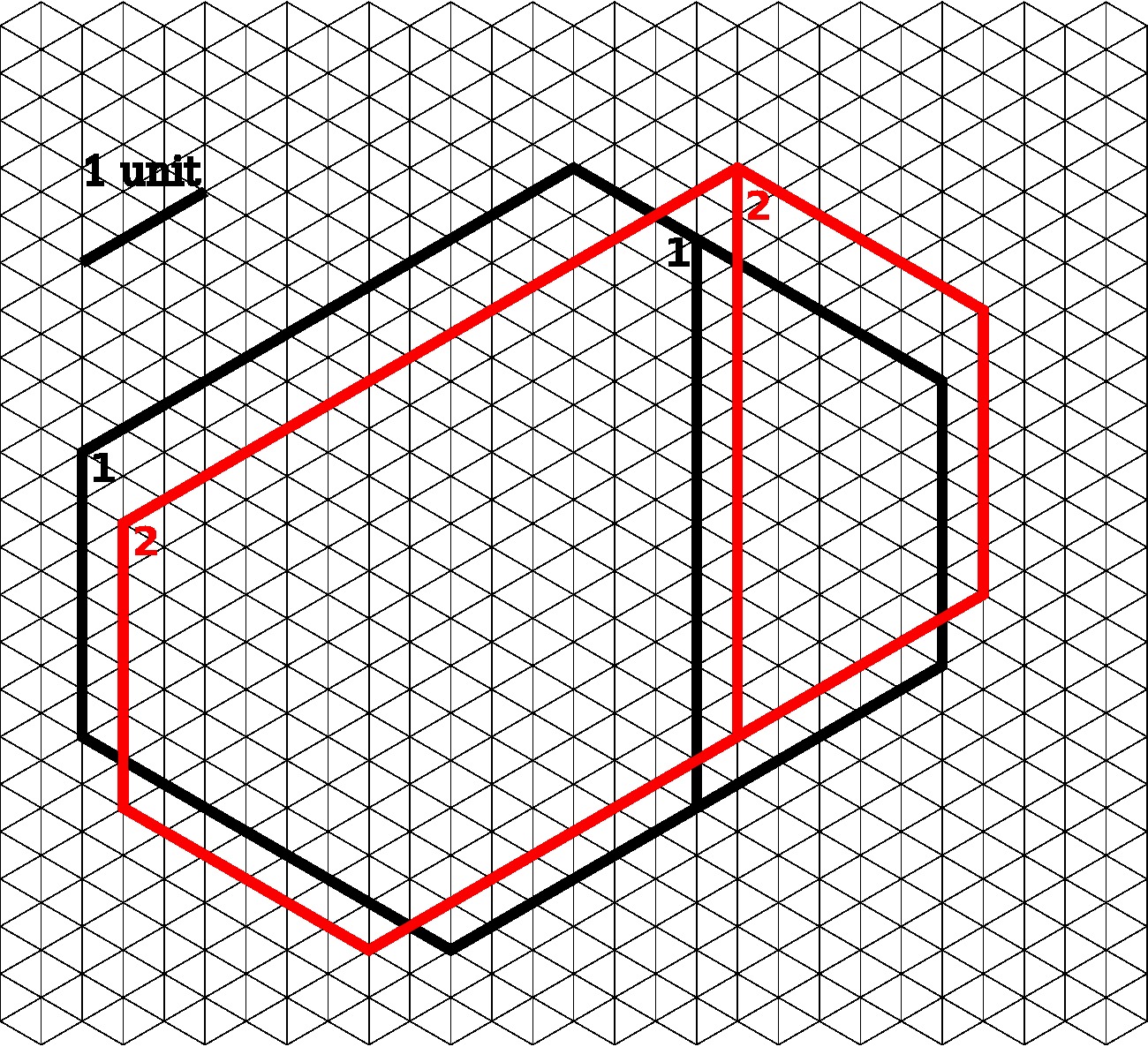}
 \caption{Action of the difference operator $\D(A,B,D)$ on a tiling of a $N=2$, $S=4$, $T=7$ hexagon drawn in canonical coordinates. The source is marked 1 and the destination 2. Only edges relevant to the model are considered: the six bordering edges and the particle line at horizontal displacement $t$ from the leftmost vertical edge. Note the slight shifting, the increase in $S$ by one, and the fact that the particle line's displacement from the left vertical edge ($=t$) is kept constant (though particle positions are shifted by a third step).}\label{ABD}
 \end{figure}

We note $_{t+}P^{S,t}_{S+}$ corresponding to the difference operator $\D(A,B,D)$ maps marked random tilings of hexagons determined by parameters $(A,B,C,D,E,F)$ to random tilings of hexagons determined by parameters $(A',B',C',D',E',F')$ (marked here refers to the particle line corresponding to parameter $t$). We figure what happens to the edges of such hexagons when parameters get shifted by $q^{\pm 1/2}$ by using equations~\eqref{canonical-edges}. Fig.~\ref{ABD} is a graphical description. In particular, we observe $_{t+}P^{S,t}_{S+}$ increases $S$ by one. Similarly for the other difference operators: they increase (decrease) $S$ or $t$ by one while leaving the other constant.

\begin{Remark} We finish by returning to the original $(q, t)$ difference operators of~\cite{EHI}, where again \textit{in this remark alone} the parameter $t$ is Macdonald's $t$. Using these operators one can construct a non-determinantal elliptic process generalizing the celebrated Macdonald processes of~\cite{bc}. A dual approach to the same construction would be to use Rains' Pieri, branching and Cauchy identities of~\cite{BCn, elliptic_littlewood}. These deserve further study. Indeed the situation is not clear even in the Hall--Littlewood limit (essentially $p, q \to 0$), as even this limit involves \emph{bounded} Cauchy/Littlewood identities, in contrast to the situation in~\cite{bc}. Moreover, in~\cite{bc}, it was the $q$-Whittaker limit that gave the authors a lot of traction. We do not know how to access a~$q$-Whittaker-like limit from the elliptic level as degenerations lead to principally specialized Macdonald polynomials, which make setting $t=0$ trivial.
\end{Remark}

\section{Perfect Markov chain sampling algorithm} \label{sec:algo}

\subsection[The $S \mapsto S+1$ step]{The $\boldsymbol{S \mapsto S+1}$ step} \label{S+1_step}

In this section, which follows closely the notation and proofs of \cite{BG, BGR}, we define a stochastic matrix
\begin{gather*}
 P^S_{S \mapsto S+1 }\colon \ \Omega(N,S,T) \times \Omega(N,S+1,T) \to [0,1],
\end{gather*}
that is measure preserving: it preserves the elliptic measure $\mu(N,S,T)$ -- the total mass of a~hexagon tiling (collection of $N$ non-intersecting lattice paths) in $\Omega(N,S,T)$. Viewed as a~Markov chain, the input for $P^S_{S \mapsto S+1 }$ is a hexagon of size $a \times b \times c$ and the output a hexagon of size $a \times (b-1) \times (c+1)$. Both the input and the output will turn out to be distributed according to $\mu(N,S,T)$ and $\mu(N,S+1,T)$ respectively.

Given a collection of non-intersecting paths $X = (X(0),\dots ,X(T)) \in \Omega(N,S,T)$, we will construct a (random) new collection $Y=(Y(0),\dots ,Y(T)) \in \Omega(N,S+1,T)$ by defining a~stochastic transition matrix $P^S_{S \mapsto S+1 }(X,Y)$. Observe that $Y(0) \in \mathpzc{X}^{S+1,0} = (0,\dots ,N-1)$ is unambiguously defined. Next we perform a \textit{sequential (inductive) update}. That is, we describe how to obtain $Y(t+1)$ given knowledge of $Y(0),\dots ,Y(t)$ and $X$. $Y(t+1)$ will be defined according to the distribution
\begin{gather*}
 \pr(Y(t+1)=Z) = \frac{P_{t+}^{S+1,t}(Y(t),Z) \cdot {}_{t+}P_{S-}^{S+1,t+1}(Z,X(t+1))}{\big(P_{t+}^{S+1,t} \cdot {}_{t+}P_{S-}^{S+1,t+1}\big)(Y(t),X(t+1))} \\
\hphantom{\pr(Y(t+1)=Z)}{} = \frac{{}_{t-}P_{S+}^{S,t+1}(X(t+1),Z) \cdot P_{t-}^{S+1,t+1}(Z,Y(t))}{\big({}_{t-}P_{S+}^{S,t+1} \cdot P_{t-}^{S+1,t+1}\big)(X(t+1),Y(t))},
\end{gather*}
 where the last equality follows from the fact that{\samepage
 \begin{gather*}
 \rho_{S+1,t+1}(A) P^{S+1,t+1}_{t-}(A,B) = \rho_{S+1,t}(B)P^{S+1,t}_{t+}(B,A)
 \end{gather*}
 (this is nothing more than the equality $\pr(A \cap B) = \pr(A) \pr(B\,|\,A) = \pr(B) \pr(A\,|\,B)$).}

We define the matrix $P_{S \mapsto S+1}\colon \Omega(N,S,T) \times \Omega(N,S+1,T) \to [0,1]$ by
\begin{gather*}
P_{S \mapsto S+1}(X,Y) =
\begin{cases}
 \displaystyle \prod\limits_{t=0}^{T-1} \frac{P_{t+}^{S+1,t}(Y(t),Y(t+1)) \cdot {}_{t+}P_{S-}^{S+1,t+1}(Y(t+1),X(t+1))}{\big(P_{t+}^{S+1,t} \cdot {}_{t+}P_{S-}^{S+1,t+1}\big)(Y(t),X(t+1))}, \\
 \displaystyle \quad \text{if \ } \prod\limits_{t=0}^{T-1} \big(P_{t+}^{S+1,t} \cdot {}_{t+}P_{S-}^{S+1,t+1}\big)(Y(t),X(t+1)) > 0, \\
 0, \ \text{otherwise}.
\end{cases}
\end{gather*}

\begin{Theorem}
The matrix $P_{S \mapsto S+1}$ is stochastic and measure preserving, in the sense that
\begin{gather} \label{meas-pres}
\mu(N,S+1,T)(Y) = \sum_{X \in \Omega(N,S,T)} P_{S \mapsto S+1} (X,Y) \mu(N,S,T)(X).
\end{gather}
\end{Theorem}

\begin{proof} (following \cite{BG})
We want to show that
\begin{gather*}
 \sum_{Y} P_{S \mapsto S+1}(X,Y) = \sum_{Y} \prod_{t=0}^{T-1} \frac{P_{t+}^{S+1,t}(Y(t),Y(t+1)) \cdot {}_{t+}P_{S-}^{S+1,t+1}(Y(t+1),X(t+1))}{\big(P_{t+}^{S+1,t} \cdot {}_{t+}P_{S-}^{S+1,t+1}\big)(Y(t),X(t+1))} = 1,
\end{gather*}
where the sum is taken over all $Y = (Y(0),\dots,Y(T)) \in \Omega(N,S+1,T)$ such that
\begin{gather} \label{cond_sum}
\prod_{t=0}^{T-1} \big(P_{t+}^{S+1,t} \cdot {}_{t+}P_{S-}^{S+1,t+1}\big)(Y(t),X(t+1)) > 0.
\end{gather}

We first sum over $Y(T)$ and because $Y(T)$ is distributed according to a singleton measure, the respective sum is one. Next we deal with the sum
\begin{gather*}
 \sum_{Y(T-1)} \frac{P_{t+}^{S+1,T-2}(Y(T-2),Y(T-1)) \cdot {}_{t+}P_{S-}^{S+1,T-1}(Y(T-1),X(T-1))}{\big(P_{t+}^{S+1,T-2} \cdot {}_{t+}P_{S-}^{S+1,T-1}\big)(Y(T-2),X(T-1))}
\end{gather*}
 over $Y(T-1)$ satisfying $\big(P_{t+}^{S+1,T-1} \cdot {}_{t+}P_{S-}^{S+1,T}\big)(Y(T-1),X(T)) > 0$ (because of \eqref{cond_sum}). Because of the quasi-commutation relations from Theorem~\ref{commrel}, we have
\begin{gather*}
 \big(P_{t+}^{S+1,T-1} \cdot {}_{t+}P_{S-}^{S+1,T}\big)(Y(T-1),X(T)) \\
 \qquad {}= \big({}_{t+}P_{S-}^{S+1,T-1} \cdot {}_{t+}P_{S-}^{S,T-1}\big)(Y(T-1),X(T)) \\
 \qquad {}\geq {}_{t+}P_{S-}^{S+1,T-1}(Y(T-1),X(T-1)) \cdot P_{t+}^{S,T-1}(X(T-1),X(T)).
\end{gather*}
 We are summing over $Y(T-1)$ such that the left-hand side above is non-vanishing, but if it vanishes, then by the above inequality so does ${}_{t+}P_{S-}^{S+1,T-1}(Y(T-1),X(T))$. This means we can drop the condition that $\big(P_{t+}^{S+1,T-1} \cdot {}_{t+}P_{S-}^{S+1,T}\big)(Y(T-1),X(T)) > 0$ and sum over all $Y(T-1)$. We obtain one for this sum (the denominator is independent of the summation variable, and summing the numerator over $Y(T-1)$ we obtain the denominator). We next sum inductively over $Y(T-2)$ and so on until we are left over with a sum over $Y(0)$. This sum only has one term, so we obtain the desired result.

To show $P_{S \mapsto S+1}$ preserves the measure $\mu$, observe first that
\begin{gather*}
 \mu(N,S,T)(X) = m_0(X(0)) \cdot P_{t+}^{S,0}(X(0),X(1))\cdots P^{S,T-1}_{t+}(X(T-1),X(T)),
\end{gather*}
where $m_0$ is the unique probability measure on any singleton set (in this case $\mathpzc{X}^{S,0}$). Then the right-hand side of~\eqref{meas-pres} becomes
\begin{gather}
 \sum_{X} m_0(X(0)) \prod_{t=0}^{T-1} P_{t+}^{S,t}(X(t),X(t+1)) \nonumber\\
\qquad {}\times \prod_{t=0}^{T-1} \frac{P_{t+}^{S+1,t}(Y(t),Y(t+1)) \cdot {}_{t+}P_{S-}^{S+1,t+1}(Y(t+1),X(t+1))}{\big(P_{t+}^{S+1,t} \cdot {}_{t+}P_{S-}^{S+1,t+1}\big)(Y(t),X(t+1))}.\label{sum1}
\end{gather}
 Pulling out factors independent of the summation variables, replacing $1=m_0(X(0))$ with $1=m_0(Y(0))$, using ${}_{t+} P_{S-}^{S+1,T}(Y(T), X(T)) = {}_{t+} P_{S-}^{S+1,0}(Y(0), X(0)) = 1$ and $P_{t+}^{S,t} \cdot {}_{t+}P_{S-}^{S,t+1} = {}_{t+}P_{S-}^{S,t} \cdot P_{t+}^{S-1,t}$, we transform~\eqref{sum1} into
\begin{gather*}
 m_0(Y(0)) \prod_{t=0}^{T-1} P_{t+}^{S+1,t}(Y(t),Y(t+1)) \sum_{X} \prod_{t=0}^{T-1} \frac{{}_{t+} P_{S-}^{S+1,t}(Y(t),X(t)) \cdot P_{t+}^{S,t}(X(t),X(t+1))} {\big( {}_{t+}P_{S-}^{S,t} \cdot P_{t+}^{S+1,t}\big)(Y(t),X(t+1))}.
\end{gather*}
 Now we sum first over $X(T)$, then over $X(T-1)$ and so on like in the previous argument to finally obtain on the left-hand side the desired result
\begin{gather*}
 m_0(Y(0))\prod_{t=0}^{T-1} P_{t+}^{S+1,t}(Y(t),Y(t+1)) = \mu(N,S+1,T)(Y). \tag*{\qed}
\end{gather*}\renewcommand{\qed}{}
\end{proof}

\subsection[Algorithmic description of the $S \mapsto S+1$ step]{Algorithmic description of the $\boldsymbol{S \mapsto S+1}$ step}

As before, whenever possible, we try to keep the notation similar to~\cite{BG}. For $x \in \N$ we define
\begin{gather*}
\mathpzc{p}(x) = \frac{q\thp\big(q^{x-t-S+T-1}, q^{x-t-T-1} v_1, q^{x+t+1} v_2, q^{x-t-S-1} v_1 v_2\big)}{\thp\big(q^{x+1},q^{x-2t-S-1} v_1, q^{x-S+T+1} v_2, q^{x-T+1} v_1 v_2\big)} \frac{\thp\big(q^{2x-t-S+1} v_1 v_2\big)}{\thp\big(q^{2x-t-S-1} v_1 v_2\big)}.
\end{gather*}

Note $\mathpzc{p}$ also depends on $S$, $T$, $v_1$, $v_2$, $q$, $p$, but we will omit these for simplicity of notation. Also note $p$ is an elliptic function of $q$, $q^S$, $q^T$, $q^t$, $v_1$, $v_2$, $q^x$. Consider (again omitting most parameter dependence)
\begin{gather*}
P(x;s) = \prod_{i=1}^{s} \mathpzc{p}(x+i-1).
\end{gather*}
$P$ is just a ratio of five length-$s$ theta-Pochhammer symbols over five others (multiplied by~$q^{s-1}$ to make everything elliptic). We define the following probability distribution on the set $\{0,1,{\dots},n\}$:
\begin{gather} \label{split_dist}
\pr(s) = D(x;n)(s) = \frac{P(x;s)}{\sum\limits_{j=0}^{n} P(x;j)}.
\end{gather}

For the exact sampling algorithm, given $X = (X(0),\dots,X(T)) \in \Omega(N,S,T)$, we will construct $Y = (Y(0),\dots,Y(T)) \in \Omega(N,S+1,T)$ by first observing that $Y(0) = (0,\dots,N-1)$ is uniquely defined. We then perform $T$ sequential updates. At step $t+1$ we obtain $Y(t+1)$ based on~$Y(t)$ and $X(t+1)$. Suppose $X(t+1) = (x_1,\dots,x_N) \in \mathpzc{X}^{S,t+1}$ and $Y(t) = (y_1,\dots,y_N) \in \mathpzc{X}^{S+1,t}$. We want to define/sample $Y(t+1) = (z_1,\dots,z_N) \in \mathpzc{X}^{S+1,t+1}$. $Y(t)$ and $X(t+1)$ satisfy $x_i - y_i \in \{0,-1,1\}$ (follows by construction from $(P_{t+}^{S+1,t} \cdot {}_{t+}P_{S-}^{S+1,t+1})(Y(t),X(t+1)) > 0$). We thus have three cases, and in each case we describe how to choose $z_i$.

\begin{itemize}\itemsep=0pt
\item \textbf{Case 1:} Consider all $i$ such that $x_i - y_i = 1$. Then $z_i = x_i$ is forced;
\item \textbf{Case 2:} Consider all $i$ such that $x_i - y_i = -1$. Then $z_i = y_i$ is forced;
\item \textbf{Case 3:} For the remaining indices, group them in blocks and consider one such called a~$(k,l)$-block (where $k$ is the smallest particle location in the block, and $l$ is the number of particles in the block). That is, we have $y_{i-1}< k-1$, $y_{i+l} > k+l$ and the block consists of
\begin{gather*}
 x_i = y_i = k,\ x_{i+1} = y_{i+1} = k+1, \ \dots, \ x_{i+l-1} = y_{i+l-1} = k+l-1.
\end{gather*}
For each such block independently, we sample a random variable $\xi$ according to the distribution $D(k;l)$. We set $z_i = x_i$ for the first $\xi$ consecutive positions in the block, and we set $z_i = x_i+1$ for the remainder of the $l- \xi$ positions. We provide an example in Fig.~\ref{algorithmic_step} below.

\begin{figure}[t] \centering
 \begin{tikzpicture}[scale=0.45]
 \node at (0, 0) {$ \bullet$};
 \node at (0, 1) {$ \circ$};
 \node at (0, 2) {$ \circ$};
 \node at (0, 3) {$ \bullet$};
 \node at (0, 4) {$ \circ$};
 \node at (0, 5) {$ \bullet$};
 \node at (0, 6) {$ \bullet$};
 \node at (0, 7) {$ \bullet$};
 \node at (0, 8) {$ \circ$};
 \node at (0, 9) {$ \bullet$};
 \node at (0, 10) {$ \circ$};

 \node at (2, 0) {$ \bullet$};
 \node at (2, 1) {$ \circ$};
 \node at (2, 2) {$ \circ$};
 \node at (2, 3) {$ \circ$};
 \node at (2, 4) {$ \bullet$};
 \node at (2, 5) {$ \bullet$};
 \node at (2, 6) {$ \bullet$};
 \node at (2, 7) {$ \bullet$};
 \node at (2, 8) {$ \circ$};
 \node at (2, 9) {$ \circ$};
 \node at (2, 10) {$ \bullet$};
 \node at (2, 11) {$ \circ$};

 \node at (6, 0) {$ \circ$};
 \node at (6, 1) {$ \bullet$};
 \node at (6, 2) {$ \circ$};
 \node at (6, 3) {$ \circ$};
 \node at (6, 4) {$ \bullet$};
 \node at (6, 5) {$ \bullet$};
 \node at (6, 6) {$ \bullet$};
 \node at (6, 7) {$ \bullet$};
 \node at (6, 8) {$ \circ$};
 \node at (6, 9) {$ \bullet$};
 \node at (6, 10) {$ \circ$};

 \node at (8, 0) {$ \circ$};
 \node at (8, 1) {$ \bullet$};
 \node at (8, 2) {$ \circ$};
 \node at (8, 3) {$ \circ$};
 \node at (8, 4) {$ \bullet$};
 \node at (8, 5) {$ \bullet$};
 \node at (8, 6) {$ \circ$};
 \node at (8, 7) {$ \bullet$};
 \node at (8, 8) {$ \bullet$};
 \node at (8, 9) {$ \circ$};
 \node at (8, 10) {$ \bullet$};
 \node at (8, 11) {$ \circ$};

 \draw[thick, smooth] (0, 0)--(2, 0);
 \draw[thick, smooth] (0, 3)--(2, 4);
 \draw[thick, smooth] (0, 5)--(2, 5);
 \draw[thick, smooth] (0, 6)--(2, 6);
 \draw[thick, smooth] (0, 7)--(2, 7);
 \draw[thick, smooth] (0, 9)--(2, 10);

 \draw[thick, smooth] (6, 1)--(8, 1);
 \draw[thick, smooth] (6, 4)--(8, 4);
 \draw[thick, smooth] (6, 5)--(8, 5);
 \draw[thick, smooth] (6, 6)--(8, 7);
 \draw[thick, smooth] (6, 7)--(8, 8);
 \draw[thick, smooth] (6, 9)--(8, 10);

 \node at (1, -1) {$S$};
 \node at (7, -1) {$S+1$};

 \node[rotate=90] at (0, 13) {$X(t)$};
 \node[rotate=90] at (2, 13) {$X(t+1)$};
 \node[rotate=90] at (6, 13) {$Y(t)$};
 \node[rotate=90] at (8, 13) {$Y(t+1)$};

 \draw[thick, smooth,->] (11, 10)--(10, 10);
 \node[right] at (11, 10) {could not jump (first case)};
 \draw[thick, smooth,->] (11, 6)--(10, 6);
 \node[right] at (11, 6) {split point determined by third case};
 \draw[thick, smooth,->] (11, 1)--(10, 1);
 \node[right] at (11, 1) {was forced to jump (second case)};

 \draw[thick, smooth,<->] (9, 4)--(9, 6);
 \node at (9.5, 5) {$2$};

 \node[rotate=90] at (4, 5.5) {$(k,l)$-block};

 \draw (2.1, 7.3)--(2.4, 7.3)--(2.4, 3.7)--(2.1, 3.7);
 \draw (5.9, 7.3)--(5.6, 7.3)--(5.6, 3.7)--(5.9, 3.7);

 \draw[thick, smooth, ->] (3.3, 4.5)--(2.7, 4.5);
 \draw[thick, smooth, ->] (3.3, 5.5)--(2.7, 5.5);
 \draw[thick, smooth, ->] (3.3, 6.5)--(2.7, 6.5);

 \draw[thick, smooth, <-] (5.3, 4.5)--(4.7, 4.5);
 \draw[thick, smooth, <-] (5.3, 5.5)--(4.7, 5.5);
 \draw[thick, smooth, <-] (5.3, 6.5)--(4.7, 6.5);
 \end{tikzpicture}
 \caption{Sample block split.} \label{algorithmic_step}
\end{figure}
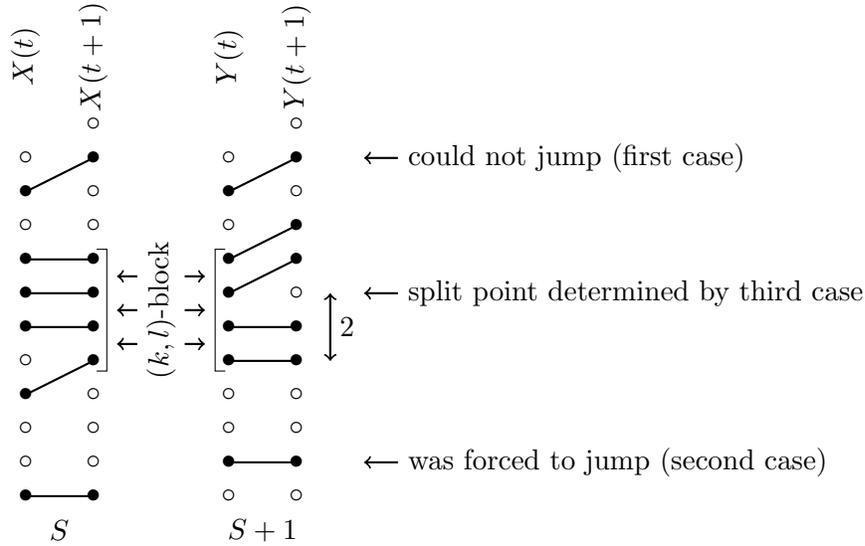
\end{itemize}

\begin{Theorem} \label{correctness}
By constructing $Y$ this way, we have simulated a $S \mapsto S+1$ step of the Markov chain $P_{S \mapsto S+1}$.
\end{Theorem}

\begin{proof}
We perform the following computation (and are interested in Case 3 described above, that is on how to split a $(k,l)$-block; note $x_i = y_i$ in the case of interest):
\begin{gather}
 \pr(Y(t+1) \!=\! Z) = \frac{P_{t+}^{S+1,t}(Y(t),Z) \cdot {}_{t+}P_{S-}^{S+1,t+1}(Z,X(t+1))}{\big(P_{t+}^{S+1,t} \cdot {}_{t+}P_{S-}^{S+1,t+1}\big)(Y(t),X(t+1))} \!=\! (\text{factors independent of $Z$}) \nonumber\\
\qquad{} \times \prod_{i\colon z_i = y_i} q^{-y_i-N+1} \frac{\thp\big(q^{y_i+T-S-t-1},q^{y_i-T-S-t-1}v_1,q^{y_i+t+1} v_2, q^{y_i+N-t} v_1 v_2\big)}{\thp\big(q^{2 y_i-t-S} v_1 v_2\big)} \nonumber\\
\qquad{} \times \prod_{i\colon z_i = y_i+1} q^{-y_i} \frac{\thp\big(q^{y_i-S-N},q^{y_i-2t-S-1}v_1,q^{y_i+T+1} v_2, q^{y_i-T+1} v_1 v_2\big)}{\thp\big(q^{2 y_i-t-S} v_1 v_2\big)} \nonumber\\
\qquad{} \times \prod_{i\colon z_i = x_i} q^{-x_i} \frac{\thp\big(q^{x_i-S-N},q^{x_i-t-T-1}v_1,q^{x_i+T+1} v_2, q^{x_i-t-S-1} v_1 v_2\big)}{\thp\big(q^{2 x_i-t-S-1} v_1 v_2\big)} \nonumber\\
\qquad{} \times \prod_{i\colon z_i = x_i+1} q^{-x_i-N} \frac{\thp\big(q^{x_i+1},q^{x_i-T-S-t-1}v_1,q^{x_i-S+T+1} v_2, q^{x_i+N-t} v_1 v_2\big)}{\thp\big(q^{2 x_i-t-S+1} v_1 v_2\big)}.\label{split_prob}
\end{gather}

We thus see the blocks split independently due to the evident product structure. The probability that the first $j$ particles in a~$(k,l)$-block stay put from $Y(t)$ to $Y(t+1)$ (and the rest of $l-j$ jump by one) is, by using the above formula
\begin{gather*}
\prod_{i=0}^{j-1} \frac{q\thp\big(q^{k+i-t-S+T-1}, q^{k+i-t-T-1} v_1, q^{k+i+t+1} v_2, q^{k+i-t-S-1} v_1 v_2\big)}{\thp\big(q^{2k+2i-t-S-1} v_1 v_2\big)} \\
{} \times\!\prod_{i=j}^{l-1}\! \frac{\thp\big(q^{k+i+1},q^{k+i-2t-S-1} v_1, q^{k+i-S+T+1} v_2, q^{k+i-T+1} v_1 v_2\big)}{\thp\big(q^{2k+2i-t-S+1} v_1 v_2\big)} \!\times\! (\text{factors independent of $j$}),
\end{gather*}
where in \eqref{split_prob} we have gauged away everything independent of the split position~$j$. This probability is nothing more than the distribution $D$ we defined in~\eqref{split_dist}. This finishes the proof.
\end{proof}

\subsection[Algorithmic description of the $S \mapsto S-1$ step]{Algorithmic description of the $\boldsymbol{S \mapsto S-1}$ step}

Similar to the $P_{S \mapsto S+1}$ matrix described in the previous two sections, we can construct a $P_{S \mapsto S-1}$ measure preserving Markov chain that takes random tilings in $\Omega(N,S,T)$ and maps them to random tilings in $\Omega(N,S-1,T)$. We proceed exactly as in Section~\ref{S+1_step} and will omit most details and theorems as they transfer verbatim from Section~\ref{S+1_step}. Given $X \in \Omega(N,S,T)$ and $Y(0),Y(1),\dots,Y(t)$ already defined inductively, we choose $Y(t+1)$ from the distribution:
\begin{gather*}
\pr (Y(t+1) = Z) = \frac{P_{t+}^{S-1,t}(Y(t),Z) \cdot {}_{t+}P_{S+}^{S-1,t+1}(Z,X(t+1))}{\big(P_{t+}^{S-1,t} \cdot {}_{t+}P_{S+}^{S-1,t+1}\big) (Y(t),X(t+1))}.
\end{gather*}

We define
\begin{gather*}
P_{S \mapsto S-1}(X,Y) =
\begin{cases}
 \displaystyle \prod\limits_{t=0}^{T-1} \frac{P_{t+}^{S-1,t}(Y(t),Y(t+1)) \cdot {}_{t+}P_{S+}^{S-1,t+1}(Y(t+1),X(t+1))}{\big(P_{t+}^{S-1,t} \cdot {}_{t+}P_{S+}^{S-1,t+1}\big)(Y(t),X(t+1))}, \\
\displaystyle \quad \text{if \ } \prod\limits_{t=0}^{T-1} \big(P_{t+}^{S-1,t} \cdot {}_{t+}P_{S+}^{S-1,t+1}\big)(Y(t),X(t+1)) > 0, \\
 0, \ \text{otherwise}.
\end{cases}
\end{gather*}

We will also sketch the algorithm for sampling using $P_{S \mapsto S-1}$. We need to define the equivalent for $\mathpzc{p}$ from the previous section. For $x \in \N$ we define
\begin{gather*}
\mathpzc{p}'(x) = \frac{q\thp\big(q^{x-t-N-1}, q^{x-t-2S} v_1, q^{x+t} v_2, q^{x-t+N-1} v_1 v_2\big)}{\thp\big(q^{x-S-N+1},q^{x-2t-S} v_1, q^{x+S} v_2, q^{x-S+N+1} v_1 v_2\big)} \frac{\thp\big(q^{2x-t-S+1} v_1 v_2\big)}{\thp\big(q^{2x-t-S-1} v_1 v_2\big)}.
\end{gather*}

As before, $\mathpzc{p}'$ is an elliptic in $q$, $q^S$, $q^N$, $q^t$, $v_1$, $v_2$, $q^x$. We also define $P'(x;s) = \prod\limits_{i=1}^{s} \mathpzc{p}'(x+i-1)$ and the following distribution on $\{0,1,\dots,n\}$:
\begin{gather*} 
\pr(s) = D'(x;n)(s) = \frac{P'(x;s)}{\sum\limits_{j=0}^{n} P'(x;j)}.
\end{gather*}

Assuming we have $X \in \Omega(N,S,T)$ with $X(t+1) = (x_1<\dots<x_N)$ and inductively $Y(0),\dots,Y(t) = (y_1<\dots<y_N)$, we sample $Y(t+1) = (z_1<\dots<z_N)$ by first observing that $x_i - y_i \in \{0,1,2 \}$ (because $(P_{t+}^{S-1,t} \cdot {}_{t+}P_{S+}^{S-1,t+1})(Y(t),X(t+1)) > 0$) and then performing appropriate updates for the following three simple cases:

\begin{itemize}\itemsep=0pt
\item \textbf{Case 1:} For all $i$ with $x_i - y_i = 0$ we set $z_i = x_i$;
\item \textbf{Case 2:} For all $i$ with $x_i - y_i = 2$ we set $z_i = y_i+1$;
\item \textbf{Case 3:} For the remaining indices (for which $x_i - y_i=1$), group them in blocks and consider one such called a $(k,l)$-block. That is, we have $y_{i-1}< k-1$, $y_{i+l} > k+l$ and the block consists of
\begin{gather*}
 x_i = y_i+1 = k,\ x_{i+1} = y_{i+1}+1 = k+1, \ \dots, \ x_{i+l-1} = y_{i+l-1}+1 = k+l-1.
\end{gather*}
For each such block independently, we sample a random variable $\xi$ according to the distribution $D'(k;l)$. We set $z_i = y_i$ for the first $\xi$ consecutive positions in the block, and we set $z_i = y_i+1$ for the remainder of the $l - \xi$ positions. See Fig.~\ref{algorithmic_step}.
\end{itemize}

An analogous of Theorem~\ref{correctness} exists and is proved in a similar way to show the above three steps are all that is necessary to simulate the Markov chain $P_{S \mapsto S-1}$.

\section{Correlation kernel and determinantal representations} \label{sec:corr}

In this section we will show the process $X(t)$ corresponding to a tiling of the hexagon is determinantal with correlation kernel given in terms of the elliptic biorthogonal functions of Spiridonov and Zhedanov \cite{BCn,SZ1}. We start by a brief overview of the necessary facts about biorthogonal functions, and continue with the heart of the proof: an application of the Eynard--Mehta theorem.

\subsection{A brief overview of elliptic biorthogonal functions}

We will first gather together a few results about univariate discrete elliptic biorthogonal functions. The notation and exposition will mostly be following \cite{BCn}. We will need to make brief use of univariate \textit{interpolation abelian functions}. They were introduced in \cite{BCn,EHI} and are, for a fixed integer $l$, $BC_1$-symmetric ratios of $BC_1$-symmetric theta functions of degree $l$ with prescribed poles and zeros. To wit
\begin{gather*}
R^*_l(x;a,b) = \frac{\thp\big(a x^{\pm 1};q\big)_{l}}{\thp\big(b q^{-l} x^{\pm 1};q\big)_{l}}.
\end{gather*}

Observe $R^*_l$ has zeros at finitely many $q$-shifts of $a$ and poles at finitely many $q$-shifts of~$b$ (up to taking reciprocals and shifting by $p$). The univariate biorthogonal functions $R_l(x;t_0\colon t_1,t_2,t_3;$ $u_0,u_1)$ of \cite{SZ1} can be defined in terms of the interpolation functions following~\cite{BCn}. Fix $|p|<1$,~$q$ as well as six parameters $t_0$, $t_1$, $t_2$, $t_3$, $u_0$, $u_1$ such that $t_0 t_1 t_2 t_3 u_0 u_1 = pq$. Then (dependence on~$p$,~$q$ implied but not written)
\begin{gather*}
R_l(x;t_0\colon t_1,t_2,t_3;u_0,u_1) = \sum_{0 \leq k \leq l} d_k R^*_k(x;t_0,u_0) = d_l R^*_l + \text{lower order terms},
\end{gather*}
where the formula for the $d_k$'s is explicitly given in \cite{BCn} and is independent of $x$ (but of course depends on $t_0$, $t_1$, $t_2$, $t_3$, $u_0$, $u_1$, $q$, $p$ and $k$). These functions have poles at shifts of $u_0^{\pm 1}$ (we will say $u_0$ \textit{controls the poles} of $R_l(x;t_0\colon t_1,t_2,t_3;u_0,u_1)$). They are elliptic in the six parameters provided the balancing condition is satisfied, as well as in the variable~$x$. Furthermore, if in addition to the balancing condition, one also has
\begin{gather*}
t_0 t_1 = q^{-m}
\end{gather*}
for some $m > 0$ an integer, the functions with poles controlled by $u_0$ and those with poles controlled by $u_1$ satisfy the following discrete biorthogonality relation on $\{0,\dots,m\}$
\begin{gather*}
 \sum_{0 \leq s \leq m} R_l\big(t_0 q^s;t_0\colon t_1,t_2,t_3;u_0,u_1\big) R_k\big(t_0 q^s;t_0\colon t_1,t_2,t_3;u_1,u_0\big) \\
 \qquad {}\times \Delta_s\big(t_0^2\,|\,q,t_0 t_1,t_0 t_2, t_0 t_3, t_0 u_0, t_0 u_1\big) = \delta_{l,k} c_l,
\end{gather*}
where $\Delta_s$ is the univariate delta symbol defined in the Introduction and
\begin{gather*}
 c_l = {\rm const} \cdot \Delta_l\big(1/u_0 u_1\,|\,q,t_0 t_1, t_0 t_2, t_0 t_3, 1/t_0 u_0, 1/t_0 u_1\big)^{-1} \\
 \hphantom{c_l}{} = {\rm const} \cdot \Delta\big(\hat{t}_0^2\,|\,q, \hat{t}_0 \hat{t}_1,\hat{t}_0 \hat{t}_2,\hat{t}_0 \hat{t}_3, \hat{t}_0 \hat{u}_0,\hat{t}_0 \hat{u}_1\big)^{-1}.
\end{gather*}

The ``hat'' parameters are defined by the relations
\begin{gather*} 
\hat{t}_0 = \sqrt{\frac{t_0 t_1 t_2 t_3}{pq}}, \qquad \hat{t}_0 \hat{t_i} = t_0 t_i, \qquad \frac{\hat{u}_j}{\hat{t}_0} = \frac{u_j}{t_0}
\end{gather*}
for $i=1,2,3$ and $j=0,1$. The ``hat'' is an involution and the hat parameters satisfy the same balancing conditions as the original parameters. They are important because by hatting we can exchange the variable and the index of the biorthogonal functions as follows:
\begin{gather*} 
 R_l\big(t_0 q^s; t_0\colon t_1,t_2,t_3;u_0,u_1\big) = R_s\big(\hat{t}_0 q^l; \hat{t}_0\colon \hat{t}_1,\hat{t}_2,\hat{t}_3;\hat{u}_0,\hat{u}_1\big).
\end{gather*}

The biorthogonal functions described above have $t_0$ as a special normalization parameter, distinguished among the $t_i$'s. That is, $R_l(t_0;t_0\colon t_1,t_2,t_3;u_0,u_1) = 1$. The normalized difference operators of Section~\ref{sec:diff_op} act on the biorthogonal functions as follows
\begin{gather}
 \D(u_0,t_0,t_1) R_l\big(\big(q^{1/2}t_0\big) q^s; q^{1/2}t_0\colon q^{1/2}t_1,q^{-1/2}t_2,q^{-1/2}t_3;q^{1/2}u_0,q^{-1/2}u_1\big) \nonumber\\
 \qquad {} = R_l\big(t_0 q^s; t_0\colon t_1,t_2,t_3;u_0,u_1\big). \label{action}
\end{gather}

Finally, we can exchange $t_0$ with another $t_j$ at the choice of picking up a factor (this is in essence a renormalization so that $R$ takes value one at $t_j$ rather than at $t_0$):
\begin{gather} \label{renorm}
R_l(x;t_1\colon t_0,t_2,t_3;u_0,u_1) = \frac{R_l(x;t_0\colon t_1,t_2,t_3;u_0,u_1)}{R_l(t_1;t_0\colon t_1,t_2,t_3;u_0,u_1)}.
\end{gather}

\subsection{Determinantal representations}

We now show the processes $t \mapsto t \pm 1$ are determinantal point processes. For a review of such processes we direct the reader to~\cite{borodin_det}. We will do the calculation for the $t \mapsto t-1$ Markov process as it leads to less complicated formulas, but analogous results hold for $t \mapsto t+1$.

For the remainder, it is convenient to change the set of parameters $A$, $B$, $C$, $D$, $E$, $F$ to the set $t_0$, $t_1$, $t_2$, $t_3$, $u_0$, $u_1$ in order for certain symmetries to become more prominent using the following:
\begin{gather} \label{relabel}
 A=t_2, \qquad q^{N-1} B = u_1, \qquad C=t_3, \qquad D = t_1, \qquad q^{N-1} E = u_0, \qquad F = t_0 .
\end{gather}
Note these parameters depend on $t$ (the time parameter), and such dependence will be made more explicit when it becomes important. Notation is as in the previous section. Note \smash{$u_0 u_1 t_0 t_1 t_2 t_3 \!=\! q$}. Since the balancing condition for the biorthogonal functions requires a~$pq$ on the right-hand side, we will again multiply~$u_1$ by~$p$.

We state the Eynard--Mehta theorem, in a ``decreasing-time'' form convenient for our computations (see \cite{borodin_det,eynardmehta} for a review and \cite{borodinrains} for an elementary proof):

\begin{Theorem} \label{eynardmehta}
 Assume we are given the following:
\begin{itemize}\itemsep=0pt
 \item a discrete biorthonormal system $\big(f_l^t,g_l^t\big)_{l \geq 0}$ on $l_2(\{0,1,\dots,L\})$ for each time $t = 0,\dots,T$;
 \item a matrix
\begin{gather*}
 v_{t \to t-1}(x,y) = \sum_{l \geq 0} f_l^{t-1}\big(t_0^{t-1} q^x\big) g_l^{t}\big(t_0^{t} q^y\big),
\end{gather*}
for $n \geq 0$, $t = 1,\dots,T$ and a parameter $t_0$ changing with time;
 \item a discrete time Markov chain $X(t)$ $($with time decreasing from $T$ to $0)$ taking values in state spaces $\mathpzc{X}^t$ $($set of possible particle positions at time~$t)$ with one-dimensional distributions proportional to
\begin{gather*}
 \det_{1 \leq k,l \leq N} \big(f^t_{k-1}\big(t_0^t q^{x_l}\big)\big) \det_{1 \leq k,l \leq N} \big(g^t_{k-1}\big(t_0^t q^{x_l}\big)\big)
\end{gather*}
and transition probabilities proportional to
\begin{gather*}
 \frac{\det_{1 \leq k,l \leq N} (v_{t \to t-1} (x_k, y_l)) \det_{1 \leq k,l \leq N} \big(f^{t-1}_{k-1}\big(t_0^{t-1} q^{y_l}\big)\big)} {\det_{1 \leq k,l \leq N} \big(f^{t}_{k-1}\big(t_0^t q^{x_l}\big)\big)}.
\end{gather*}
\end{itemize}
Then
\begin{gather*}
 \pr (x_1 \in X(\tau_1),\dots,x_s \in X(\tau_s)) = \det_{1 \leq k,l \leq s} (K(\tau_k,x_k;\tau_l,x_l))
\end{gather*}
where
\begin{gather*}
K(\tau_1, x_1; \tau_2, x_2) = \begin{cases}
 \displaystyle \sum\limits_{s \geq 0} f_s^{\tau_1}\big(t_0^{\tau_1} q^{x_1}\big) g_s^{\tau_2}\big(t_0^{\tau_2} q^{x_2}\big), & \text{if \ } \tau_1 \geq \tau_2, \\
\displaystyle -\sum\limits_{s \geq N} f_s^{\tau_1}\big(t_0^{\tau_1} q^{x_1}\big) g_s^{\tau_2}\big(t_0^{\tau_2} q^{x_2}\big), & \text{if \ } \tau_1 < \tau_2.
 \end{cases}
\end{gather*}
\end{Theorem}

The first step in showing the required determinantal formulas needed to apply the Eynard--Mehta theorem is the following determinantal formula due to Warnaar~\cite{War}:

\begin{Lemma} \label{warnaardet}
 We have
\begin{gather*}
\det_{1 \leq k,l \leq n} R_{l-1}(z_k;t_0\colon t_1,t_2,t_3; u_0, p u_1) = {\rm const} \cdot \prod_{k<l} \varphi(z_k,z_l) \prod_{k} \frac{1}{\thp\big(q^{1-n} u_0 z_k^{\pm 1};q\big)_{n-1}},
\end{gather*}
where $z_k = t_0 q^{x_k}$, the constant is independent of the $z_k$'s and nonzero.
\end{Lemma}

\begin{proof}
This proof is essentially the same as that of Lemma~5.3 in~\cite{War}, but is reproduced here for clarity. A first observation is that the constant in front of the right-hand side will not matter much, and because it is ignored, the proof is somewhat simpler (of course, something has to be said about it not being zero). If we denote the left-hand side by $L$ and the right-hand side by $R$, we notice both $L$ and $R$ are elliptic in the $z_k$'s (for $R$ this is a direct calculation, and for $L$ the biorthogonal functions inside the determinant are elliptic as mentioned in the previous section though one can just see this from the definition in terms of abelian interpolation functions). Fixing a variable $z_k$, we see poles for $L/R$ come from the zeros of $R$ or the poles of~$L$. For the latter, the poles are controlled by~$u_0$ but are exactly canceled by the zeros of $1/R$ appearing in the univariate product (one can see this from the definition of biorthogonal functions in terms of abelian interpolation functions). For the former the zeros of $R$ possibly leading to poles are $z_k = z_l$, $z_k = 1/z_l$ for $l \ne k$ (and $p$ shifts thereof). Plugging in $z_k = z_l$ into $L$ makes two columns the same, so $L$ vanishes. Since univariate biorthogonal functions are $BC_1$-symmetric in the variable, $L$ also vanishes if $z_k z_l = 1$ for some $l \ne k$. Hence all the poles of $L/R$ are removable, and since $L/R$ is elliptic, it must be constant. To show the constant is nonzero, we notice that the functions inside the determinant are linearly independent, so the columns of the determinant are linearly independent. This concludes the proof.
\end{proof}

\begin{Remark}We arrived at the above formula noticing the right-hand side appears in Corollary~5.4 of~\cite{War}. What appear in the determinant on the left are the abelian interpolation functions $R^*_l$ discussed in the previous section
\begin{gather*}
 \det_{1 \leq k,l \leq n} \frac{\thp\big(az_k^{\pm 1};q\big)_{n-l}}{\thp\big(bz_k^{\pm 1};q\big)_{n-l}} = a^{\binom{n}{2}} q^{\binom{n}{3}} \prod_{k<l} \varphi(z_k,z_l) \prod_{k} \frac{\thp\big(b/a,abq^{2n-2k};q\big)_{k-1}}{\thp\big(bz_k^{\pm 1};q\big)_{n-1}}.
\end{gather*}

The above formula in fact allows us to compute the constant explicitly by expanding the biorthogonal functions in terms of abelian interpolation functions. Only the leading coefficient is of interest for the determinant, and it is explicitly given in~\cite{BCn}.
\end{Remark}

To simplify notation hereinafter we let
\begin{gather*}
\Phi_l^t\big(t_0q^{s}\big):=R_l\big(t_0 q^{s};t_0\colon t_1,t_2,t_3;u_0, pu_1\big), \qquad \Psi_{l}^t\big(t_0q^{s}\big):=R_l\big(t_0 q^{s};t_0\colon t_1,t_2,t_3; pu_1, u_0\big).
\end{gather*}
The $t$ superscript for these functions stands for the fact their arguments, as it will become apparent in the next proposition, are essentially locations of the particles at time $t$. Likewise the parameters depend on~$t$ ($t_i$ and $u_j$ are implicit for~$t_i^t$, $u_j^t$ respectively; see~\eqref{relabel} and~\eqref{parmatching}). We will also denote
 \begin{gather}
\tilde{\Psi}_l\big(t_0 q^s\big) = \Psi_l\big(t_0 q^s\big) \Delta_s\big(t_0^2\,|\,q,t_0 t_1,t_0 t_2, t_0 t_3, t_0 u_0, p t_0 u_1\big)/c_l,\label{psi_tilde}
 \end{gather}
so that
 \begin{gather*}
\sum_{s \geq 0} \Phi_k\big(t_0 q^s\big) \tilde{\Psi}_l\big(t_0 q^s\big) = \delta_{k,l}.
 \end{gather*}

Thus Lemma~\ref{warnaardet} along with \eqref{tline2} and \eqref{relabel} yields the following.
\begin{Proposition}
 We have
\begin{gather*}
 \pr(X(t) = (x_1,\dots,x_N)) = {\rm const} \cdot \det_{1 \leq k,l \leq n} \Phi_{l-1}^t \big(t_0q^{x_k}\big) \cdot \det_{1 \leq k,l \leq n} \Psi_{l-1}^t\big(t_0q^{x_k}\big) \cdot \prod_{k} \Delta_{x_k} \\
\hphantom{\pr(X(t) = (x_1,\dots,x_N)) }{} = {\rm const} \cdot \det_{1 \leq k,l \leq n} \Phi_{l-1}^t \big(t_0q^{x_k}\big) \cdot \det_{1 \leq k,l \leq n} \tilde{\Psi}_{l-1}^t\big(t_0q^{x_k}\big).
\end{gather*}
\end{Proposition}

\begin{Proposition} \label{transmatrix}
We have
\begin{gather*}
 v_{t \to t-1} (k,l) := \sum_{s \geq 0} \Phi_s^{t-1}\big(t_0^{t-1} q^k\big) \tilde{\Psi}_s^{t}\big(t_0^{t} q^l\big) = \frac{1}{Z}\big(w_0' \delta_{k,l} + w_1' \delta_{k+1,l}\big)
\end{gather*}
with $w_0'$ and $w_1'$ as in Theorem~{\rm \ref{thmt-}} and $Z = \frac{1}{\thp\big(u_0^{t-1} t_0^{t-1}, u_0^{t-1} t_1^{t-1}, t_0^{t-1} t_1^{t-1}\big)}$.
\end{Proposition}

\begin{proof}We observe that
\begin{gather*}
\sum_{s \geq 0} \Phi_s^t \big(t_0^t q^k\big) \tilde{\Psi}_s^t \big(t_0^t q^l\big) = \delta_{k,l},
\end{gather*}
which expresses the relation $BA = 1$ where $A(k,l) = \Phi_k^t\big(t_0 q^l\big)$, $B(k,l) = \tilde{\Psi}_l^t\big(t_0 q^k\big)$ and we know $AB=1$ by definition (see~\eqref{psi_tilde}). We now apply the difference operator $\D\big(u_0^{t-1},t_0^{t-1},t_1^{t-1}\big)$ corresponding to the Markov transition $t \mapsto t-1$ to both sides and observe the parameters at time $t$ are the required $q$ shifts of the parameters at time $t-1$ (see~\eqref{action}). Finally on the right-hand side we have a delta function which is acted upon by the difference operator to produce the desired result (see Proposition~\ref{deltafunction}).
\end{proof}

\begin{Remark} In \cite{BG,BGR} formulas as in Proposition~\ref{transmatrix} were proven via the three term recurrence relation satisfied by the orthogonal polynomial ensembles considered ($q$-Racah and Hahn respectively). Such a relation exists for biorthogonal functions as well \cite{SZ1} and in conjunction with arguments from say \cite{BG} provides an alternative proof of Proposition~\ref{transmatrix}.
\end{Remark}

\begin{Remark} \label{remt+}
 A similar result holds if we apply the transition $t \mapsto t+1$ which corresponds to the operator $\D(u_1,t_2, t_3)$. For that though, we have to renormalize the biorthogonal functions at either $t_2$ or $t_3$ (see~\eqref{action} and~\eqref{renorm}), so the bidiagonal matrix that will appear on the right-hand side will be of the above form conjugated by two diagonal matrices (coming from the renormalization coefficients). This is an artifact of our choice of coordinates (we are counting particles going up from the bottom left edge of the hexagon).
\end{Remark}

Finally, in applying Theorem~\ref{eynardmehta} to the $t \to t-1$ Markov chain $X(t)$ we need to check that the transition probabilities have the required determinantal form. This is a consequence of Theorem~\ref{thmt-}, Lemma~\ref{warnaardet} and the following computation (the proof of which is immediate from Theorem~\ref{thmt-} and Proposition~\ref{transmatrix}). Using the notation from Theorem~\ref{thmt-} for $w_0'$, $w_1'$, $X$, $Y$)
\begin{gather*}
\det_{1 \leq k,l \leq N} (v_{t \to t-1} (x_k, y_l)) = {\rm const} \cdot \prod_{k\colon y_k=x_k-1} w'_1(x_k) \prod_{k\colon y_k=x_k} w'_0(x_k).
\end{gather*}

We thus obtain the following.
\begin{Proposition} We have
\begin{gather*}
\pr(X(t-1) = Y\,|\,X(t) = X)= {\rm const} \cdot \det_{1 \leq k,l \leq N} (v_{t \to t-1} (x_k, y_l)) \frac{\det_{1 \leq k,l \leq N} \big(\Phi_{k-1}^{t-1} \big(t_0^{t-1} q^{y_l}\big)\big) }{ \det_{1 \leq k,l \leq N} \big(\Phi_{k-1}^{t} \big(t_0^{t} q^{x_l}\big)\big) }.
\end{gather*}
\end{Proposition}

It finally leads to
\begin{Theorem} The Markov processes $t \mapsto t \pm 1$ discussed in Section~{\rm \ref{sec:dist}} meet the assumptions of Theorem~{\rm \ref{eynardmehta}} and are therefore determinantal.
\end{Theorem}

\begin{proof} This follows from all the results gathered in this Section for the $t-$ Markov chain with $f=\Phi$ and $g = \tilde{\Psi}$ in the notation of Theorem~\ref{eynardmehta}. For $t+$ see Remark~\ref{remt+}.
\end{proof}

\begin{Remark}For obtaining quantitative arctic boundary-type results about our measures, we can try to look at the asymptotics of the diagonal of the correlation kernel of the process (the probability that a particle is present at that site)
\begin{gather*}
K(x,x) = \sum_{i=0}^{S+N-1} R_i^t\big(t_0 q^x\,|\,t_0\colon t_1,t_2,t_3;u_0,pu_1\big) R_i^t\big(t_0 q^x\,|\,t_0\colon t_1,t_2,t_3;pu_1,u_0\big) \\
\hphantom{K(x,x) =}{}
 \times \Delta_x\big(t_0^2\,|\,q,t_0 t_1,t_0 t_2,t_0 t_3,t_0 u_0,p t_0 u_1\big) \\
\hphantom{K(x,x) =}{}
 \times \Delta_i\big(1/(pu_0u_1)\,|\,q,t_0t_1,t_0 t_2,t_0 t_3,1/(t_0u_0),1/(pt_0 u_1)\big),
\end{gather*}
but said asymptotics appear complicated and we do not pursue them here.
\end{Remark}

\appendix

\section{Symmetric lozenge weights} \label{app:sym_wts}

In this appendix we show how to assign $S_3$-invariant weights to the three types of rhombi (lozenges) that make up a tiling of a hexagon. We start with the $2 \times 2 \times 2$ triangle $\includegraphics[scale=0.05]{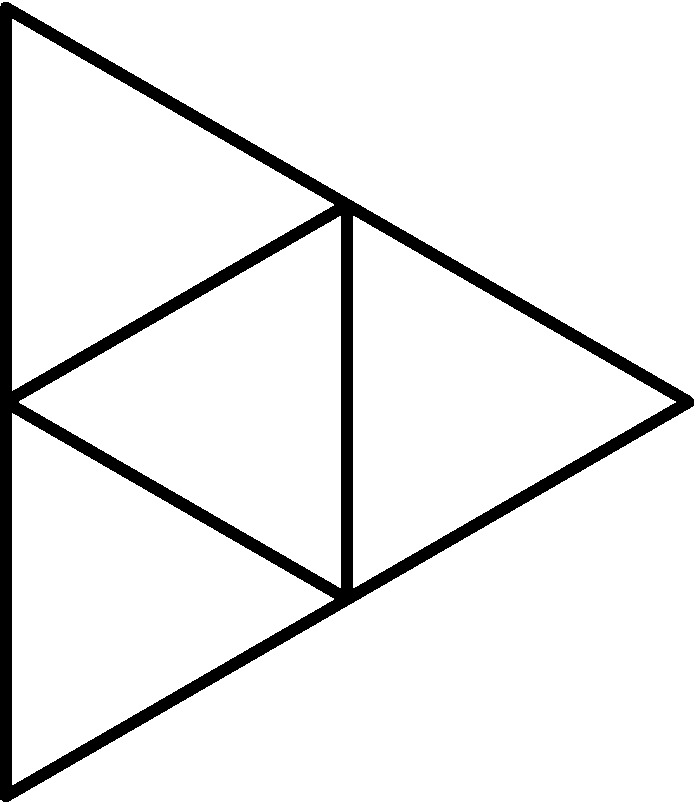}$ that contains an overlap of the three types of rhombi considered. To the three different types of rhombi in this triangle we assign labels $\tu_1$, $\tu_2$, $\tu_3$ that multiply to one $\tu_1 \tu_2 \tu_3 = 1$ using the convention depictued here: $\includegraphics[scale=0.1]{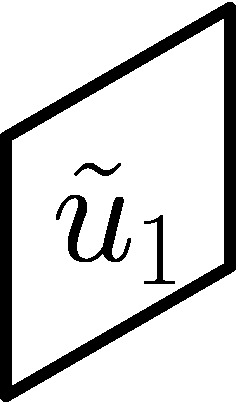} \ \ \includegraphics[scale=0.1]{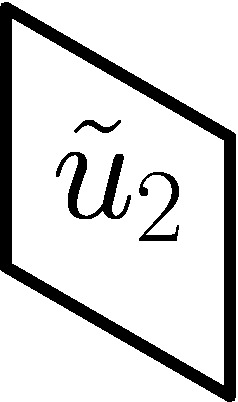} \ \ \includegraphics[scale=0.1]{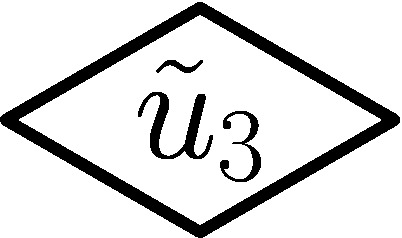}$.

\begin{figure}[t]\centering
\includegraphics[scale=0.13]{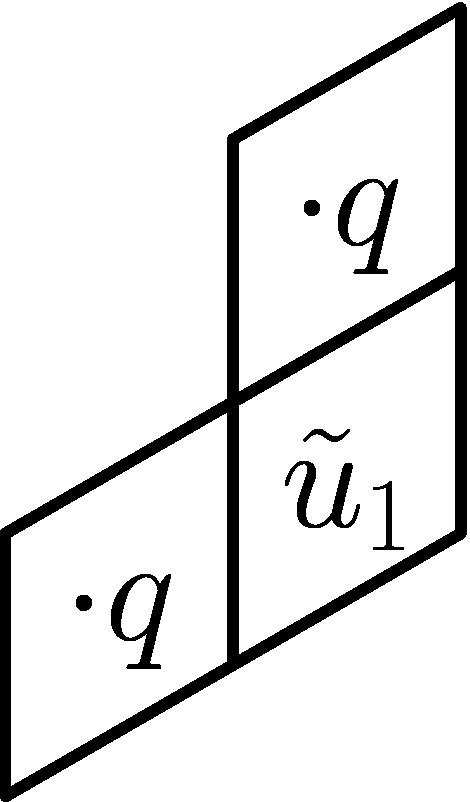} \ \ \ \ \ \ \includegraphics[scale=0.13]{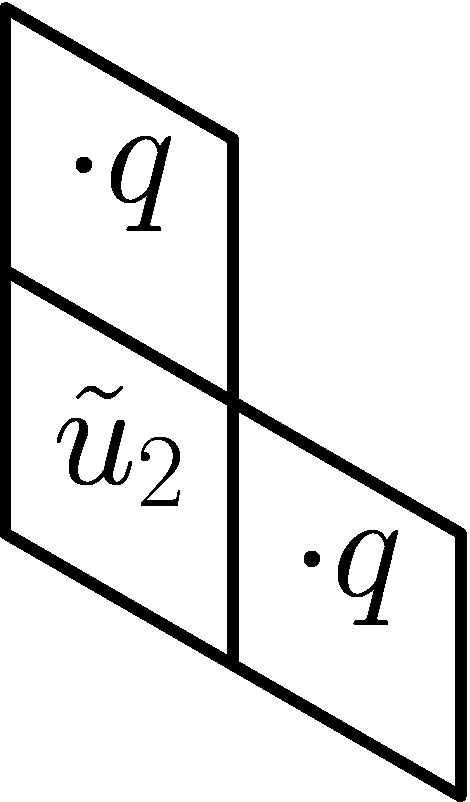} \ \ \ \ \ \ \includegraphics[scale=0.13]{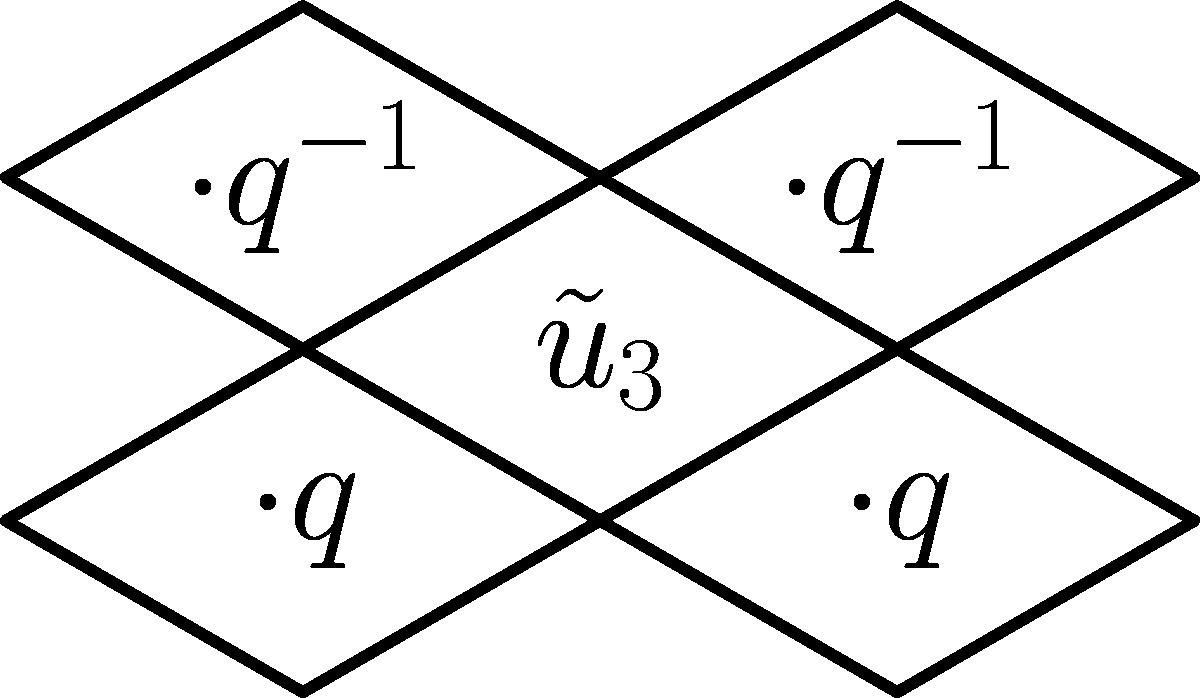}
\caption{Shifting lozenges in the triangular lattice, we multiply the labels by $q$ or $q^{-1}$ as depicted.}
\label{shifts}
\end{figure}

Each $\tu_i$ will eventually be a power of $q$ times $u_i$, see Section~\ref{prob_model}. First, we can obviously shift any such rhombus along the directions given by its edges, either upwards or downwards. If we shift the horizontal lozenge labeled $\tu_3$ upwards-right or upwards-left, the label of the new lozenge will be multiplied by~$q^{-1}$. If we shift it downwards-right/left, the label will get multiplied by $q$. Naturally, if we shift directly upwards, the label will be multiplied by~$q^{-2}$ (a composite of an upwards-right and upwards-left shift). A similar rule is used for lozenges with labels~$\tu_2$ and~$\tu_3$. The process is depicted in Fig.~\ref{shifts}. Translating any lozenge along its long diagonal does not change its label.

\begin{figure}[t]\centering
 \includegraphics[scale=0.15]{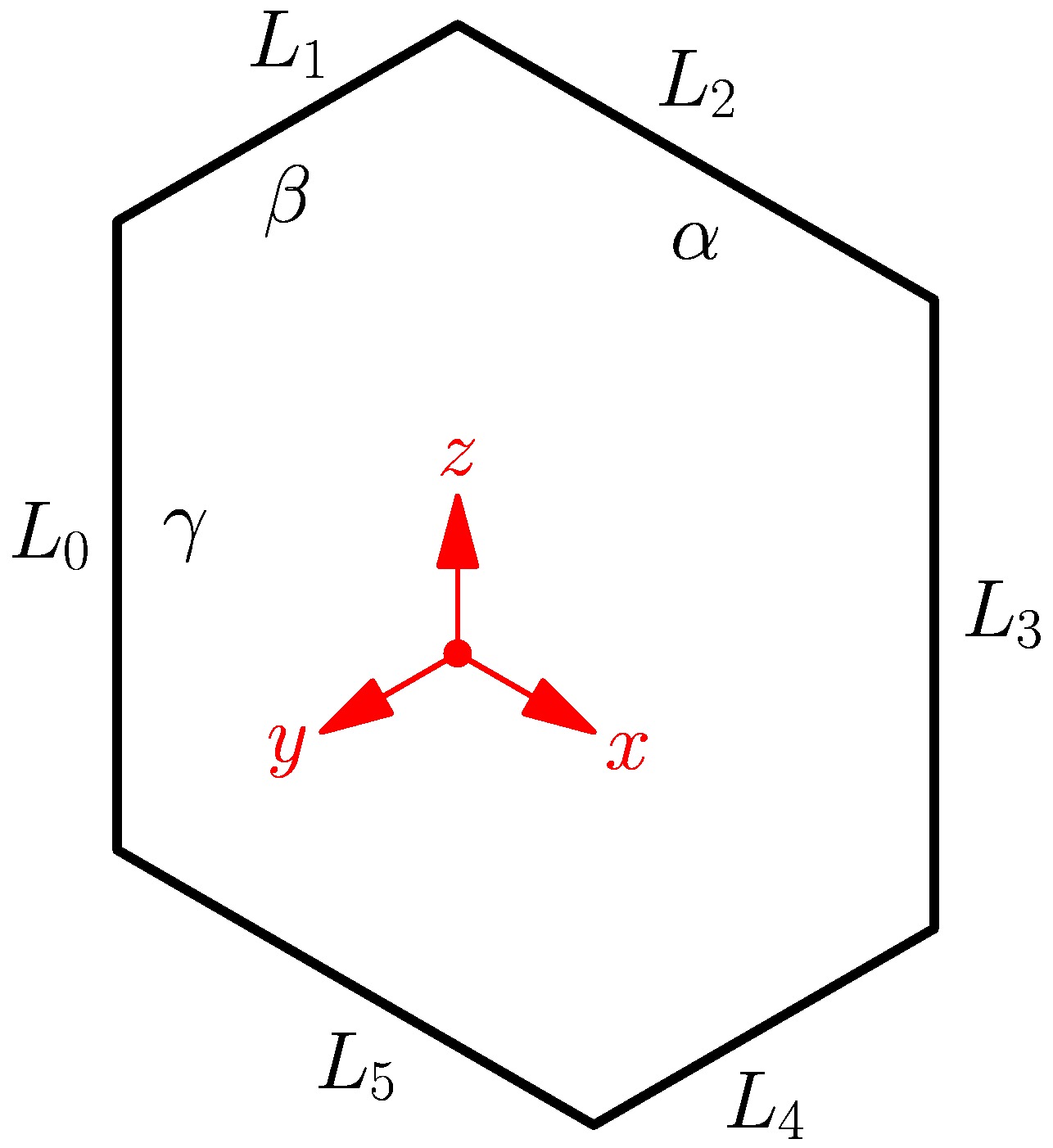}
 \caption{An $\alpha \times \beta \times \gamma$ hexagon with canonical coordinates of the edges on the outside and edge lengths on the inside.}
 \label{abc_hex_3D}
\end{figure}

To a lozenge with label $\tu_i$ ($i=1,2,3$) we assign the following weight
\begin{gather*}
 wt(\text{lozenge with label $\tu_i$}) = \tu_i^{-1/2} \thp(\tu_i), \qquad i=1,2,3,
\end{gather*}
where
\begin{gather*}
 \tu_1 = q^{y+z-2x} u_1,\qquad \tu_2 = q^{x+z-2y} u_2, \qquad \tu_3 = q^{x+y-2z} u_3, u_1 u_2 u_3 = 1,
\end{gather*}
$u_1$, $u_2$, $u_3$ are three complex numbers that multiply to one and $(x,y,z)$ is the three-dimensional coordinate of the center of a lozenge. At this point we need to fix a choice of square roots: $\sqrt{q}$, $\sqrt{u_1}$, $\sqrt{u_2}$, $\sqrt{u_3}$ such that $\sqrt{u_1} \sqrt{u_2} \sqrt{u_3} = 1$. Further note the three-dimensional coordinates are only defined up to the diagonal action of $\Z$. The three lozenges with labels~$u_i$ ($x=y=z=0$) have their centers at the hidden corner of the hexagon (the origin in Fig.~\ref{abc_hex_3D}).

This way of assigning weights is manifestly $S_3$-invariant. The weight of a tiling of the hexagon is the product of weights of all lozenges comprising the tiling. Furthermore, as a probability measure, we recover the same probability distribution as in Section~\ref{prob_model}. To see this, one can simply check the weight ratio of a full $1 \times 1 \times 1$ box to an empty $1 \times 1 \times 1$ box and observe the result is the same as in~\eqref{wtratio}.

The $S_3$-invariance can be viewed at the level of the partition function (the sum of weights of all tilings in a hexagon written in this gauge) as follows. We start with an $\alpha \times \beta \times \gamma$ hexagon. The origin is at the hidden corner of the 3D box. In canonical coordinates
\begin{gather*}
 (\tu_1, \tu_2, \tu_3) = \big(q^{y+z-2x} u_1, q^{x+z-2y} u_2, q^{x+y-2z} u_3\big)
\end{gather*}
the six bounding edges have the following equations (see Fig.~\ref{abc_hex_3D} for correspondence between edges and $L_i$'s)
\begin{alignat*}{3}
& L_0= \tu_1 / \tu_2 = q^{3 \beta} u_1/u_2, \qquad && L_1= \tu_3 / \tu_1 = q^{-3 \gamma} u_3/u_1,& \\
& L_2= \tu_2 / \tu_3 = q^{3 \gamma} u_2/u_3, \qquad && L_3= \tu_1 / \tu_2 = q^{-3 \alpha} u_1/u_2,& \\
& L_4= \tu_3 / \tu_1 = q^{3 \alpha} u_3/u_1, \qquad && L_5= \tu_2 / \tu_3 = q^{-3 \beta} u_2/u_3.&
\end{alignat*}
With these weights we have the following.

\begin{Proposition} The partition function for an $\alpha \times \beta \times \gamma$ hexagon is equal to
\begin{gather*}
 P \times \lim_{\rho \to 1} \frac{\gpqq\big(q^{1+\alpha+\beta+\gamma}\rho,q^{1+\alpha}\rho,q^{1+\beta}\rho,q^{
1+\gamma}\rho\big)}{\gpqq\big(q \rho,q^{1+\alpha+\beta}\rho,q^{1+\alpha+\gamma}\rho,q^{1+\beta+\gamma}\rho\big)} \\
 \quad{} \times \frac{\gpqq\big(q^{1-\alpha+\beta+\gamma}u_1,q^{1-\alpha}u_1,q^{1-\beta+\alpha+\gamma}u_2,q^{1-\beta}u_2,q^{1-\gamma+\alpha+\beta}u_3,
 q^{1-\gamma}u_3\big)}{\gpqq\big(q^{ 1-\alpha+\beta}u_1,q^{1-\alpha+\gamma}u_1,q^{1-\beta+\alpha}u_2, q^{1-\beta+\gamma}u_2,q^{ 1-\gamma+\alpha}u_3,q^{1-\gamma+\beta}u_3\big)} \\
 = P \times \lim_{\rho \to 1} \frac{\gpqq\big(q (L_0 L_2 L_4)^{\frac{1}{3}} \rho,q (L_0 L_4 L_5)^{\frac{1}{3}}
\rho,q (L_0 L_1 L_2)^{\frac{1}{3}} \rho, q (L_2 L_3 L_4)^{\frac{1}{3}} \rho\big)}{\gpqq\big(q \rho, q
(L_0/L_3)^{\frac{1}{3}}\rho,q (L_4/L_1)^{\frac{1}{3}}\rho,q (L_2/L_5)^{\frac{1}{3}}\rho\big)} \\
\quad {}\times \frac{\gpqq\big(q (L_0 L_2 L_3)^{\frac{1}{3}}\!, q (L_0 L_3 L_5)^{\frac{1}{3}}\!,q (L_2 L_4 L_5)^{\frac{1}{3}}\!,q (L_1 L_2 L_5 )^{\frac{1}{3}}\!,q (L_0 L_1 L_4)^{\frac{1}{3}}\!,q
(L_1 L_3 L_4)^{\frac{1}{3}}\big)}{\gpqq\big(q (L_0/L_4)^{\frac{1}{3}}\!,q (L_3/L_1)^{\frac{1}{3}}\!, q (L_5/L_3)^{\frac{1}{3}}\!,q (L_2/L_0)^{\frac{1}{3}}\!,q (L_4/L_2)^{\frac{1}{3}},q (L_1/L_5)^{\frac{1}{3}}\big)},
\end{gather*}
where
\begin{gather*}
P= q^{\alpha \beta \gamma - \frac{\alpha \beta^2+\beta \alpha^2+\alpha
\gamma^2 + \gamma \alpha^2+\beta \gamma^2+\gamma \beta^2}{4}} u_1^{-\frac{\beta
\gamma}{2}} u_2^{-\frac{\alpha \gamma}{2}} u_3^{-\frac{\alpha \beta}{2}},\\
\Gamma_{p,q,t}(x) = \prod_{i,j,k \geq 0} \big(1-p^{i+1} q^{j+1} t^{k+1} / x\big)\big(1-p^{i} q^{j} t^{k} x\big).
\end{gather*}

It is left invariant by $S_3$ permuting the coordinates $\tu_i$. Furthermore, this invariance can be expanded to the group $W(G_2) = S_3 \rtimes \Z_2 = \operatorname{Dih}_6$ $($the symmetry group of a regular hexagon$)$ with the missing involution being the transformation
\begin{gather*}
(u_1,u_2,u_3) \mapsto \left( \frac{1}{q^A u_1},\frac{1}{q^B u_2},\frac{1}{q^C u_3} \right),
\end{gather*}
where $A=-2 \alpha+\beta+\gamma$, $B=\alpha-2 \beta+\gamma$, $C=\alpha+\beta-2 \gamma$.
\end{Proposition}

\begin{proof}We start with the elliptic MacMahon identity derived in the Appendix of \cite{BGR}
\begin{gather*}
 \frac{\sum_{\text{tilings $T$}} wt(T,G)}{wt(0,G)} = q^{\alpha \beta \gamma}\prod_{\substack{ 1\leq x \leq
\alpha \\ 1\leq y \leq \beta \\ 1\leq z\leq \gamma }}
\frac{\thp\big(q^{x+y+z-1},q^{y+z-x-1}u_1,q^{x+z-y-1}u_2,q^{x+y-z-1}u_3\big)}{\thp\big(q^{
x+y+z-2},q^{y+z-x}u_1,q^{x+z-y}u_2,q^{x+y-z}u_3\big)},
\end{gather*}
where $0$ denotes the empty tiling (box) and $G$ is any gauge equivalent to the ones used in this paper (that is to say, both sides are gauge-independent). For $G$ the $S_3$-invariant gauge herein discussed, the formula for the empty tiling multiplied by the right-hand side above simplifies the partition function via straightforward computations. We arrive at the desired result using the following transformations for gamma functions
\begin{gather*}
\Gamma_{p,q} (qx) = \thp(x) \Gamma_{p,q}(x), \qquad \Gamma_{p,q,t} (tx) = \Gamma_{p,q}(x) \Gamma_{p,q,t}(x).
\end{gather*}

The limit $\rho \to 1$ is needed for technical reasons to avoid zeros of triple gamma functions.

For $S_3$-invariance, it suffices to show how edges transform under the 3-cycle $(\tu_1,\tu_2,\tu_3) \to (\tu_2,\tu_3,\tu_1)$ (a~$120^{\circ}$ clockwise rotation) and the transposition $\tu_1 \leftrightarrow \tu_2$ (a reflection in the $z$ axis). For the 3-cycle, the new edges (denoted with primes) have equations
\begin{gather*}
 L_i'=L_{i+2},
\end{gather*}
where $+2$ is taken mod~6, while for the transposition we have
\begin{gather*}
 L_0'=1/L_3, \qquad\! L_1'=1/L_2, \qquad\! L_2'=1/L_1, \qquad\! L_3'=1/L_0, \qquad\! L_4'=1/L_5, \qquad\! L_5'=1/L_4.
\end{gather*}

Both these transformations leave the partition function invariant. The extra involution giving the group $W(G_2)$ is a reflection through the centroid of the hexagon having coordinates
\begin{gather*}
\big(q^{A/2} u_1,q^{B/2} u_2,q^{C/2} u_3\big),
\end{gather*}
so that the edges transform as
\begin{gather*}
 L_i'=1/L_{i+3},
\end{gather*}
where $+3$ is taken mod 6. We look at the first form of the partition function written in the statement. We use the following two difference equations to simplify the calculations and arrive at the original form
\begin{gather*}
\gpqq(q/x) = \gpqq(pqx) = \Gamma_{q,q}(qx) \gpqq(qx), \qquad\!\! \frac{\Gamma_{q,q} \big(q^l q^mx,x\big)}{\Gamma_{q,q}\big(q^l x,q^m x\big)} = (-x)^{ml} q^{-l \binom{m}{2} - m \binom{l}{2}}. \!\!\!\!\!\!\tag*{\qed}
\end{gather*}\renewcommand{\qed}{}
\end{proof}

\section{Computer simulations} \label{app:sim}

In this section we present computer simulations of the exact sampling algorithm from Section~\ref{sec:algo}. We are (with one exception) looking at $200\times200\times200$ hexagons, and parameters are chosen so the elliptic measure sampled is positive throughout the range of the algorithm (recall that the algorithm starts with a $200 \times 400 \times 0$ box and increases $c$ while decreasing $b$ by one, until it reaches the desired size~-- after 200 iterations in our case). Under each figure we list the values of the four parameters $p$, $q$, $v_1$, $v_2$. Computations and simulations are done using double precision, the \smash{$S \mapsto S+1$} algorithm polynomial algorithm described above, and a custom program written in Java that can handle large hexagons (in excess of $N=1000$ particles) fast enough on modern CPUs.

\begin{figure}[t]\centering
 \includegraphics[scale=0.06]{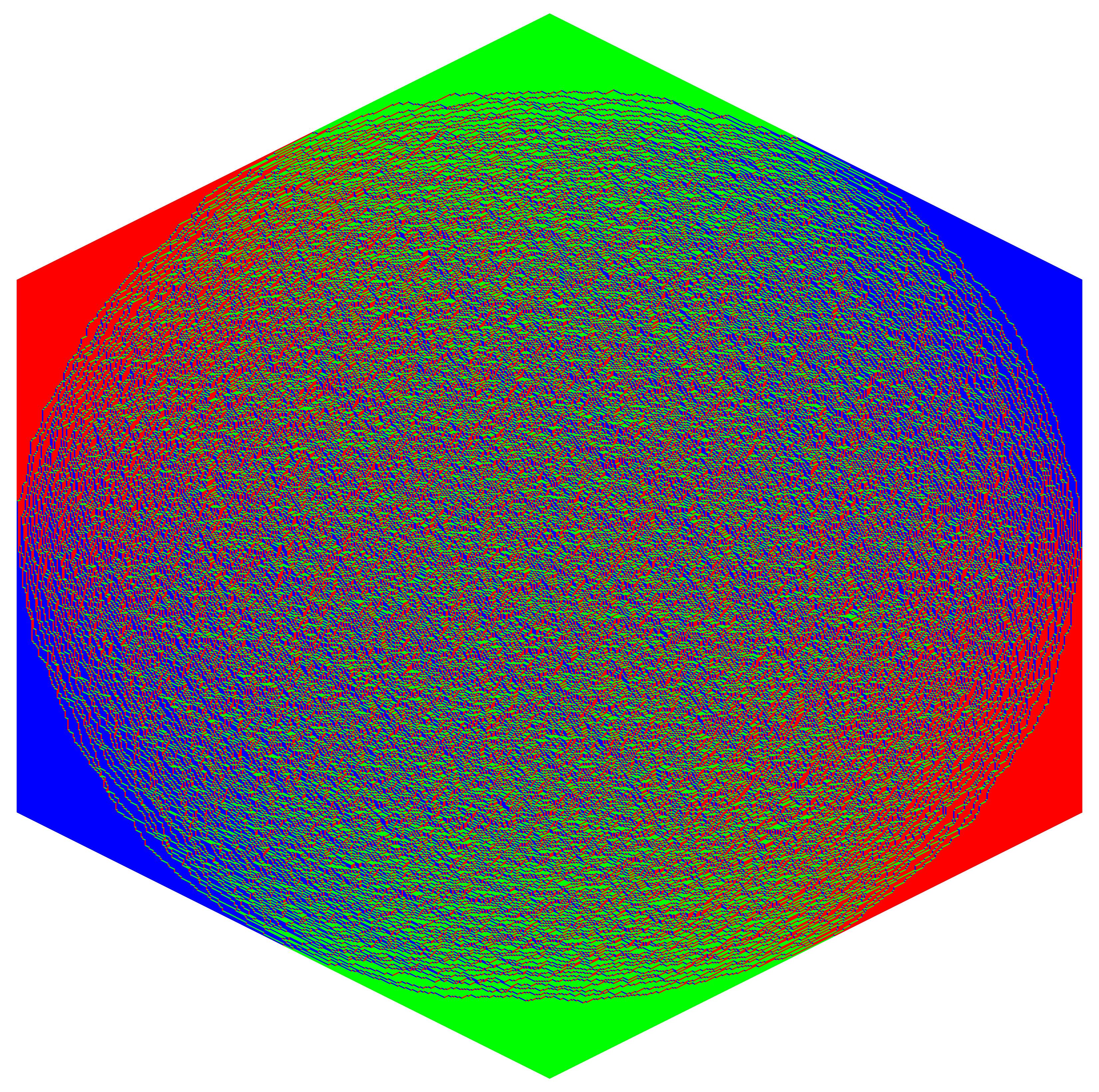}
 \caption{$p=10^{-7}$, $q=0.999999995$, $v_1 = 0.0000214$, $v_2 = 1.00675$. $400 \times 400 \times 400$. Because $q$ is \textit{very} close to~1, the limit shape looks uniform (recall that $q=1$ gives rise to the uniform measure).} \label{uniform}
 \end{figure}

\begin{figure}[t] \centering
\includegraphics[scale=0.12]{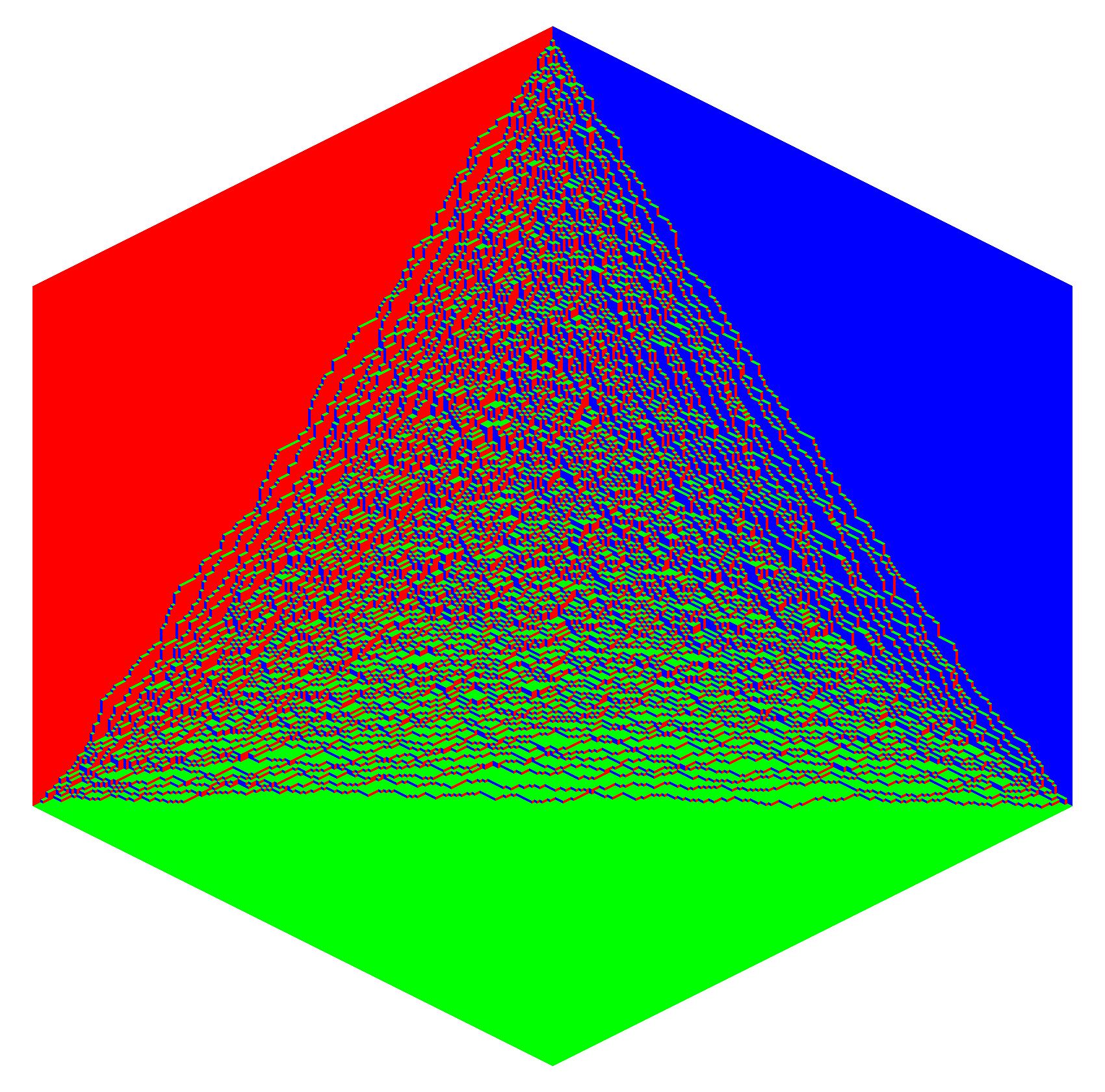}
\caption{An instance of a trinodal arctic boundary. $p=0.00186743$, $\arg q=0.000835422$, $\arg v_1 = 0.667502$, $\arg v_2 = -0.000835422$.}
\label{triangle}
\end{figure}

\begin{figure}[t] \centering
\includegraphics[scale=0.12]{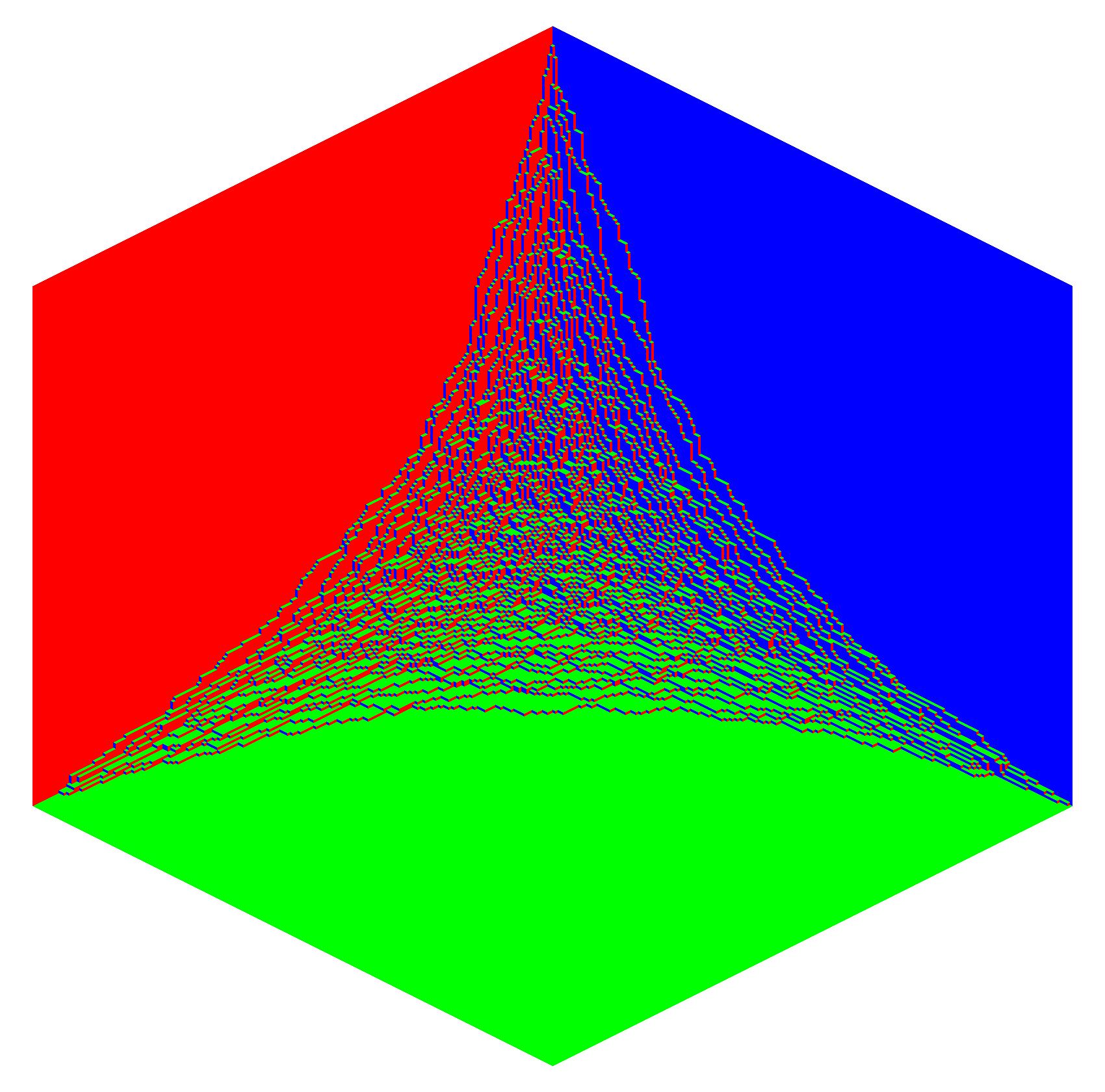}
\caption{Another instance of a trinodal arctic boundary. $p=0.2$, $\arg q=0.000835422$, $\arg v_1 = 0.667502$, $\arg v_2 = -0.000835422$. Note $p$ is larger in this case than in the previous.}\label{triangle_hyp}
\end{figure}

\begin{figure}[t] \centering
\includegraphics[scale=0.12]{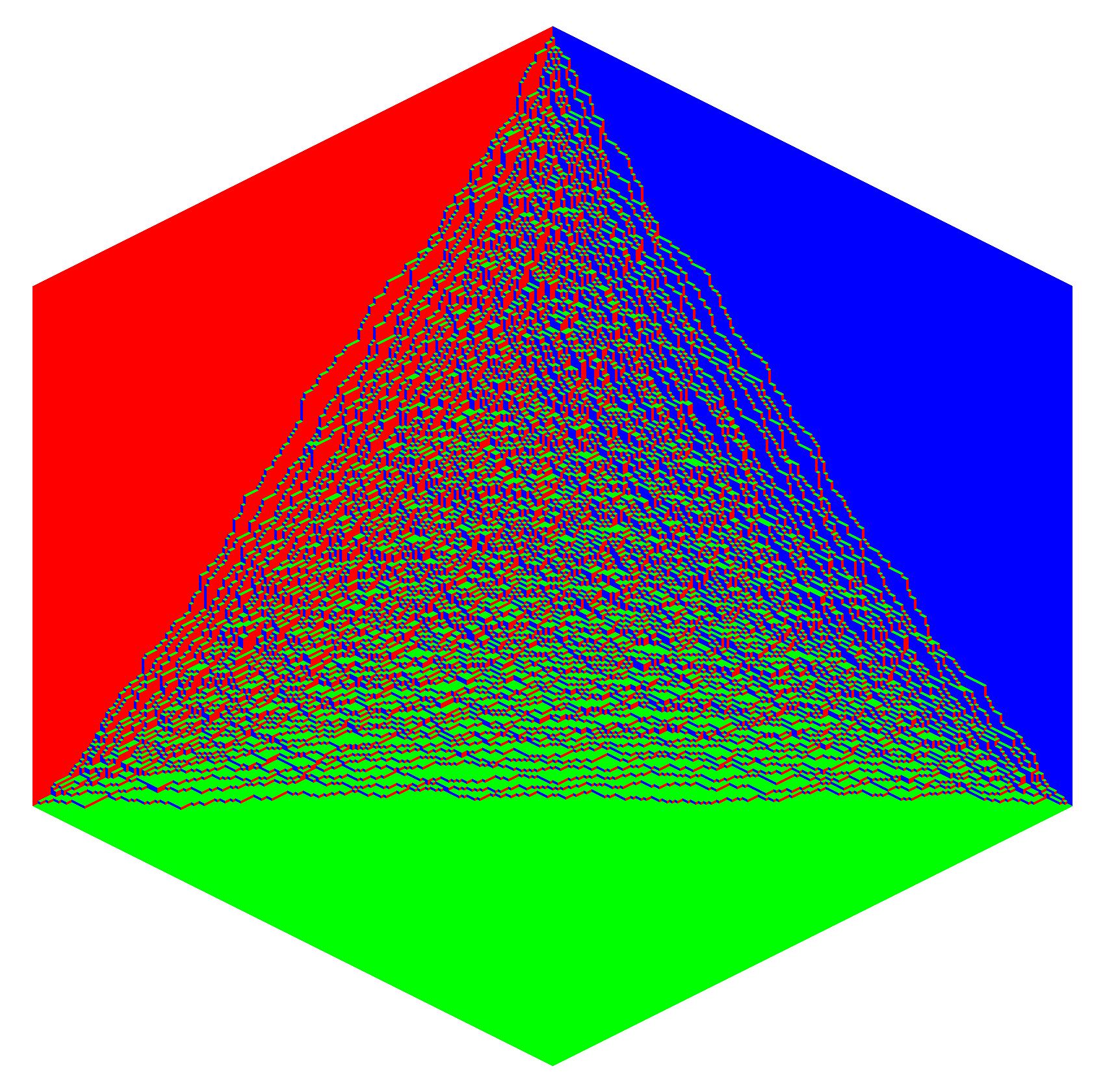}
\caption{Top level trigonometric $p=0$ case. As above, $\arg q=0.000835422$, $\arg v_1 = 0.667502$, $\arg v_2 = -0.000835422$.}
\label{top_level}
\end{figure}

In Fig.~\ref{uniform} we observe that the sample looks like one from the uniform measure with the arctic ellipse theoretically predicted in~\cite{CLP} clearly visible. Figs.~\ref{triangle} and~\ref{triangle_hyp} exhibit a new behavior for the arctic circle: the curve seems to acquire three conjectural nodes at the three vertices of the hexagon seen in the pictures. To obtain these shapes, the parameters have been tweaked so that the elliptic weight ratio vanishes (or $= \infty$) at the respective corners. In other words, the weight ratio~\eqref{wtratio} is ``barely positive'' as described in Section~\ref{positivity}. To be more precise, we have $q=e^{\frac{2 \pi i}{T-1}}$, $v_1 = q^{2T-1}$, $v_2 = 1/q$. This fixes three of the four parameters of the measure and we have the extra degree of freedom $p$ and so we obtain a one parameter family of conjecturally trinodal arctic boundaries. All simulations are taken from the trigonometric positivity case ($q$, $v_1$, $v_2$ are of unit modulus~-- see Section~\ref{positivity}). While the first arctic boundary looks like an equilateral ``flat'' triangle, the second looks ``thinner''. The change from Fig.~\ref{triangle} to~\ref{triangle_hyp} is an increase in $p$. Indeed if we increase $p$ further the triangle will get thinner and thinner, until it will degenerate into a union of the three coordinate axes as $p \to 1$. The limit $p \to 0$ yields the same ``thinning behavior'' in the real positivity case. Finally in Fig.~\ref{top_level} we exhibit a trinodal case in the top level trigonometric case $p=0$ when $q$, $v_1$, $v_2$ are of unit modulus.

\subsection*{Acknowledgements} The author would like to thank Alexei Borodin, Fokko van de Bult, Vadim Gorin, and Eric Rains for their help through numerous conversations. He is also indebted to Igor Pak and Greta Panova for putting the tiling picture herein described into perspective, and to three anonymous referees for improving the clarity of the manuscript. This article was written while the author was a~graduate student in the Department of Mathematics at the California Institute of Technology to which many remerciements are due for all its support during the five years the author spent there.

\pdfbookmark[1]{References}{ref}
\LastPageEnding

\end{document}